\documentclass[twocolumn]{aastex62}
\usepackage{savesym}
\usepackage{amsmath,amstext}
\usepackage{gensymb}
\let\splitbox\relax 
\usepackage{blindtext}
\usepackage{rotating}
\usepackage[caption=false]{subfig}
\usepackage{array}
\usepackage[export]{adjustbox}
\usepackage{float}
\usepackage{longtable}
\usepackage{CJKutf8}
\usepackage{lineno}

\def\as{\farcs}

\def\C18O{C$^{18}$O}
\def\3CO{$^{13}$CO}
\def\2CO{$^{12}$CO}

\def\N2Dp{N$_2$D$^+$}

\graphicspath{{./}{figures/}}

\submitjournal{ApJ}
\accepted{2023.2.5}

\shorttitle{ALMA Protostellar Outflow Survey}
\shortauthors{Hsieh et al.}

\begin{document}

\title{{ The Evolution of Protostellar Outflow Cavities, Kinematics, and Angular Distribution of Momentum and Energy in Orion A: Evidence for Dynamical Cores}
}

\correspondingauthor{Cheng-Han Hsieh}
\email{cheng-han.hsieh@yale.edu}

\author[0000-0003-2803-6358]{Cheng-Han Hsieh \begin{CJK}{UTF8}{bsmi}(謝承翰)\end{CJK}}
\affiliation{Department of Astronomy, Yale University, New Haven, CT 06511, USA}

\author{H{\'e}ctor G. Arce}
\affiliation{Department of Astronomy, Yale University, New Haven, CT 06511, USA}

\author{Zhi-Yun Li}
\affiliation{Department of Astronomy, University of Virginia, Charlottesville, VA 22904, USA}

\author{Michael Dunham}
\affiliation{Department of Physics, State University of New York at Fredonia, NY 14063, USA}

\author{Stella Offner}
\affiliation{Department of Astronomy, University of Texas at Austin, TX 78712-1710, USA}

\author{Ian W. Stephens}
\affiliation{Department of Earth, Environment, and Physics, Worcester State University, Worcester, MA 01602, USA}

\author{Amelia Stutz}
\affiliation{Departamento de Astronomía, Universidad de Concepción, Concepción, Chile}

\author{Tom Megeath}
\affiliation{Department of Astronomy, University of Toledo,  OH 43606, USA}

\author{Shuo Kong}
\affiliation{Department of Astronomy and Steward Observatory, University of Arizona, Tucson, AZ 85721, USA}

\author{Adele Plunkett}
\affiliation{National Radio Astronomy Observatory, 520 Edgemont Rd, Charlottesville, VA 22903, USA}

\author{John J. Tobin}
\affiliation{National Radio Astronomy Observatory, 520 Edgemont Rd, Charlottesville, VA 22903, USA}

\author{Yichen Zhang}
\affiliation{Department of Astronomy, University of Virginia, Charlottesville, VA 22904, USA}
\affiliation{RIKEN Cluster for Pioneering Research, Wako, Saitama 351-0198, Japan}

\author{Diego Mardones}
\affiliation{Departamento de Astronomía, Universidad de Chile, Región Metropolitana, Chile}

\author{Jaime E. Pineda}
\affiliation{The Center for Astrochemical Studies, Max Planck Institute for Extraterrestrial Physics, Garching bei München 85748, Germany}

\author{Thomas Stanke}
\affiliation{Max Planck Institute for Extraterrestrial Physics, Gie\ss enbachstra\ss e 1, 85748 Garching}

\author{John Carpenter}
\affiliation{Joint ALMA Observatory (JAO), Santiago, Chile}






\begin{abstract}

We present Atacama Large Millimeter/submillimeter Array observations of the $\sim$10 kAU environment surrounding 21 protostars in the Orion A molecular cloud tracing outflows. Our sample is composed of Class 0 to flat-spectrum protostars, spanning the full $\sim$1 Myr lifetime. We derive the angular distribution of outflow momentum and energy profiles and obtain the first two-dimensional instantaneous mass, momentum, and energy ejection rate maps using our new approach: the Pixel Flux-tracing Technique (PFT). Our results indicate that by the end of the protostellar phase, outflows will remove $\sim$2 to 4 M$_\odot$ from the surrounding $\sim$1 M$_\odot$ low-mass core. These high values indicate that outflows remove a significant amount of gas from their parent cores and continuous core accretion from larger scales is needed to replenish core material for star formation.  This poses serious challenges to the concept of ``cores as well-defined mass reservoir", and hence to the simplified core to star conversion prescriptions. Furthermore, we show that cavity opening angles, and momentum and energy distributions all increase with protostar evolutionary stage. This is clear evidence that even garden-variety protostellar outflows: (a) effectively inject energy and momentum into their environments on $10~$kAU scales, and (b) significantly disrupt their natal cores, ejecting a large fraction of the mass that would have otherwise fed the nascent star. Our results support the conclusion that protostellar outflows have a direct impact on how stars get their mass, and that the natal sites of individual low-mass star formation are far more dynamic than commonly accepted theoretical paradigms.

\end{abstract}


\section{Introduction} 
\label{sec:intro}


Protostellar outflows are one of the most important { ingredients of } star formation for a variety of reasons. { In particular, outflows} are believed to be one of the main mechanisms that help disperse the protostellar core and terminate the infall phase, { and assist in the removal of disk angular momentum} { which drives accretion} \citep{2013MNRAS.431.1719M,2014ApJ...784...61O,2014prpl.conf..451F,2016ARA&A..54..491B}. Through these processes, outflows impact the mass-assembly process, the final masses of stars, and the initial mass function (IMF) \citep{2007prpl.conf..245A,2007A&A...462L..17A,2021MNRAS.502.3646G,2022arXiv220310068P}. 

Comparing the core-mass function (CMF) with the IMF { suggests} that the mass of stars strongly depends on the mass of their parent cores with a core to star efficiency of about 30\% (e.g., \citet{1998A&A...336..150M,2001A&A...372L..41M,1998ApJ...508L..91T,2006A&A...447..609S,2007A&A...462L..17A,2008ApJ...684.1240E,2010A&A...518L.102A,2015A&A...584A..91K,2021arXiv211208182M} and references therein).  If gas dispersal by outflows is the main cause of this relatively low efficiency, then this would imply that outflows are capable of dispersing about 70\% of the cores' mass. Some observations have shown outflows have enough energy and momentum to disperse the material in the surrounding parent core within 1 Myr \citep{2014ApJ...783...29D,2016ApJ...832..158Z}. In addition, { a number of} studies have found { that the} outflow opening angle { broadens} with time \citep{2006ApJ...646.1070A,2014ApJ...783....6V,2017AJ....153..173H}, thereby facilitating the gas clearing process in the surrounding envelope. 

On the other hand, other studies have shown that outflow mass-loss rates are too low to be the main agent of core dispersal \citep{2010MNRAS.408.1516C,2007A&A...472..187H}. A recent study of outflow cavities using HST NIR images of the scattered light around protostars  in Orion found no evidence for the widening of outflow cavities as protostars evolve \citep{2021ApJ...911..153H}. Their results indicate that gas dispersal by outflows cannot be the main process that produces the { low star formation efficiency of 30\%.}  
However, there are no other known mechanisms that can explain the low star formation efficiency in cores. 

Resolving this problem, { requires characterizing} outflows based on their structure and morphology (e.g., opening angle), as well as other physical properties (e.g., mass, momentum, and energy loss rate) and linking them { to their evolutionary status. Although outflows have been studied since the  early 1980s \citep{1980IAUS...87..173S,1981ApJ...245L..19S}, they} remain complicated systems for which accurate measurements of their physical properties and their evolutionary stage are hard to ascertain without high angular resolution, high dynamic range observations of sizeable samples of sources.



{ 
In this paper,
to address the problems above, we use a method to obtain accurate outflow properties. This  involves a new method that we call the Pixel Flux-tracing Technique as well as a modified version of an annulus method developed by \citet{1996A&A...311..858B}.}
We also use our sensitive observations to derive outflow momentum and energy opening angles. With these { information} we establish a powerful new technique for observing outflow-core interaction  that
{ complements traditional opening angles in determining} how much core material is impacted by the outflow. 



{ To study the protostellar outflow-core interactions, we use the Atacama Large Millimeter/submillimeter Array (ALMA) to survey protostars at different evolutionary stage in the Orion A Molecular Cloud.} The Orion A Giant Molecular Cloud (GMC) is the closest star-forming region { that} harbors both high and low-mass star formation. This GMC is at a distance of about 400\,pc and has a gas mass of about $10^5\,M_\odot$ \citep{2008hsf1.book..459B}. 
The cloud and the young stellar sources in it have been studied extensively with large, multi-wavelength, surveys covering multiple scales, 
(e.g.\ \citealt{2012AJ....144..192M,2013ApJ...767...36S,2016AJ....151....5M,2016ApJS..224....5F,2018ApJS..236...25K,2020ApJ...890..130T,2020ApJ...905..119F,2022A&A...658A.178S}). 
The availability of these rich datasets makes { Orion A} a perfect target to study outflow evolution and its impact on protostellar { cores}. 

{ In} Section \ref{sec:sample} we { describe} the sample selection for this study. In Section \ref{sec:obs} we { describe} the observational data, the calibration, and the imaging process. In Section \ref{sec:results} we describe the methods we use to obtain outflow properties and  show { the results of our observations.} In Section \ref{sec:discussion}, we analyze and discuss the evolution of the outflow properties (opening angle, mass, momentum, and energy ejection rate) and the impact on cores. In Section \ref{sec:conclusion} we { summarize} the main findings and give our conclusions.


\section{Sample selection} 
\label{sec:sample}

\begin{figure*}[tbh]
\centering
\includegraphics[width=\hsize]{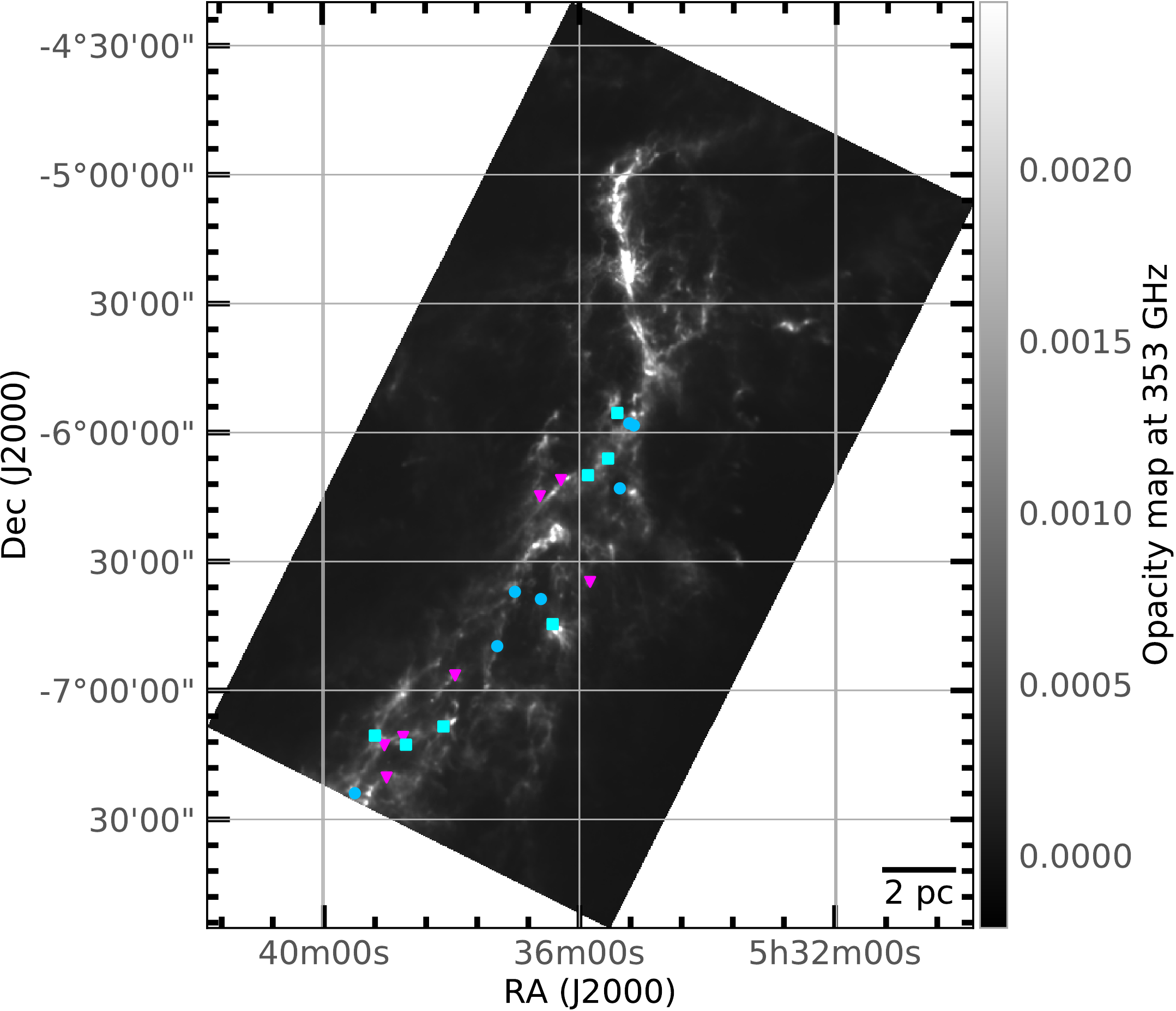}
\caption{ Sources from this survey over-plotted on the Orion A Herschel-Planck dust opacity map at 353 GHz \citep{2014A&A...566A..45L}. The blue circles, red triangles and green squares mark the position of Class 0, Class I and flat-spectrum sources, respectively. 
}
\label{fig:source_position}
\end{figure*}

Our survey consists of 21 protostellar sources evenly distributed in 3 different evolutionary stages: Class 0, Class I, and flat-spectrum. The sources were all observed by the Herschel Orion Protostar Survey (HOPS), from which { we} obtain { the} derived properties of the protostars, including evolutionary class, bolometric temperature and luminosity  \citep{2016ApJS..224....5F}. 
All sources are located in the northern part of the L1641 region,  a relatively low-density star-forming region south of the Orion Nebula Cluster, that is far  (at least $\sim$8\,pc) { from} high mass protostars and photodissociation regions. 
{ The positions of the sources are over-plotted on the Herschel-Planck dust opacity map at 353 GHz from \citet{2014A&A...566A..45L} in \autoref{fig:source_position}. The blue circles, red triangles and green squares mark the position of Class 0, Class I and flat-spectrum sources.} 

We chose targets { that} were thought to have no binary or multiple companions by the time we submitted the observing proposal, and to { be} relatively isolated with no neighboring young stellar objects closer than 0.12\,pc (60\as0), in order to avoid possible contamination from nearby outflows. Only one source in our sample (HOPS-150) was recently discovered to be part of a binary system { with a projected binary separation of $\sim 3\as$ }    \citep[$\sim 1200$~au][]{2020ApJ...890..130T,2022ApJ...925...39T}.
All sources have been observed in the near-infrared by the VISTA Orion A Survey and at  $1.6-870\mu$\,m with the Hubble, Spitzer, and Herschel space telescopes  \citep[e.g.,][]{2012AJ....144..192M,2016ApJS..224....5F,2016ApJ...821...52K}. { In addition, the sample spans a limited range in $L_{bol}$ ($\sim 0.4 - 15$) and this range is comparable in all three evolutionary classes.}
{ A} summary of { the} source properties { is} shown in \autoref{table:1}.

\begin{table*}
\setlength{\tabcolsep}{6pt} 
\caption{Summary of source properties}             
\label{table:1}      
\centering                          
\begin{tabular}{c c c c c c c c c c c c}        
\hline\hline                 
Source  & RA & DEC &  Distance\tablenotemark{a}  & $V_{syst}$\tablenotemark{b}& $T_{bol}$\tablenotemark{c} &Class\tablenotemark{c} & $L_{bol}$\tablenotemark{c} & { ALMA Core} & { JCMT Core}\\    
&(J2000) &(J2000) & (pc) & (km s$^{-1}$) & (K) &  &(L$_\odot$) & { Mass [M$_\odot$]}\tablenotemark{d} & { Mass [M$_\odot$]}\tablenotemark{e} \\ 
\hline                        
HOPS-10  & 05:35:09 & -05:58:26 & 388.2 & 8.19 & 46.2 & 0 & 3.33 & 1.33 & 0.90\\ 
\hline                                   
HOPS-11  & 05:35:13 & -05:57:57 & 388.3 & 7.18& 48.8 & 0 & 9.0 & 1.09 & 1.34\\ 
\hline
HOPS-13  & 05:35:24 & -05:55:33 & 388.7 & 6.00& 383.6 & flat & 1.15 & 0.27 & 0.50\\
\hline
HOPS-127  & 05:39:00 & -07:20:22 & 394.5 & 4.68 & 133.3 & I & 0.39 & 0.27 & 0.26\\
\hline
HOPS-129  & 05:39:11 & -07:10:34 & 393.8 & 3.40 & 191.3 & flat & 1.67 & 0.57 & 0.42\\
\hline
HOPS-130  & 05:39:02 & -07:12:52 & 393.2 & 3.46 & 156.7 & I & 1.48 & 0.18 & 0.26\\
\hline
HOPS-134  & 05:38:42 & -07:12:43 & 391.2 & 6.05& 781.9 & flat & 7.77 & 0.32 & 0.19\\
\hline
HOPS-135  & 05:38:45 & -07:10:56 & 391.1 & 4.65& 130.3 & I & 1.14 & 0.24 & 0.47\\
\hline
HOPS-150A  & 05:38:07 & -07:08:29 & 388.5 & 3.81& 245.2 & flat & 3.77 & 0.46 & 0.10\\
\hline
HOPS-150B\tablenotemark{f}  & 05:38:07 & -07:08:32 & 388.5 & 3.81& 245.2 & flat & 3.77 & 0.46 & 0.10\\
\hline
HOPS-157  & 05:37:56 & -06:56:39 & 387.7 & 5.35& 77.6 & I & 3.82 & 0.38 & 0.57\\
\hline
HOPS-164  & 05:37:00 & -06:37:10 & 385.0 & 5.84 & 50.0 & 0 & 0.58 & 1.26 & 0.84\\
\hline
HOPS-166  & 05:36:25 & -06:44:41 & 383.5 & 8.78& 457.1 & flat & 15.47 & 2.33 & 0.73\\
\hline
HOPS-169  & 05:36:36 & -06:38:54 & 384.0 & 7.01& 32.5 & 0 & 3.91 & 0.43 & 1.68\\
\hline
HOPS-177  & 05:35:50 & -06:34:53 & 383.6 & 8.97& 84.7 & I & 0.43 & 0.92 & 1.11\\
\hline
HOPS-185  & 05:36:36 & -06:14:58 & 386.1 & 7.59& 96.9 & I & 1.04 & 0.95 & 0.66\\
\hline
HOPS-191  & 05:36:17 & -06:11:11 & 386.6 & 8.11& 196.7 & I & 0.58 & 0.49 & 0.11\\
\hline
HOPS-194  & 05:35:52 & -06:10:01 & 386.6 & 8.53& 645.0 & flat & 12.72 & 1.12 & 0.64\\
\hline
HOPS-198  & 05:35:22 & -06:13:06 & 386.0 & 5.56& 61.4 & 0 & 0.85 & 0.44 & 0.68\\
\hline
HOPS-200  & 05:35:33 & -06:06:09 & 387.1 & 8.55& 244.4 & flat & 0.29 & 0.69 & 0.10\\
\hline
HOPS-355  & 05:37:17 & -06:49:49 & 385.8 & 6.61 & 44.9 & 0 & 1.18 & 0.18 & 0.77\\
\hline
HOPS-408  & 05:39:30 & -07:23:59 & 398.8 & 3.73 & 37.9 & 0 & 0.52 & 0.47 & 0.58\\
\hline

\end{tabular}
\tablenotetext{a}{Distances are from \citet{2020ApJ...890..130T}, who use Gaia data of YSOs near each protostar to estimate their distances.}\tablenotetext{b}{We measured the system velocity by fitting a Gaussian to our C$^{18}$O data. For details  see { Section \ref{3.11}.}}\tablenotetext{c}{Results come from integration over the Spectral Energy Distributions (SEDs) in \citet{2016ApJS..224....5F}.}\tablenotemark{d}{ The ALMA core mass are measured from our Cycle 6 ALMA C$^{18}$O data. The selected core size is $\sim$6500-7500\,au.}\tablenotetext{e}{ The JCMT core mass are from JCMT 850 $\mu$m dust continuum data. The cores are identified by \citet{2016ApJ...833...44L} and have a typical core size of $\sim$6000\,au.}
\tablenotetext{f}{HOPS-150 was revealed to be a binary system, with a separation of $\sim  120$~au, by  \citet{2020ApJ...890..130T,2022ApJ...925...39T}. SED measurements of HOPS-150 by \citet{2016ApJS..224....5F} were centered on HOPS-150A. However, the outflow cavity seen in HST IR images shown in \citet{2021ApJ...911..153H} is clearly centered on HOPS-150B, but contain the combined fluxes of both sources. The molecular outflow data for HOPS-150 discussed in this paper,  likely arises from  HOPS-150B instead of HOPS-150A.}
\end{table*}

\section{Observations} 
\label{sec:obs}

The results presented here come from Cycle 6 
ALMA multi-line observations of the environment around protostars in the the Orion A cloud 
(Project ID: 2018.1.00744.S, PI: Arce, H.). 
The observations were conducted using ALMA Band 6 which simultaneously observed the 1.29 mm dust continuum emission and the following six molecular lines which trace different density and kinematic regimes:  
$^{12}$CO(2-1), $^{13}$CO(2-1), C$^{18}$O(2-1), H$_2$CO($3_{0,3}$-$2_{0,2}$), SiO(5-4), and \N2Dp (3-2). 

In this paper, we concentrate on the $^{12}$CO(2-1), $^{13}$CO(2-1), and C$^{18}$O(2-1) lines, which have rest frequencies of { 230.53800 GHz, 220.39868 GHz, and 219.56035 GHz}, respectively. 
{ After H$_2$, $^{12}$CO is the most abundant molecule in molecular clouds, and is typically the most easily observed since H$_2$ lacks dipole rotational transitions.}
Together with the less abundant $^{13}$CO and C$^{18}$O  they trace  regions across a wide range of (column) densities  ($5\le A_v \le 15$) \citep{2021A&A...645A..27G}. 
{ We use} these three CO isotopologues to probe the outflow properties. 
The other lines we observed, which trace the infalling gas, energetic jets, and cold and warm core regions will be presented in forthcoming papers.

Each source in our sample was observed using a 12m array { 7-pointing} mosaic (in the { C43-2} configuration), and a { three-pointing} mosaic using the 7m array.  
This allowed us to cover the environment surrounding each source over a region of about 0.1 pc or { $\sim$50\as0}  and ensured the maps had an approximately uniform sensitivity within { the} primary beam of the 12m array at the observed frequencies, about 25\arcsec ($\sim 0.05$ pc), centered at the source. 

The 12\,m array observations were made between 2018 December 21,  and 2019 January 15, and consisted of 15 executions, with an on-source integration of 35.4\,minutes  
for each of the { 7-pointing} mosaics used to cover each source.  J0423-0120 was used for bandpass and flux calibration and J0542-0913 was used for phase calibration. Between 45 to 50 { antennas} were used on different observation dates, which included baselines ranging from about 15.1\,m\, to 500.2\,m, which sample scales between 1\as1 to 16\as2 (440\,au to 6480\,au).
The 7m array observations, which were conducted using 70 executions between 2018 October 3, and 2018 November 23, used J0522-3627 for bandpass and flux calibration and J0501-0159 for the phase calibration. 
The average 7m array integration time for each { 3-pointing} mosaic around each source in our sample was 2 hours and 20.5 minutes.
These observations used between 9 to 12 antennas that included baselines ranging from approximately 8.9\,m to 48.9\,m , sampling scales between 5\as5 to 29\as0 (2200\,au to 11,600\,au).

We also obtained Total Power (TP) array data for our observations to trace the large-scale emission structure and recover all flux from the molecular line emission toward our sources of interest.
The TP observations were conducted between 2018 September 19 and 2019 January 8, and provided 98\as0 by 96\as0 on-the-fly maps centered on each source. The on-source time for each of these maps was 6 hours and 51.6 minutes.

The pipeline in the Common Astronomy Software Applications (CASA) version 5.4.0-70 was used to calibrate the ALMA data. Additional data flagging was performed by the North American ALMA Science Center to remove artifacts in the 12\,m continuum data caused by insufficient sampling at the shortest baselines. 

We used the CASA task \textit{tclean} to 
image the combined 12m and 7m array interferometer data.
We ran this task with automasking, which automatically defines different cleaning regions for each channel, and may vary for different iterations.
The deconvolver option was set to hogbom, that proved to be more stable than the multi-scale option which produced artifacts in many cases. We also set the velocity resolution parameter to -1, which ensures the output cleaned maps { have} the same velocity resolution as the original uv data: 0.079 km~s$^{-1}$ for $^{12}$CO (2-1), and 0.083 km~s$^{-1}$ for $^{13}$CO and C$^{18}$O (2-1). We set the cyclefactor parameter to 2.0 to terminate the minor cycle (deconvolving the observed image from the telescope beam) faster and trigger the next major cycle. The criteria for stopping a minor cycle in \textit{tclean}  is set by the product of peak residual level, maximum point spread function sidelobe level, and the value of the cyclefactor parameter. 
Using the default value for the cyclefactor parameter (1.0) caused the minor cycle to generate artifacts at the low sensitivity edge of our mosaics. { Such} artifacts did not appear (i.e., the algorithm would be stable) only when both the ``cyclefactor" was set to a value equal to or greater than 2, and ``automasking" was turned on. For imaging, we used natural weighting to maximize  the signal-to-noise ratio while maintaining a similar beam size of about 1.1\arcsec \/  for all line maps. { Primary beam correction was applied to each map resulting in  higher noise at the edge of the mosaic maps.}
We then used the CASA task \textit{Feather} to combine the array data with the total power data. The resulting synthesized beam size, velocity resolution, and the average rms noise level  of the 
CO, $^{13}$CO and C$^{18}$O maps are shown in \autoref{table:2}. 




\begin{center}
\begin{longtable*}{c c c c c c c c }
\caption{Summary of ALMA data used in analysis}\\ %
\label{table:2}
&  &    & \\
\hline
\hline
&  &    & Beam   & Average RMS  & RMS in  \\
Source  & Line  &Beam Size & Position    & in  field of view\tablenotemark{a} &  central region\tablenotemark{b}\\
&  &(\as) & Angle ($^{\circ}$) &(mJy\,beam$^{-1}$)& (mJy\,beam$^{-1}$)\\
\hline
\endfirsthead
\multicolumn{6}{c}%
{\tablename\ \thetable\ -- \textit{Continued from previous page}} \\
\hline
&  &    & Beam   & Average RMS  & RMS in \\
Source  & Line  &Beam Size & Position    & in field of view\tablenotemark{a} & central region\tablenotemark{b}\\
&  &(\as) & Angle ($^{\circ}$) &(mJy\,beam$^{-1}$)& (mJy\,beam$^{-1}$)\\
\hline
\endhead
\hline \multicolumn{4}{r}{\textit{Continued on next page}} \\
\endfoot
\hline
\endlastfoot
\hline                        
  & $^{12}$CO 2-1  & $1\as26 \times 1\as00$ & 82.1  & 9.7 &6.8\\
HOPS-10  & $^{13}$CO 2-1  & $1\as30 \times 1\as$04 & 81.9  & 13.0 &9.2 \\
  &C$^{18}$O 2-1   & $1\as31 \times 1\as04$ & 83.7 &  10.1 &7.1\\   
\hline                                   
  & $^{12}$CO 2-1  & $1\as26 \times 0\as99$ & 79.3  & 4.0 &7.2\\
HOPS-11  & $^{13}$CO 2-1  & $1\as30 \times 1\as04$ & 80.7  & 13.0 &9.1\\
  &C$^{18}$O 2-1  & $1\as31 \times 1\as04$ & 82.4 & 10.2 &7.1\\   
\hline     
  & $^{12}$CO 2-1 & $1\as26 \times 0\as99$ & 79.9   & 12.3 &7.4\\
HOPS-13  & $^{13}$CO 2-1  & $1\as30 \times 1\as04$ &80.7  & 13.0 &9.3\\
  &C$^{18}$O 2-1  & $1\as30 \times 1\as04$ & 82.4 &  10.2 &7.0\\   
\hline   
  & $^{12}$CO 2-1 & $1\as25 \times 0\as98$ & -87.5  & 10.3 &7.2\\
HOPS-127  & $^{13}$CO 2-1 & $1\as29 \times 1\as03$ & 89.9 & 13.6 &9.5\\
  &C$^{18}$O 2-1  & $1\as30 \times 1\as02$ & -88.9 & 11.8 &9.4\\   
\hline   
  & $^{12}$CO 2-1 & $1\as24 \times 0\as98$ & -86.3 & 7.1 &5.0\\
HOPS-129  & $^{13}$CO 2-1 & $1\as29 \times 1\as03$ & -89.6 & 13.5 &9.4\\
  &C$^{18}$O 2-1  & $1\as30 \times 1\as02$ & -88.5 & 10.5 &7.3\\   
\hline   
  & $^{12}$CO 2-1 & $1\as25 \times 0\as98$  & 88.6 & 7.0 & 4.9\\
HOPS-130  & $^{13}$CO 2-1 & $1\as29 \times 1\as03$ & 86.1 & 13.3 & 9.3\\
  &C$^{18}$O 2-1  & $1\as30 \times 1\as03$ & 87.1 & 10.3 & 7.2\\   
\hline   
  & $^{12}$CO 2-1 & $1\as26 \times 0\as99$ & 89.6 & 7.3 & 5.3\\
HOPS-134  & $^{13}$CO 2-1 & $1\as30 \times 1\as04$ & 87.3 & 13.2 & 9.3\\
  &C$^{18}$O 2-1 & $1\as31 \times 1\as03$ & 88.5 & 10.5 & 7.3\\
\hline   
  & $^{12}$CO 2-1 & $1\as26 \times 0\as99$ & -89.7 & 7.1 &5.0\\
HOPS-135  & $^{13}$CO 2-1 & $1\as30 \times 1\as03$ & 88.2 & 13.6 &9.4\\
  &C$^{18}$O 2-1  & $1\as31 \times 1\as03$ & 89.5 &  13.6 & 11.8\\   
\hline   
  & $^{12}$CO 2-1 & $1\as26 \times 0\as99$ & -89.0 & 10.6 & 7.1\\
HOPS-150  & $^{13}$CO 2-1 & $1\as29 \times 1\as03$ & 89.0 & 13.6 & 9.5\\
  &C$^{18}$O 2-1  & $1\as30 \times 1\as03$ & -89.7 & 10.6 & 7.4\\   
\hline   
  & $^{12}$CO 2-1 & $1\as24 \times 0\as98$ & 84.7 & 6.9 & 4.9\\
HOPS-157  & $^{13}$CO 2-1 & $1\as29 \times 1\as03$ & 84.6 & 13.5 & 9.7\\
  &C$^{18}$O 2-1   & $1\as30 \times 1\as03$ & 86.0 & 10.6 & 7.5\\   
\hline  
  & $^{12}$CO 2-1 & $1\as25 \times 0\as98$ & 84.2 & 9.5 & 7.4\\
HOPS-164  & $^{13}$CO 2-1 & $1\as29 \times 1\as03$ & 84.4 & 14.4 & 12.1\\
  &C$^{18}$O 2-1   & $1\as30 \times 1\as03$ & 85.8 &  10.3 & 7.3\\   
\hline   
  & $^{12}$CO 2-1 & $1\as25 \times 0\as98$ & 86.3 & 9.2 & 6.5\\
HOPS-166  & $^{13}$CO 2-1 & $1\as29 \times 1\as03$ & 85.3 & 13.5 & 9.6\\
  &C$^{18}$O 2-1 & $1\as30 \times 1\as03$ & 86.7 & 10.4 & 7.0\\   
\hline 
& $^{12}$CO 2-1 & $1\as25 \times 0\as99$  & 88.5  &  9.9 & 7.1\\
HOPS-169  & $^{13}$CO 2-1 & $1\as29 \times 1\as03$ & 85.9   & 13.2 & 9.3\\
  &C$^{18}$O 2-1  & $1\as30 \times 1\as03$ & 87.1 &  4.1 & 2.9\\   
\hline 
  & $^{12}$CO 2-1 & $1\as26 \times 0\as99$ & 83.6  & 6.8 & 4.8\\
HOPS-177  & $^{13}$CO 2-1  & $1\as29 \times 1\as03$ & 84.1  & 13.0 & 9.1\\
  &C$^{18}$O 2-1  & $1\as30 \times 1\as03$ & 85.3 & 10.2 & 7.2\\   
\hline 
  & $^{12}$CO 2-1 & $1\as26 \times 1\as00$ & 82.1  & 6.7 & 4.8\\
HOPS-185  & $^{13}$CO 2-1 & $1\as30 \times 1\as04$ &82.3  & 13.1 & 9.1\\
  &C$^{18}$O 2-1 & $1\as31 \times 1\as04$ & 83.9 & 10.1 & 7.2\\   
\hline 
  & $^{12}$CO 2-1 & $1\as26 \times 0\as99$ & 79.5  & 6.8 & 4.8\\
HOPS-191  & $^{13}$CO 2-1 & $1\as30 \times 1\as04$ & 80.9   & 13.1 & 9.2\\
  &C$^{18}$O 2-1 & $1\as31 \times 1\as04$ & 82.6 & 10.2  & 7.1\\   
\hline 
  & $^{12}$CO 2-1 & $1\as25 \times 0\as99$ & 83.9  & 10.1 & 7.1\\
HOPS-194  & $^{13}$CO 2-1 & $1\as29 \times 1\as03$ & 84.3  & 13.1 & 9.3\\
  &C$^{18}$O 2-1  & $1\as30 \times 1\as03$ & 85.5 & 10.2 & 7.1\\   
\hline 
  & $^{12}$CO 2-1 & $1\as26 \times 0\as99$ & 83.3 & 6.7 & 4.7\\
HOPS-198  & $^{13}$CO 2-1 & $1\as30 \times 1\as03$ & 83.7 & 13.2 & 9.6\\
  &C$^{18}$O 2-1  & $1\as31 \times 1\as04$ & 85.0 & 10.2 & 7.1\\   
\hline 
  & $^{12}$CO 2-1 & $1\as26 \times 0\as99$ & 82.9  & 6.7 & 4.7\\
HOPS-200  & $^{13}$CO 2-1 & $1\as30 \times 1\as04$ & 83.0  & 13.1 & 9.2\\
  &C$^{18}$O 2-1 & $1\as31 \times 1\as04$ & 84.5 & 10.2 & 7.1\\   
\hline 
  & $^{12}$CO 2-1 & $1\as26 \times 0\as99$ & -89.8  & 7.0 & 4.9\\
HOPS-355  & $^{13}$CO 2-1 & $1\as30 \times 1\as03$ & 88.0  & 13.6 & 9.7\\
  &C$^{18}$O 2-1 & $1\as31 \times 1\as03$ & 89.2 & 10.4  & 7.3\\   
\hline 
  & $^{12}$CO 2-1 & $1\as25 \times 0\as98$ & -88.2 & 7.1 & 5.0\\
HOPS-408  & $^{13}$CO 2-1 & $1\as29 \times 1\as03$ & 89.3 & 13.6 & 9.6\\
  &C$^{18}$O 2-1 & $1\as30 \times 1\as03$ & -89.5 & 10.6 & 7.4\\   
\hline
\hline
\end{longtable*}

\tablenotetext{a}{Average rms noise out to a distance where the sensitivity is 40\% that of the phase center ($\sim$ 25\as0 from the center ).}
\tablenotetext{b}{Average rms noise out to a distance where the sensitivity is 80\% that of the phase center ($\sim$ 15\as0 from the center).}
\end{center}

\section{Results} 
\label{sec:results}

\begin{figure*}[tbh]
\begin{adjustbox}{addcode={\begin{minipage}{1.0\width}}{\caption{
    Protostellar outflows and envelopes in our survey, which include 21 sources at different evolutionary stages. $^{12}$CO $J = 2-1$ integrated intensity maps are trace the blue-shifted (blue) and red-shifted (red) molecular outflow lobes. C$^{18}$O $J = 2-1$ integrated intensity maps, shown in green,  trace the dense circumstellar envelope. The upper, middle, and bottom rows show Class 0, Class I, and flat-spectrum protostars, respectively, and are arranged in order of increasing Tbol (which we use as a proxy for age) from left to right. Source names are shown above each map. Each map has a diameter of about 50\as4, which is set by the 40\% primary beam power response (compared to the center of the map). The coordinates of the center of the panels are given in \autoref{table:1}.
    }\end{minipage}},rotate=90,center}
    \includegraphics[width=1.3\textwidth]{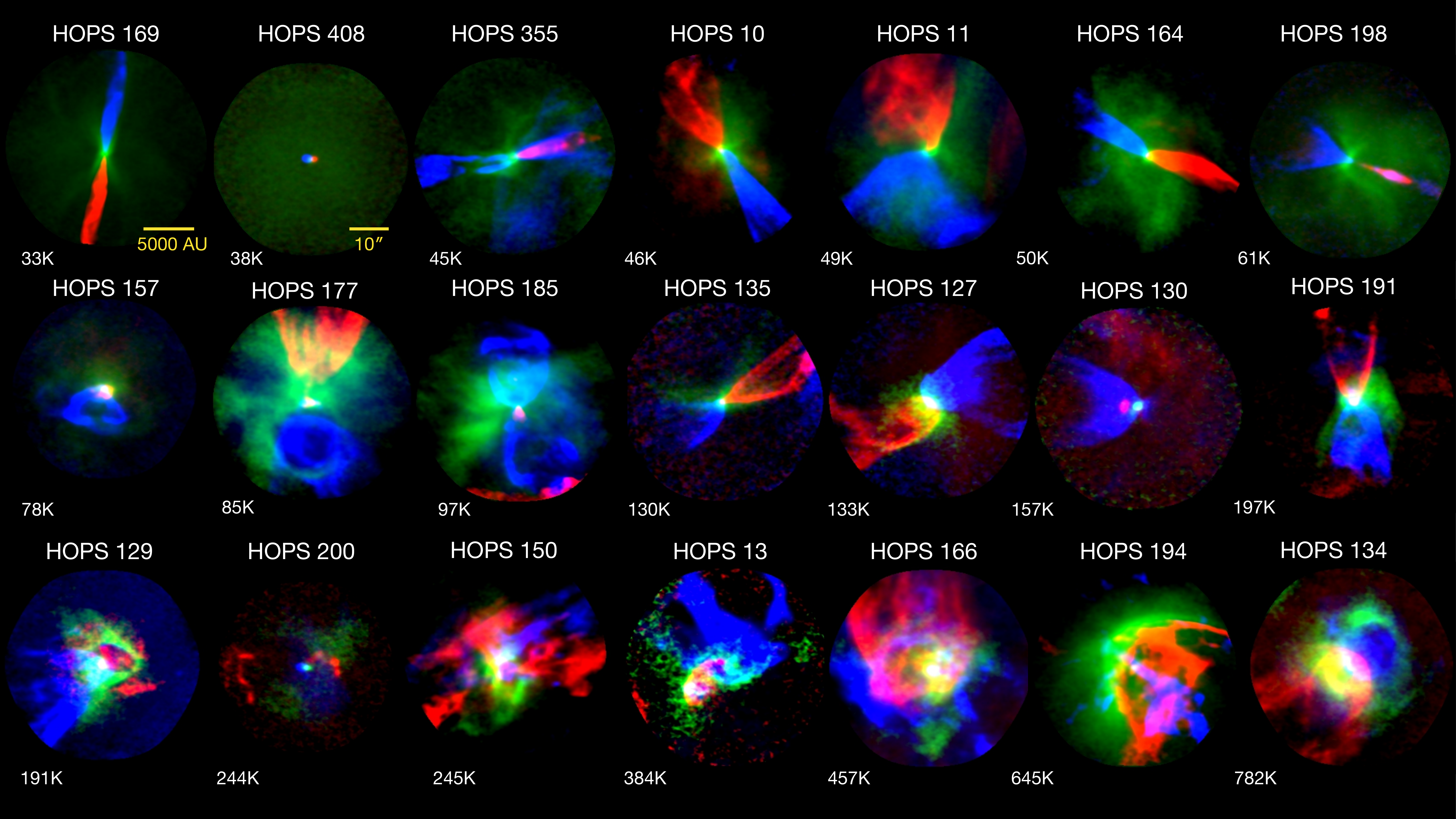}
\end{adjustbox}
\label{fig:outflow_gallery}
\end{figure*} 

The outflows of our 21 HOPS sources are shown in \autoref{fig:outflow_gallery}, where the red and blue colors represent the red- and blue-shifted outflow lobes, respectively, traced by the $^{12}$CO $J = 2-1$. 
The green color represents the dense cores traced by C$^{18}$O $J = 2-1$. The source name and the bolometric temperature, an indicator of protostellar age \citep{1993ApJ...413L..47M,1995ApJ...445..377C}, are also labeled for each source.  { The upper, middle, and bottom rows show Class 0, Class I, and flat-spectrum sources respectively. Most of the Class 0 and Class I sources show clear bipolar outflows, and flat-spectrum sources show complex and messy structures. HOPS-408 shows a very compact outflow possibly indicating its young age. The outflows from HOPS-157, HOPS-185, and HOPS-130 mostly show emission from blue-shifted velocities. For HOPS-157 the peak of the  C$^{18}$O emission (shown in green) is coincident with the very compact red-shifted lobe. The outflow from HOPS-185 is clearly bipolar, but we detect mostly blue-shifted emission for both lobes. This could be due to the fact that for this system, the outflow axis is relatively close to the plane of the sky (see \S~4.1.4).} 
{ Besides the unipolar outflows, asymmetrical features can be seen in many of the bipolar outflows in \autoref{fig:outflow_gallery}. The outflow in HOPS-169 shows a clear ``S-shape" indicating the source is precessing. The axis of the red-shifted outflow lobe in HOPS-11 changes at the position where it appears to interact with the dense core traced by the C$^{18}$O. The lobes of the outflows from HOPS-355 and HOPS-164 appear to be misaligned. In addition, the outflow from HOPS-198 shows a clear difference in width of its two lobes. In this section, we will derive the outflow properties for all the sources shown in \autoref{fig:outflow_gallery}.  }


\subsection{Determining Outflow Properties}

Molecular outflow properties are key for assessing the impact { protostars} have on their surrounding medium. One of their most important properties is the  mass, as many of the other properties depend on it. This is calculated using the $^{12}$CO line emission, which in most cases is optically thick. Hence, it is essential to correct the line opacity (which usually depends on the outflow velocity) to obtain an accurate mass estimate (as explained in \S~\ref{sec:opacity_correction}) \citep{2014ApJ...783...29D, 2015ApJ...802...86B}.
Another issue in estimating the total outflow mass is that, in many cases, it is very difficult to separate the low-velocity outflow component from the cloud emission \citep{, 2001ApJ...554..132A, 2020ApJ...896...11F,  2022ApJ...926...19X}. In general, the mass of the low-velocity component of a molecular outflow is significantly greater than the high-velocity outflow mass. Without including the low-velocity outflow component, the outflow mass can be severely underestimated by factors of about 2 to 10 \citep{2011ApJ...743...91O, 2014ApJ...783...29D}. 
 
\begin{figure*}[tbh]
\centering
\subfloat{%
  \includegraphics[width=0.48\textwidth]{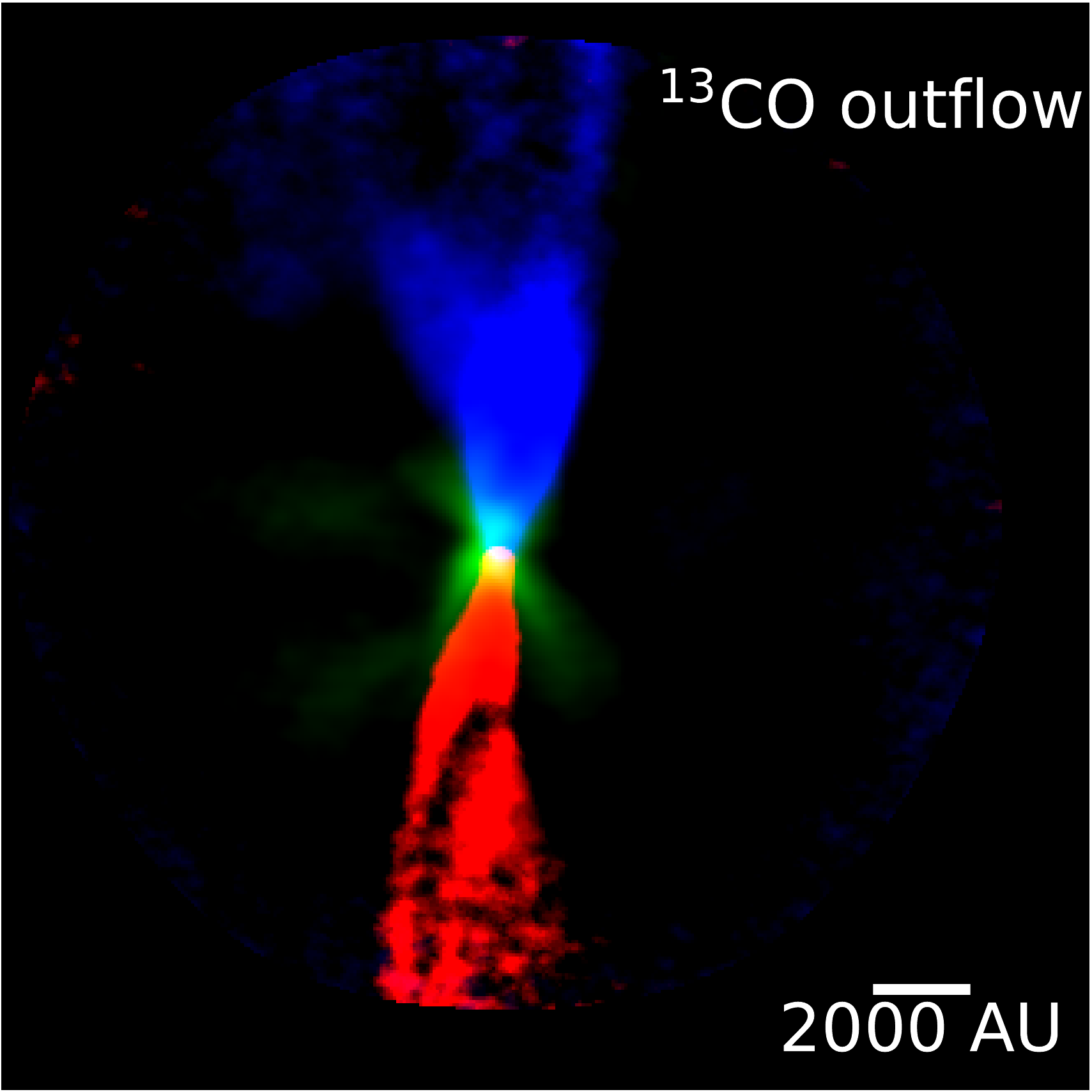}%
}
\subfloat{%
  \includegraphics[width=0.48\textwidth]{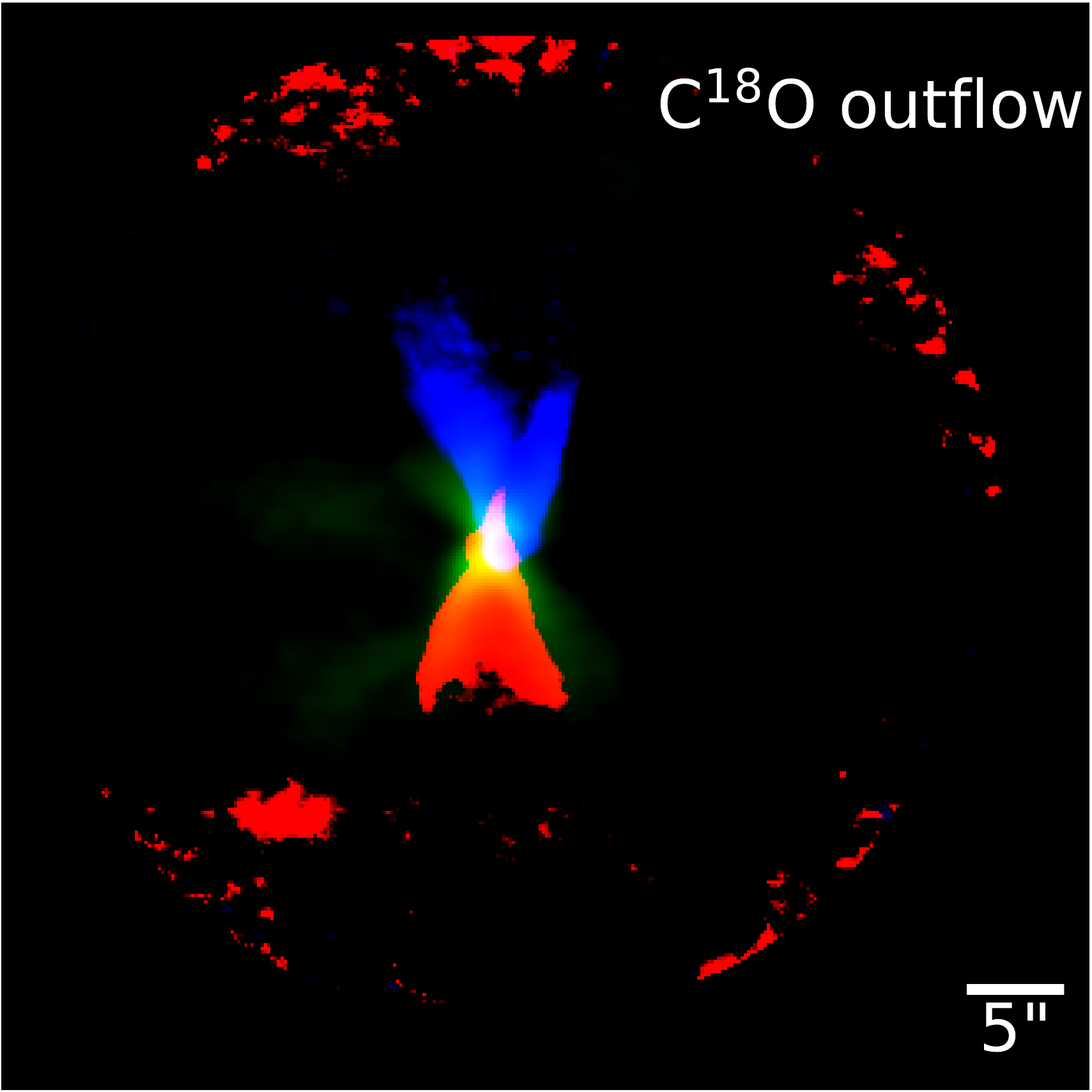}
}

\caption{HOPS-169 outflow traced by $^{13}$CO (left) and C$^{18}$O (right) centered at $\alpha (J2000)=5\textsuperscript{h}36\textsuperscript{m}36\fs1, \delta(J2000)= 06\degree38\arcmin54\as2$). Left:  $^{13}$CO $J= 2-1$ integrated intensity maps are used to trace the blue-shifted (blue) and red-shifted (red) outflow lobes. Right: Same outflow but traced by the C$^{18}$O $J= 2-1$ emission. In both plots, the C$^{18}$O $J= 2-1$ integrated intensity map (integrated over velocities close to the system velocity) traces the dense envelope, and it is shown in green.
}
\label{fig:c18o_outflow}
\end{figure*}

Sensitive ALMA observations can provide the means { to obtain} more accurate mass estimations than was previously possible \citep[see, e.g.,][]{2016ApJ...832..158Z}. Our ALMA observations clearly show low-velocity outflow structures in the low-opacity (or optically thin) $^{13}$CO and C$^{18}$O(2-1) lines (see \autoref{fig:c18o_outflow}), which trace the outflow down to velocities of 1\,km\,s$^{-1}$, or less, from the system velocity. Emission from these rarer isotopologes also allows us to correct for optical depth effects 
(as explained in \S~\ref{sec:opacity_correction}) needed to obtain accurate 
outflow mass estimates \citep[e.g.,][]{2016ApJ...832..158Z}.

Other important outflow properties, such as the momentum and kinetic energy, as well as the mass, momentum and energy injection rates, depend on the outflow velocity. In order to obtain an accurate outflow velocity one needs to correct the observed 
velocity (which only probes the component of the velocity along the line of sight) 
by the inclination of the outflow axis with respect 
to the plane of the sky. Recent complementary ALMA observations allow us to obtain fairly accurate estimates of the inclination of the { disk plane, which is expected to be perpendicular to the outflow}
(see \S~\ref{sec:inclination_correction}).

In the subsections below we describe the procedure we use to obtain various outflow properties. In some cases the methods are based on the techniques used by other recent studies (as it is the case for the outflow mass and line opacity correction). In other cases, we introduce new methods, such as one way we use to derive the outflow mass ejection rate.

\subsubsection{System velocity}
\label{3.11}
The envelope central velocity, also referred to as the system velocity ($v_{syst}$), is needed to derive the outflow (radial) velocity, and hence the outflow momentum and energy. To obtain $v_{syst}$ we use the central velocity determined from a Gaussian fit to the C$^{18}$O  line averaged over a 
$\sim18"\times18"$ region centered on the protostar. If there are multiple components in the C$^{18}$O average spectrum, then we only fit the component with the highest intensity. 
In order to corroborate our estimate of $v_{syst}$, we create position-velocity  diagrams perpendicular to the outflow direction for each source and check whether the envelope emission is centered at the system velocity, as expected from models of circumstellar envelopes with infall and rotation \citep[e.g.,][]{2022PASP..134i4301O}. Detailed PV-diagram modeling with an infall-rotation model will be presented in a future paper.
The estimated system velocities for all sources are shown in \autoref{table:1}.  The outflow radial velocity ($v_{out}$) is then the difference between the observed LSR velocity ($V_{LSR}$) and the system velocity. That is, 
$v_{out} = | V_{LSR}  - v_{syst} |$.

\subsubsection{$^{12}$CO and $^{13}$CO opacity correction}
\label{sec:opacity_correction}

\begin{figure}
\centering
\includegraphics[width=\hsize]{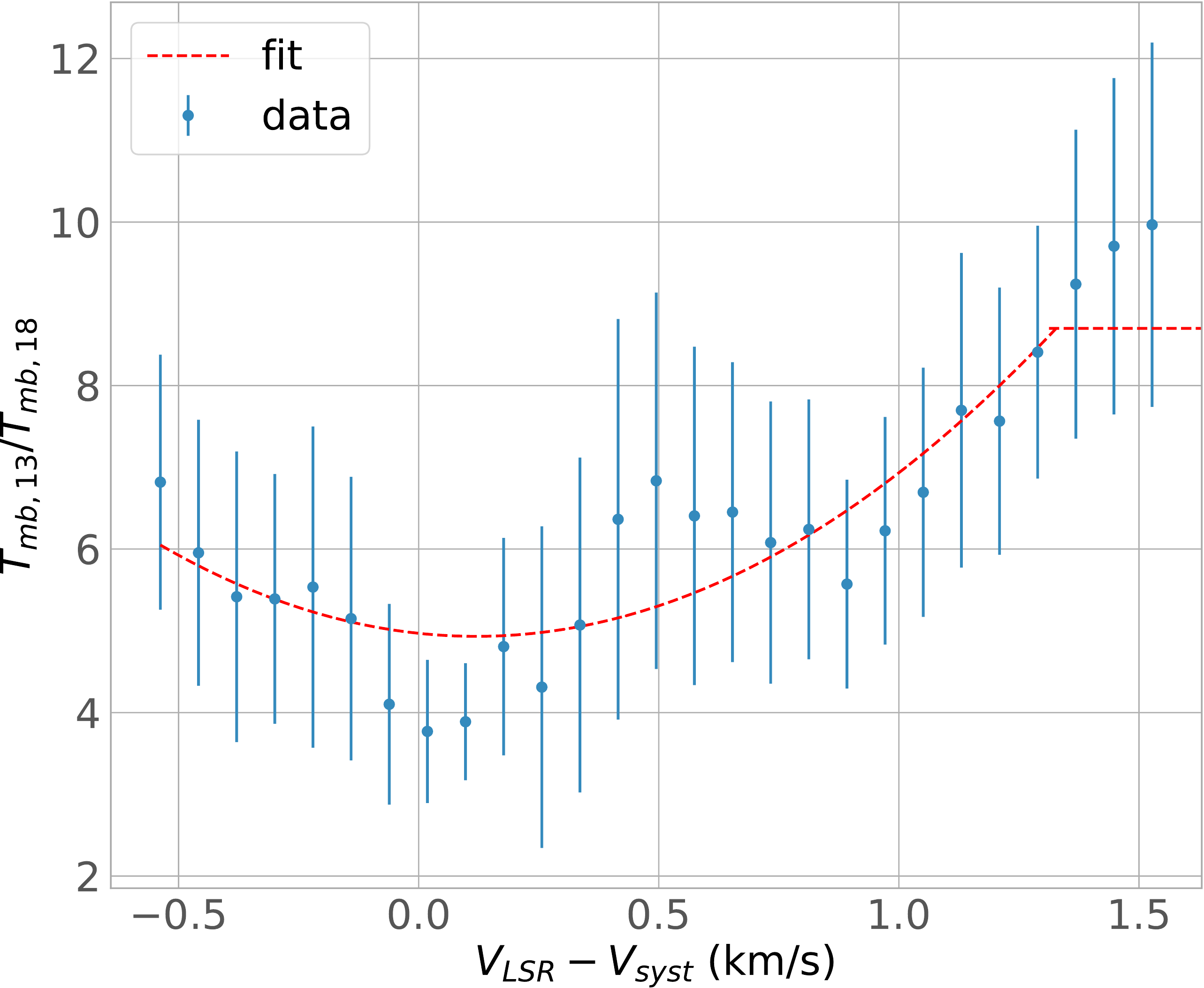}
\caption{Average $^{13}$CO(2-1)/C$^{18}$O(2-1) brightness temperature ratio in the red-shifted outflow lobe of HOPS-169. The ratio is calculated for every velocity channel using data above 3$\sigma$ within the outflow mask region. The blue points represent the data and the red line is the best fit second-degree polynomial. The polynomial is truncated at the isotopic ratio of 8.7. The error bars represent the dispersion about the mean of the brightness temperature ratio across the red-shifted outflow lobe. }
\label{fig:optical_depth_correction}
\end{figure}


In Orion A, as in most other star-forming molecular clouds, the $^{12}$CO(2-1) is optically thick \citep{2018ApJS..236...25K}.
In our high-resolution ALMA data, we found that for many sources the $^{13}$CO(2-1) is also { optically thick} near the system velocity.
Thus, to obtain accurate mass { determination from the} emission maps of these species, one needs to correct for the opacity of both the $^{12}$CO and $^{13}$CO (2-1) lines. 
Moreover, the opacity of these lines increases towards velocities closer to the system velocity  \citep{2014ApJ...783...29D,2016ApJ...832..158Z}.  A proper opacity correction should take this into consideration.

We start by identifying the channels where there is clear outflow emission in the $^{12}$CO maps, avoiding velocity channels near the system velocity as they are dominated by ambient cloud emission.
We use these channels to produce integrated intensity maps that serve to delineate the boundaries of the { red-shifted} and blue-shifted outflow lobes. These outflow masks are created to help isolate the outflow emission from ambient gas. All outflow properties discussed here are measured within these outflow masks.



The optical depth correction factor is outlined by \citet{2014ApJ...783...29D}, who { used} $^{13}$CO to correct for the opacity of the $^{12}$CO line. An extension of this method, using C$^{18}$O to correct for the opacity of the $^{13}$CO line, is described by  \citet{2016ApJ...832..158Z}. We followed their procedure, which we summarize below.  

In general, the  C$^{18}$O is thought to be optically thin \citep{2018ApJS..236...25K}. 
We also assume that 
the $^{12}$CO, $^{13}$CO and C$^{18}$O emission associated with the molecular outflow are in local thermal equilibrium (LTE) with the same excitation temperature of about $50$\,K \citep{2014ApJ...783...29D}. We assume that the $^{12}$CO, $^{13}$CO, and C$^{18}$O emission from the outflow are emitted from the same volume and different isotopic species have the same excitation temperature. 
$^{12}$CO can easily be thermalized even at densities $\le 100$\,cm\,$^{-3}$ \citep{2009tra..book.....W}. In the dense protostellar cores with a number density of the order of $10^6-10^8$\,cm\,$^{-3}$, rarer CO isotopes would also be thermalized. This is a very good approximation especially for the lower J transitions because they have smaller Einstein A coefficients that lead to a slower depopulation rate. Thus, in our case, the assumption of LTE conditions is satisfactory. 

The brightness temperature ratio for $^{13}$CO and C$^{18}$O can be expressed as:
\begin{gather}
\frac{T_{R,13}(\nu)}{T_{R,18}(\nu)}=\frac{1-exp(-\tau_{\nu,13})}{1-exp(-\tau_{\nu,18})}
=\frac{1-exp(-\tau_{\nu,13})}{\tau_{\nu,18}},\\
\approx  X_{13,18} \frac{1-exp(-\tau_{\nu,13})}{\tau_{\nu,13}},
\end{gather}
where $T_R$ is the brightness temperature and the subscripts 13 and 18 represent $^{13}$CO and C$^{18}$O respectively. $X_{13,18}$ is the abundance ratio of { $^{13}$CO relative to C$^{18}$O} which we assume to be 8.7 \citep{1992A&ARv...4....1W}. The correction factor CF$_{13}$ for the  $^{13}$CO opacity is
\begin{gather}
 CF_{13} =\frac{\tau_{\nu,13}}{1-exp(-\tau_{\nu,13})}
\approx  X_{13,18} \frac{T_{R,18}(\nu)}{T_{R,13}(\nu)}.
\end{gather}
Then, the $^{12}$CO correction is determined using the same equation, but using $X_{12,13}=62$ \citep{1993ApJ...408..539L} instead of $X_{13,18}$, and the brightness temperature of the $^{13}$CO and $^{12}$CO emission in place of $T_{R,18}$ and $T_{R,13}$, respectively. 
To obtain the brightness temperature ratio $T_{R,18}(\nu)/T_{R,13}(\nu)$ and $T_{R,13}(\nu)/T_{R,12}(\nu)$ we fit a second order polynomial to the red-shifted and blue-shifted outflow lobe of each source independently. { The ratio is calculated using the average brightness temperature for each channel inside the outflow the mask.} An example for the HOPS-169 red-shifted outflow lobe (\autoref{fig:c18o_outflow}) is shown in \autoref{fig:optical_depth_correction}. The correction factor estimated from the red-shifted and blue-shifted outflow lobes are then used to correct for the opacity for the entire position-position-velocity (PPV) cube. 
We first correct for the  $^{13}$CO opacity using the $^{13}$CO to C$^{18}$O ratio. Then the ratio of the $^{12}$CO to the
corrected $^{13}$CO emission was used to correct the optically thick $^{12}$CO emission.   

\subsubsection{Outflow mass, momentum, energy estimation}

We follow \citet{2014ApJ...783...29D}
to derive most of the molecular outflow properties using the $^{12}$CO, $^{13}$CO, and C$^{18}$O data. 
The molecular hydrogen (H$_2$) column density of the outflowing material can be calculated from the $^{12}$CO emission, using: 
\begin{gather}
N_{H_2}=f\left(J, T, X_{\mathrm{CO}}\right)I,\\
f\left(J, T, X_{\mathrm{CO}}\right)=X_{\mathrm{CO}} \frac{3 k}{8 \pi^{3} v \mu^{2}} \frac{(2 J+1)}{(J+1)} \frac{Q(T)}{g_{J}} e^{\frac{E_{J+1}}{k T}},
\end{gather}
where { $I$} is the (opacity corrected) integrated intensity in K\,km s$^{-1}$, and  $J$ is the quantum number for total rotational angular momentum. One can replace  the frequency ($\nu$), magnetic dipole moment ($\mu$), partition function ($Q$), upper level energy ($E_{J+1}$), the energy statistical weight ($g$) and abundance ratio relative to hydrogen $X$ with the values for $^{13}$CO and C$^{18}$O, to  obtain estimates of the $H_2$ column density using the other two CO isotopologues. We assumed  abundance ratios of $10^{-4}$,  $1.6\times 10^{-6}$, and $1.7\times 10^{-7}$ for CO, $^{13}$CO and C$^{18}$O, respectively { , which are consistent with the istopologue ratios  discussed in Section \ref{sec:opacity_correction}} \citep{1982ApJ...262..590F,1993ApJ...408..539L,2019ApJ...873...16H}.

 For each channel, we computed the outflow mass  using:
\begin{gather*}
    M= \mu_{H_2}m_HA_{pixel}N_{H_2,}
\end{gather*}
where $\mu_{H_2}=2.8$ is the mean molecular weight of H$_2$ \citep{2008A&A...487..993K}, $m_H$ is the hydrogen atom mass, and $A_{pixel}=6.47\times10^{24}\times d^2$\,cm$^2$ is the pixel area at a distance of $d$\,pc. We use
the distance estimate derived by
\citet{2020ApJ...890..130T}, who used Gaia data of young stars in the Orion molecular cloud to estimate the distance to each protostar (which we include in \autoref{table:1}). 

\begin{figure}
\centering
\includegraphics[width=\hsize]{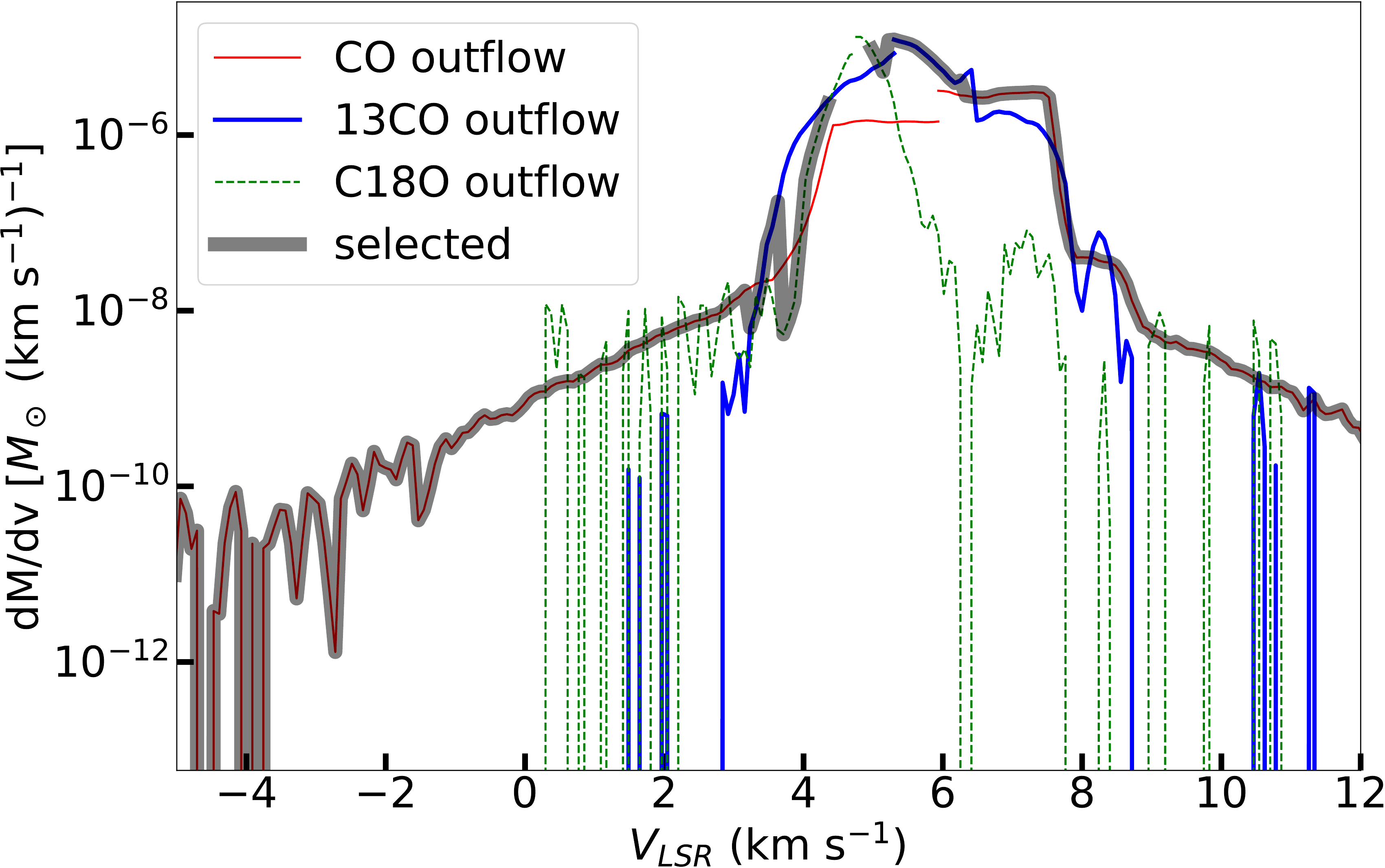}
\caption{Molecular outflow mass spectrum for the outflow from HOPS-135, estimated using $^{12}$CO, $^{13}$CO and C$^{18}$O. The black line marks the final selected mass spectrum used for estimating the outflow mass. 
}
\label{fig:mass_spectrum}
\end{figure}

We computed the mass for every pixel in every velocity channel, hereafter the mass spectrum, within the outflow mask using $^{12}$CO, $^{13}$CO, and C$^{18}$O. 
We go through the channel maps of each tracer (starting at the highest outflow velocities) and use the $^{12}$CO to derive the outflow mass for the high-velocity emission. For lower velocity outflow emission, the $^{12}$CO becomes extremely optically thick and contaminated with the ambient gas emission, so we then use the $^{13}$CO outflow emission to estimate the mass at lower velocities. For sources in which there is clear outflow emission traced by C$^{18}$O (see, e.g., \autoref{fig:c18o_outflow}), we use it to estimate the mass of the lowest outflow velocity components, near the system velocity. 
For all sources, the emission at velocities close to $v_{syst}$ becomes dominated by the large-scale core (ambient) emission. The velocity at which the emission is dominated by core emission, rather than by outflow emission, is different for each source and isotopologue. Thus we used different velocity ranges for each source, lobe, and tracer in order to try to capture as much of the outflow emission as possible.   We select the velocity range for each tracer such that the protostellar outflow shows a clear intensity contrast with the ambient  emission, and avoid channels where the structure is dominated by the cloud emission.
 The velocity range used for each tracer and source is shown in \autoref{table:velocity_range}.
 An example mass spectrum for HOPS-135 using all three isotopologues is shown in \autoref{fig:mass_spectrum}, where the final mass spectrum is highlighted in black. Note that to avoid contamination from large-scale cloud emission at outflow velocities
  we performed a cloud subtraction procedure, described in \autoref{sec:Appendix_A}. All sources were used for each analysis unless otherwise specified (see \autoref{table:analysis_sources} for  excluded sources). 


\subsubsection{Inclination correction}
\label{sec:inclination_correction}

{ The} inclination correction plays a very important role in determining the correct value of the outflow momentum and energy, as well as the outflow mass, momentum, and energy ejection rates. On average inclination can affect the outflow mass ejection rate by a factor of 3 to 5, and without the correction any possible evolution trend could be rendered undetectable in the data. { In this paper, we define  $i$ to be the inclination angle of the outflow axis with respect to the line-of-sight. That is, an outflow  with $i = 0^\circ$ has a  ``pole-on'' orientation and one with $i = 90^\circ$ is said to be ``edge-on''.}

We used the deconvolved disk major and minor axes around each of the protostars in our sample, measured by \citet{2020ApJ...890..130T} from their ALMA 0.87 mm data, to derive the disk inclination angle. Assuming  the disk is thin, flat and circular, we derive the inclination angle for each disk using the ratio of the observed major to minor axes.
We then assume the outflow axis is perpendicular to the disk, to estimate the outflow's inclination angle with respect to the line of sight sky (these are listed in \autoref{table:inclination_angle}). 
This method works well when the disk continuum emission is  well-resolved, yet  free of extended envelope emission.  This works reasonably well for all sources except for HOPS-177, for which
the inclination angle derived  using this method
($i = 81\arcdeg$) is highly uncertain, as the continuum emission is basically the size of the beam (i.e., the emission appears unresolved) and the flux density 
signal-to-noise of this source is only about 3.7 
\citep{2020ApJ...890..130T}. 
Moreover, the CO velocity structure for HOPS-177 is not consistent with that expected of a wide-angle outflow with an approximately edge-on orientation\footnote{{ For} an edge-on wide angle outflow one would expect to see both red and blue-shifted emission in both lobes.}. Hence, for this source we adopt  $i=57.3\arcdeg$
\citep[the mean inclination angle from a random uniform distribution of outflow orientations on the sky,][]{1996A&A...311..858B}.

We also estimated the inclination angle by fitting a wide-angle wind entrainment model to the $^{12}$CO data. This method has been used in the past \citep{1996ApJ...472..211L,2000ApJ...542..925L} and we wanted to assess how well it compared to the method  using the disk continuum observations described above.  
A description of the fitting procedure, and a more detail comparison of the inclination angle estimates obtained using the two different methods are included in the \autoref{Appendix:B}.


\subsubsection{Outflow mass, momentum, and energy ejection rates estimation}
\label{sec:mass_rates_results}



Currently, the most common way to measure the molecular mass outflow mass rate, the outflow momentum injection rate (also referred to as the outflow force, $F_{out}$), or the outflow energy injection rate (also referred to as the outflow mechanical luminosity, $L_{out}$)  is by dividing the outflow quantity of interest by a ``dynamical time''  ($t_{dyn}$). This $t_{dyn}$ is simply obtained by dividing the outflow lobe length ($R_{lobe}$) by a ``characteristic'' outflow velocity, which typically is chosen to be the maximum (detectable) molecular outflow velocity \citep[e.g.,][]{2013A&A...556A..76V,2014ApJ...783...29D}. 

However, this simple  method is heavily dependent on  observational parameters (see Sec.~\ref{rates_methods}) and  inaccurately assumes that all the gas in the molecular outflow is moving at a uniform (constant) velocity from the source position to the current position. 
One problem with this assumption is that the gas in a molecular outflow is mostly made of material that has been entrained by the underlying protostellar wind rather than material that was launched by the protostar-disk system \citep{2022MNRAS.510.2552R,2017ApJ...847..104O}. 

\begin{figure}
\centering
\includegraphics[width=\hsize]{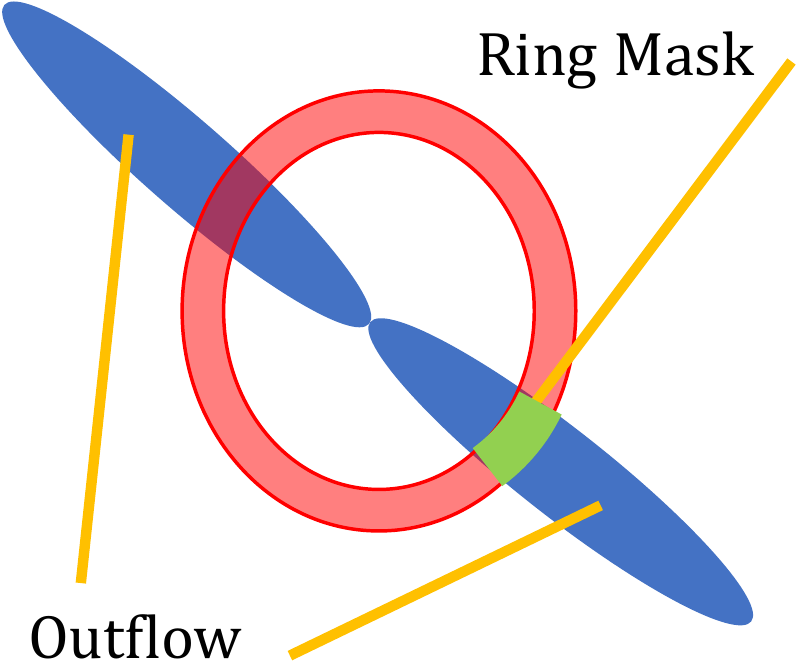}
\caption{Cartoon diagram of the ring mask used in the Ring Method for  determining the  molecular outflow mass, momentum and energy rates.}
\label{fig:RM}
\end{figure}

{ For estimating the outflow momentum injection rate, we developed a procedure based on the ``annulus method" described in \citet{1996A&A...311..858B} which here we call the ring method (RM).} For this, we first created a series of ``ring masks" with thickness ($\Delta R$) around the protostars in our sample, at different distances ($R$) from the positions of the outflow sources. We computed the outflow mass, momentum and energy as a function of velocity within the ring mask {(see \autoref{fig:RM})}. Then, to estimate the rate, we calculated the ``crossing time" across the ring mask for each channel, which is obtained by dividing $\Delta R$ by the outflow velocity, { where both are corrected for the inclination of the system, i.e.,  $v_{out,cor} = \lvert (v_{chan}-v_{syst})/{\rm cos}(i) \lvert ; \Delta R_{cor} = \Delta R/{\rm sin}(i)$.}
Then we sum up the rate estimate for each velocity channel. 

{ Our Ring Method does not use a constant area as in procedure used by \citet{1996A&A...311..858B}. In their method, the momentum flux is obtained following this equation:
\begin{gather}
F_{\text {obs }} \propto \int_{\text {wings }} \frac{T_{\mathrm{A}}^{*}\left(v_{\mathrm{rad}}\right) \mathrm{d} v_{\mathrm{rad}} \operatorname{Area}(r, \Delta r) v_{\mathrm{rad}}}{\Delta r / v_{\mathrm{rad}}},
\end{gather}
where the annulus has a radius of $r$ and width of $\Delta r$, $v_{rad}$ is the radial velocity, and Area is simply $r \Delta r \Delta \theta $, where $\Delta\theta$ is the angular extent of the annulus sector. \citet{1996A&A...311..858B} assumes the outflow fills the entire annulus for every velocity channel, and $\Delta\theta$ is the telescope beam FWHM. This is usually a good assumption for CO maps made with single dish telescopes where the outflow is not resolved. The above equation also assumes the $^{12}$CO is optically thin. Although  \citet{1996A&A...311..858B} applied 
a correction factor to correct for the optical depth of the $^{12}$CO line, their correction is the same for all sources  and does not depend on  outflow velocity. In our study we use multiple CO isotopes to construct the mass spectrum to avoid saturation and to trace the outflow mass at different velocities (See \autoref{fig:mass_spectrum}). Within the ring, the area of the outflow changes as a function of velocity. In our Ring Method, we account for this effect, and the momentum flux can be expressed as:
\begin{gather}
    F_{\text {obs }} = \int \frac{mass\left(v_{\mathrm{chan}}\right)   v_{\mathrm{out,corr}}}{\Delta R_{cor} / v_{\mathrm{out,corr}}} \mathrm{d} v_{\mathrm{chan}},
\end{gather}
where outflow mass is computed for each velocity channel within the intersection region between the ring and the outflow mask (See \autoref{fig:RM} green region). That is, the outflow width can be different for different outflow velocities. }

\begin{figure}
\centering
\includegraphics[width=\hsize]{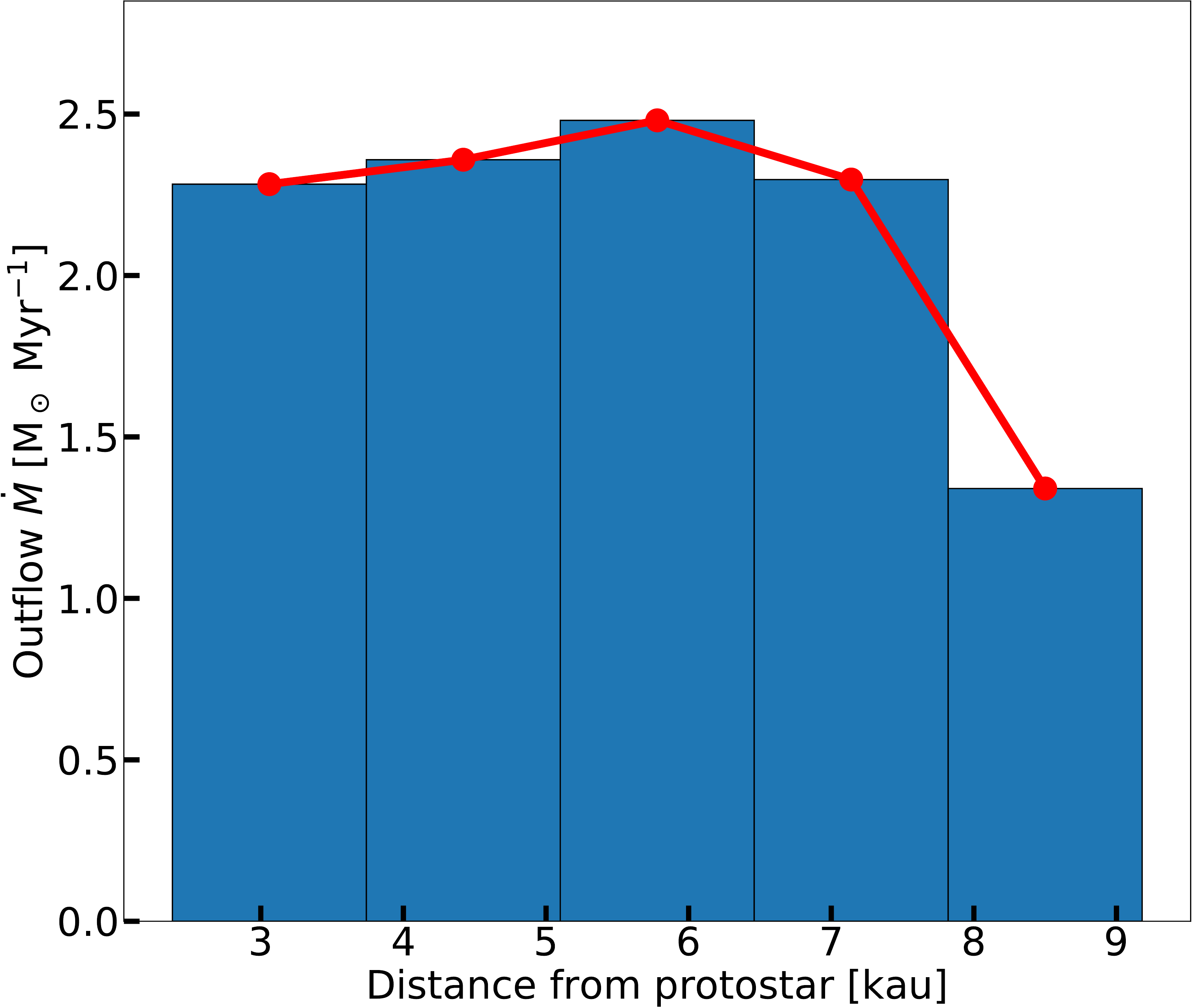}
\caption{ Molecular outflow mass rate (per lobe) for the Class 0 source HOPS-10.   The profile is corrected for inclination angle.} 
\label{fig:dM_r}
\end{figure}

This method has several advantages over previous simple methods. The main one is that it does not assume the molecular outflow moves at a uniform velocity from the launched position to the current position. In addition, the outflow rate is calculated for each velocity channel independently, rather than using one velocity for all outflow emission. Moreover, it is not dependent on the field of view (which will bias the assumed outflow length) or the assumed $v_{max}$, which will depend on the sensitivity of the observations. Thus this method provides a systematic way to obtain molecular outflow mass, momentum, and energy rates (hereafter referred to as ``molecular outflow rates'') with better accuracy compared to  previous estimates. 
We can also construct the ring mask at different { distances} from the protostar to obtain a radial profile. An example of the molecular outflow mass  rate per lobe profile for HOPS-10 is shown in \autoref{fig:dM_r}.

\begin{figure*}[tbh]
\centering
\makebox[\textwidth]{\includegraphics[width=\textwidth]{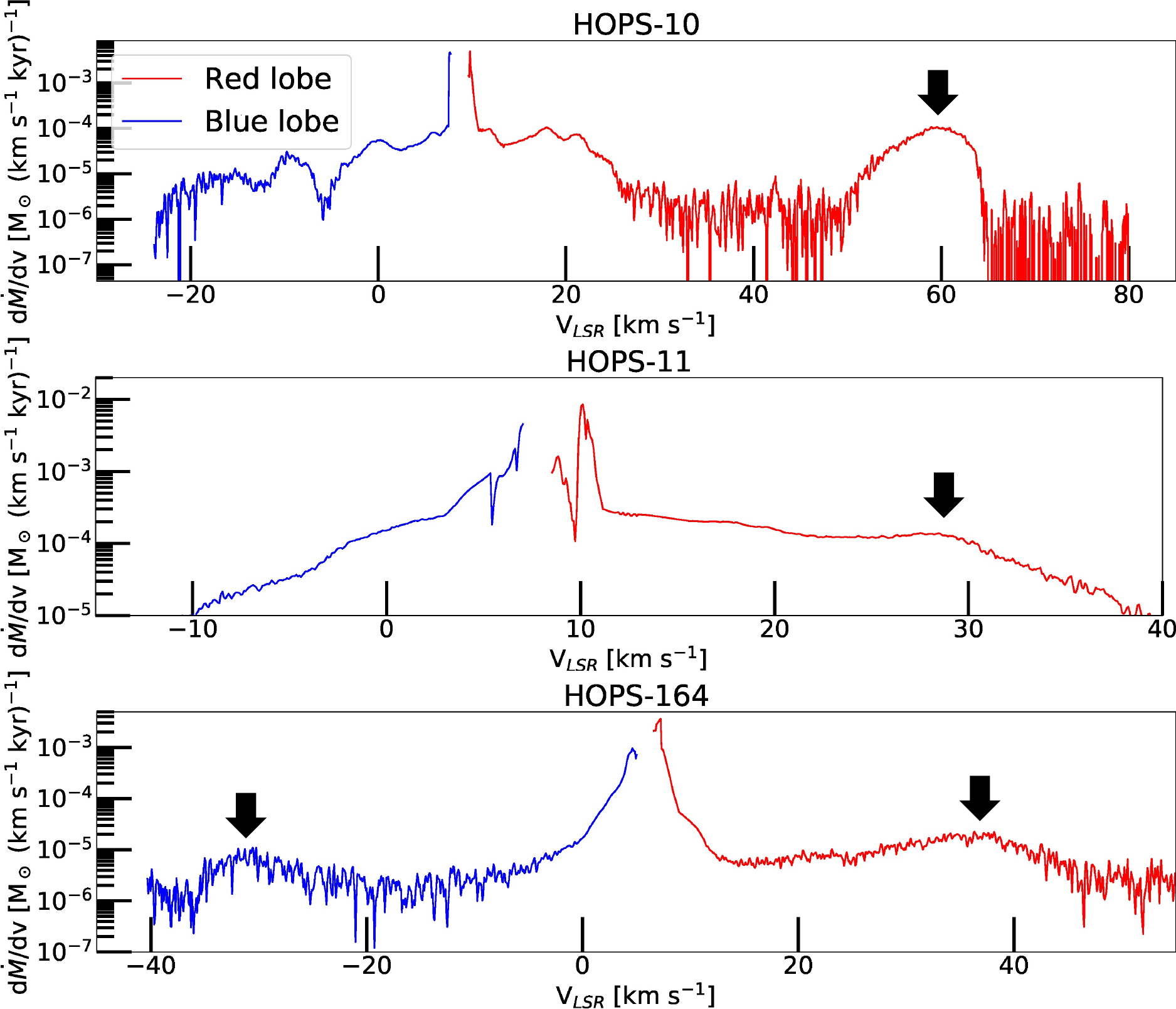}}
\caption{Example of molecular outflow { mass}  rate spectrum for HOPS-10, HOPS-11, and HOPS-164 using the Ring Method (see \autoref{fig:RM}). { The mass ejection rate spectrum is the mass spectrum dividing by the crossing time across the ring ($mass\left(v_{\mathrm{rad}}\right)/ \left(\Delta r / v_{\mathrm{rad}}\right)$).} The black arrows mark the position of high-velocity jets.
}
\label{fig:dM_v}
\end{figure*}

\begin{figure*}[tbh]
\centering
\makebox[\textwidth]{\includegraphics[width=\textwidth]{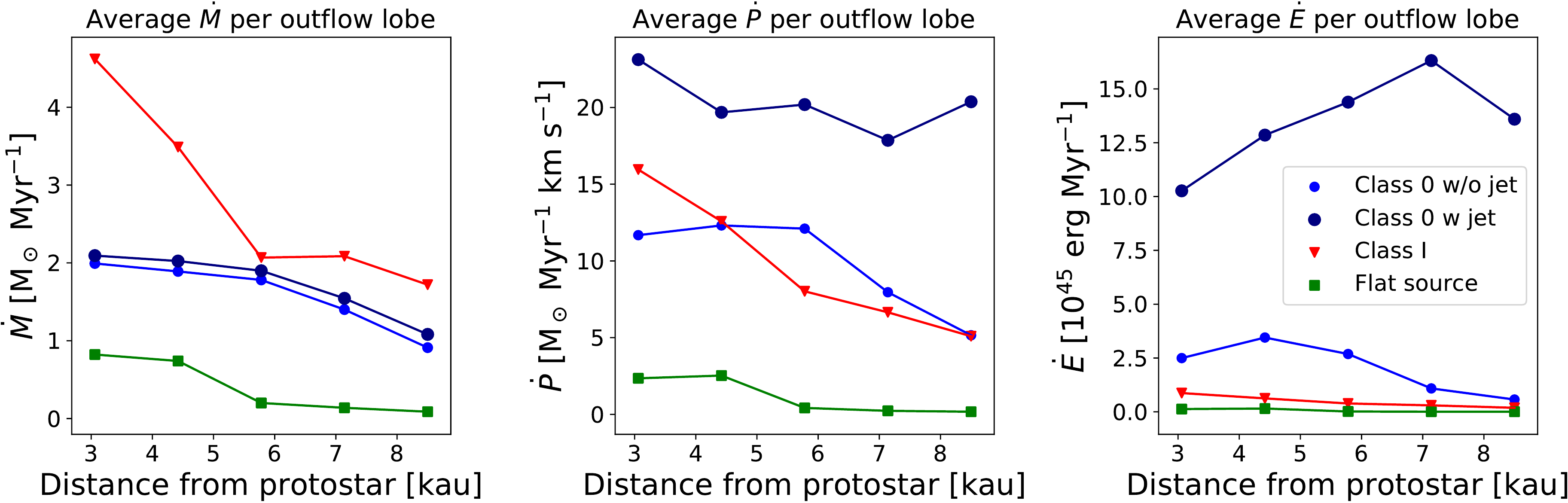}}
\caption{Profiles of the average molecular outflow mass, momentum and energy rates per outflow lobe for different protostellar evolutionary classes (Class 0, Class I and flat-spectrum). For the Class 0 sources we show the average rates with and without inclusion of the high-velocity jet component. All results shown are inclination corrected. { In creating this plots, the Class 0 source HOPS-408 was not included because its outflow is barely resolved, and the flat-spectrum sources HOPS-166 and HOPS-194 were not included  because they have significantly higher bolometric luminosity ($> 10\,L_\odot$) and significantly higher core masses  compared to other sources in this study. The dispersion of outflow rate values for each evolutionary class is approximately a factor of 2.}
}
\label{fig:dM_dP_dE_r}
\end{figure*} 




Among our sample, several Class 0 sources have high-velocity collimated jet components traced only by CO that are clearly different from the rest of the molecular outflow emission of each source.  These are: HOPS-10, HOPS-11, { and} HOPS-164. Unlike the low-velocity outflow component, these molecular jets could be mostly made of { disk-}launched material 
\citep{2020A&ARv..28....1L}.
{ To compare the entrained material with the disk-launched material, we separate out} the extremely high-velocity components which can clearly be seen in the { mass rate spectrum} of a source, as shown in \autoref{fig:dM_v}. 
{ The outflow mass rate spectrum is the outflow mass spectrum (\autoref{fig:mass_spectrum}) divided by the crossing time across the ring ($mass\left(v_{\mathrm{out,cor}}\right)/ \left(\Delta R / v_{\mathrm{out,cor}}\right)$).} We stress that the outflow mass rate we measure with our data is that of the entrained gas, and not of the mass that is directly ejected from the central disk-star system. Measuring the outflow mass rate for the entrained gas is crucial as it is  directly linked to envelope dispersal. 

To study the evolution of { the} molecular outflow   rates we averaged all sources within each Class in  our sample (see \autoref{fig:dM_dP_dE_r}). We did not include HOPS-408 in the average for Class 0 sources, as the outflow from this source is  barely resolved and the outflow rate cannot be measured using the ring mask method. { HOPS-166 and HOPS-194 are not included in the average because they have significantly higher bolometric luminosity ($> 10\,L_\odot$) and significantly higher core mass as compared to other sources in this study.} 
In \autoref{fig:dM_dP_dE_r} we plot the average molecular outflow rates per lobe for each evolutionary class. 
The derived values of these rates, at five different distances from each source are listed in the Appendix (see \autoref{table:outflow_rate_vall_p1} to \autoref{table:outflow_rate_vall_p3}), for all sources in our sample. 


From \autoref{fig:dM_dP_dE_r}, it is clear that Class I protostars have the highest total molecular outflow mass rate. This is opposite to the recent simulations showing the decrease in outflow mass-loss rate from Class 0 to Class I protostars \citep{2022MNRAS.510.2552R}. The molecular outflow mass rate per lobe of Class I protostars is around $60\%$ higher than that of Class 0 protostars and six times higher than flat-spectrum sources. The sharp decrease in mass outflow rate between Class I and flat-spectrum sources indicates that the majority of the mass entrainment occurs during the Class I phase. 


In \autoref{fig:dM_dP_dE_r},  we plot the average molecular outflow rates (per outflow lobe) with and without the high-velocity jets, for the Class 0 sources. 
The molecular outflow mass  rate profile does not change much when the  high-velocity jets are added. This indicates the high-velocity jets in Class 0  do not carry much mass compared to the wider low-velocity swept-up material, and that the majority of the entrained mass lies within the low-velocity { molecular outflow}.  
However, the momentum and energy ejection rate increase by a factor of 2 and 4 respectively for Class 0 sources, when high-velocity jets are included. Thus, the momentum in the low-velocity outflow is comparable to the high-velocity jets in Class 0, and  the dominant energy ejection rate lies in the collimated high-velocity jets.

{ We average all the distance bins in \autoref{fig:dM_dP_dE_r}, and multiply by 2 for two outflow lobes to obtain the average outflow rate for each protostellar stage.} Using our estimate of the average molecular outflow mass  rate during each protostellar stage, and assuming that the Class 0 phase lasts about $0.13-0.26$\, Myrs, the Class I phase lasts approximately $0.27-0.52$\, Myrs, and the flat-spectrum phase  also lasts
around $0.27-0.52$\, Myrs  \citep{2015ApJS..220...11D}, we can estimate
the total mass of the circumstellar material 
entrained by a ``typical'' outflow  
during each phase.
We find that on average a protostar will entrain around ${ 0.5}-0.9$\,M$_\odot$ of circumstellar material during the Class 0 phase  (using an average molecular outflow mass rate for this evolutionary phase of $\overline{\dot{M}}_{out}={ 3.5}$\,M$_\odot~$Myr$^{-1}$), $1.5-2.9$\,M$_\odot$ of material during Class I phase (using $\overline{\dot{M}}_{out}=5.6\,M_\odot$~Myr$^{-1}$), and ${ 0.2-0.4}$\,M$_\odot$ during the flat-spectrum phase (using $\overline{\dot{M}}_{out}={ 0.8}$\,M$_\odot$~Myr$^{-1}$). Thus, we expect that by the end of the protostellar phase,  a typical outflow will entrain about 2.2 to { 4.2} \,M$_\odot$ of circumstellar material (see \autoref{table:core_growth}).


Nearly all of the momentum and energy injected by the outflow to the core is concentrated during the Class 0 phase. Adopting the same time-scales for each protostellar stage as above (and including the momentum and energy injection rates from the high-velocity jet component in the calculations), we expect outflows to inject a total of about $5.3-10.5$\,M$_\odot$\,km~s$^{-1}$ of momentum and $3.5-7.0\times 10^{45}$\,erg of energy in the Class 0 phase (using average momentum and energy injection rates for this evolutionary phase of $\overline{\dot{P}}_{out}=$ 40.5 M$_\odot$~Myr$^{-1}$\,km s$^{-1}$, and $\overline{\dot{E}}_{out}=$ $2.7\times10^{46}$\,erg~Myr$^{-1}$); approximately 5.1 to 9.9 M$_\odot$\,km~s$^{-1}$ of momentum and $2.7-5.2\times 10^{44}$\,erg of energy in the Class I phase (using, $\overline{\dot{P}}_{out}=$ 19.06\,M$_\odot$~Myr$^{-1}$ km$\,$s$^{-1}$, and $\overline{\dot{E}}_{out}=$ $9.9\times10^{44}$\,erg~Myr$^{-1}$); and about ${ 0.6-1.2}$\,M$_\odot$\,km~s$^{-1}$ of momentum and ${ 3.3-6.3\times 10^{43}}$\,erg of energy in the flat-spectrum phase (using $\overline{\dot{P}}_{out}=$ { 2.3}\,M$_\odot$~Myr$^{-1}$~km s$^{-1}$, and $\overline{\dot{E}}_{out}=$ ${ 1.2\times10^{44}}$\,erg~Myr$^{-1}$). We thus estimate that by the end of the protostellar phase, the total outflow momentum and energy will be approximately  ${ 11.0-21.6}$\,M$_\odot$\,km~s$^{-1}$ and $3.8-7.6\times 10^{45}$\,erg, respectively. The results are summarized in \autoref{table:MPEdMdPdE_results}.

\begin{table*}
\setlength{\tabcolsep}{6pt} 
\caption{The total mass, momentum, energy injected/displaced by the molecular outflow, and corresponding molecular outflow rates at different evolutionary stages}            
\label{table:MPEdMdPdE_results}      
\centering                          
\begin{tabular}{c c c c c c c c c c c }        
\hline\hline                 
Class & Duration & $\overline{\dot{M}}_{out}$ & $\overline{\dot{P}}_{out}$  & $\overline{\dot{E}}_{out}$  & $M_{out}$ & $P_{out}$  & $E_{out}$  & Core Mass\tablenotemark{a} \\
 & [Myr]  &[M$_\odot$\,Myr$^{-1}$] &  [M$_\odot$\,km$\,$s$^{-1}$ &  [erg\,Myr$^{-1}$] & [M$_\odot$] & [M$_\odot$\,km$\,$s$^{-1}$] &  [erg] & [M$_\odot$] \\
&   &  & \,Myr$^{-1}$] & &  &   &   \\

\hline                        
0\tablenotemark{b} &0.13 - 0.26 & 3.5 & 40.5 & $2.7 \times 10^{46}$ & 0.5 - 0.9 & 5.3 - 10.5 & $(3.5 - 7.0) \times 10^{45}$ & (0.95, 1.04) \\
\hline                                   
I  & 0.27 - 0.52 & 5.6 & 19.1 & $9.9\times10^{44}$ & 1.5 - 2.9 & 5.1 - 9.9 & $(2.7 - 5.2)\times 10^{44}$ & (0.49, 0.49) \\
\hline   
flat\tablenotemark{c} & 0.27 - 0.52 & 0.8 & 2.3 & $1.2\times10^{44}$ & 0.2 - 0.4 & 0.6 - 1.2 & $(3.3 - 6.3)\times 10^{43}$ & (0.46, 0.26) \\
\hline   

\end{tabular}
\tablenotetext{a}{The first value is the average core mass from our ALMA C$^{18}$O data. The second value is the average core mass derived from the JCMT 850 $\mu$m dust continuum data \citep{2016ApJ...833...44L}. Both core mass estimates are measured using a similar core size of $\sim$6000\,au.}
\tablenotetext{b}{HOPS-408 is excluded from these estimates as we were not able to obtain a reliable molecular outflow mass rate for this source.}
\tablenotetext{c}{HOPS-194 and HOPS-166 are excluded from the average as these two sources have significantly higher core mass  derived from C$^{18}$O as compared to core masses derived from the dust continuum. They also have the highest bolometric luminosity ($ > 10\,L_\odot$) in our sample.} 
\end{table*}

\begin{figure*}[tbh]
\centering
\makebox[\textwidth]{\includegraphics[width=\textwidth]{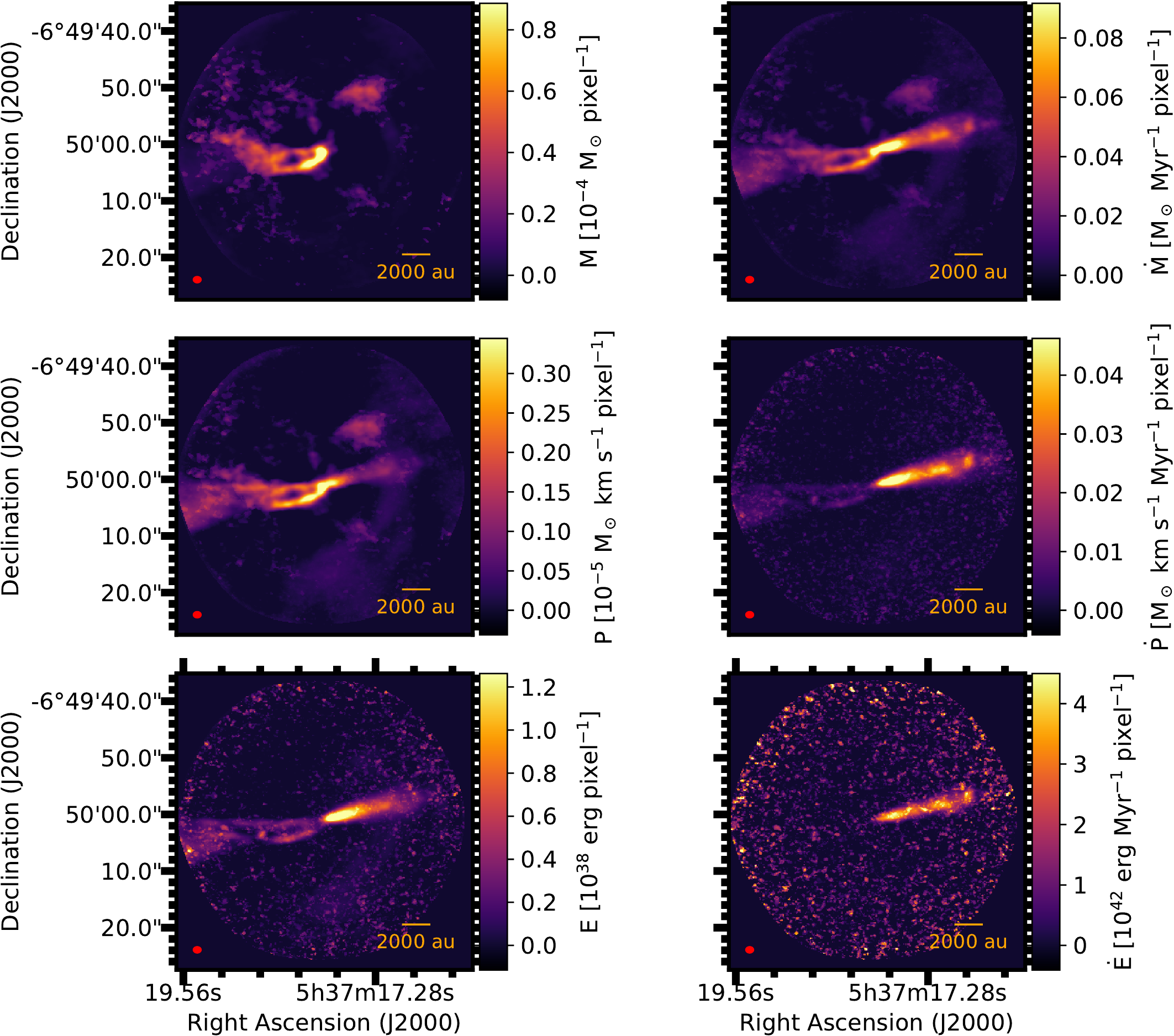}}
\caption{HOPS-355 molecular outflow mass, momentum, and energy  maps (panels in the left column), and the corresponding  rate maps (panels in right column). {  The red ellipse in the bottom left of each panel represents the synthesized beam.  Note that the pixel size is the same for all maps and is 0\as17 $\times$  0\as17.}
}
\label{fig:2D_MPE_dMdPdE}
\end{figure*}

\subsection{Pixel Flux-tracing Technique and observational instantaneous Outflow Mass, Momentum, Energy ejection rate maps }

Motivated by the Ring Method (described above), we developed another procedure to estimate  outflow mass rates, which we call the pixel flux-tracing technique (PFT). \footnote{The code is available on GitHub. https://github.com/chenghanhsieh/pixel-flux-tracing-technique/blob/main/README.md}
The basic idea of the PFT is to view each pixel as a thin ring and estimate the instantaneous outflow mass rate across each pixel. This is similar to the flux tracing commonly used in simulations (e.g.\ \citet{2012ApJS..198....7M,2014ApJS..211...19B}). We start with { calculating} the mass, momentum, or energy spectrum for each pixel. Then, we compute the crossing time ($\tau_{{ cross}_{pix}}$) for each velocity channel, as in the ring method, but we use the pixel width ($\Delta R_p$) instead of the ring width. That is, 
\begin{gather}
\tau_{{ cross}_{pix,vchan}} = \frac{\Delta R_p/sin(i)}{\lvert (v_{chan}-v_{syst})/cos(i) \lvert},
\end{gather}
where $i$ is the inclination angle, { $(v_{chan}-v_{syst})/cos(i)$} is the  outflow velocity for that particular velocity channel, corrected for inclination ($v_{{\rm out,corr}}$). { Then the outflow mass, momentum, and energy flux for a pixel located at position $(x,y)$ can be computed as:

\begin{gather}
    F_{\text {obs}}\left(x,y\right) = k \int \frac{mass\left(v_{\mathrm{chan}},x,y\right)   v_{\mathrm{out,corr}}^n}{\tau_{{ cross}_{pix,vchan}}} \mathrm{d} v_{\mathrm{chan}},
\end{gather}
where $n = 0, 1, 2$ and $k = 1.0, 1.0, 0.5$ for mass, momentum, and energy flux respectively.} To avoid cloud contamination, we developed a cloud subtraction method for the PFT
(different from that used for the RM), which we show in \autoref{sec:Appendix_A}. 


\begin{figure*}[tbh]
\centering
\makebox[\textwidth]{\includegraphics[width=\textwidth]{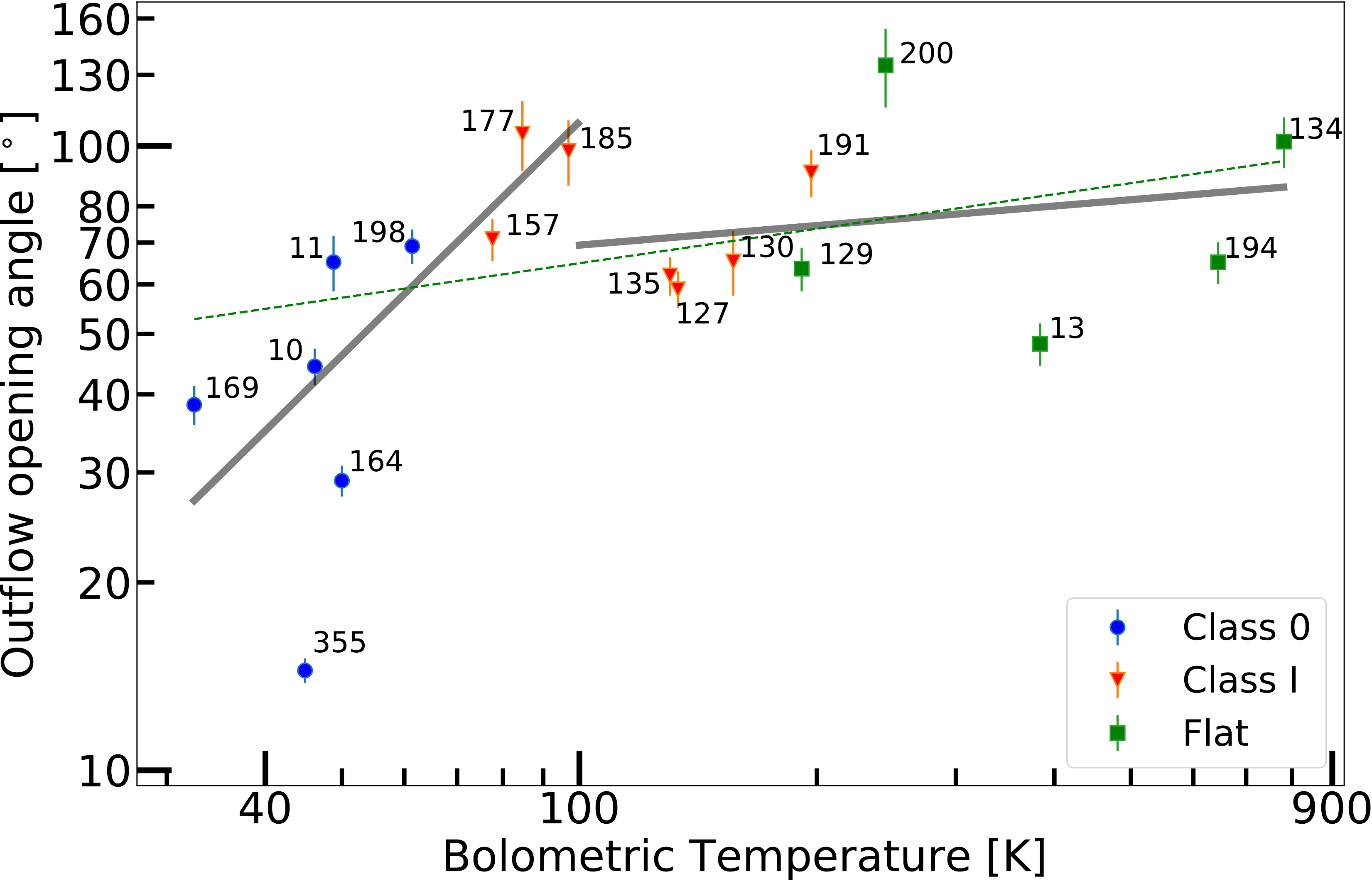}}
\caption{Outflow opening angle versus bolometric temperature for the sources in our sample. 
The opening angle for 2 flat sources (HOPS-150 and HOPS-166) and one Class 0 source (HOPS-408) are not included in the plot as their opening angles are ill-defined (see text for details). For HOPS-194, the Gaussian fits to the $^{12}$CO pixel angle distribution is very poor, so we measured the outflow opening angle by eye using the C$^{18}$O integrated intensity map. The thin dash green line represents a power-law fit of all the data described by equation $y=10^c x^m$, with $m = 0.18 \pm 0.10$ and $c = 1.45 \pm 0.22$. The two thick grey lines represent the broken power-law fits with a break  at $T_{\rm bol} = 100\,K$. For $T_{\rm bol} \le 100\,K$, $m = 1.24 \pm 0.30$ and $c = -0.45 \pm 0.57$; $T_{\rm bol} > 100\,K$, $m = 0.10 \pm 0.19$ and $c = 1.63 \pm 0.47$. Outflow opening angles have been corrected for the outflows' inclination. 
}
\label{fig:opening_angle}
\end{figure*}

In \autoref{fig:2D_MPE_dMdPdE}, we present {the outflow instantaneous mass, momentum, and energy ejection rate maps for HOPS-355.} For other sources, the maps are shown in the Appendix (\autoref{fig:HOPS10} to \autoref{fig:HOPS408}). { It is important to point out that instead of a single rate value, as obtained by previous methods (e.g.\ \citealt{2013A&A...556A..76V}), the 2D maps from PFT contain spatial information of how the rates vary within the outflow cavity.} The pixel scale in the figure sets is 66\,au (0\as17). Each map carries a different physical meaning. The instantaneous molecular outflow mass rate maps track the material entrained by the outflow. The instantaneous momentum rate maps show the force acting on the gas in each pixel. The instantaneous energy ejection rate maps, which highly weigh the high-velocity components, show the distribution of the molecular outflow mechanical luminosity and typically reveal the collimated high-velocity components which are closely linked to the accretion-driven jet launched by the protostar. We will use the instantaneous energy ejection rate maps to identify protostellar jets and explore the relationship between jet mass-loss rate and accretion rate in a future paper.  

\subsection{Outflow opening angles}

To { assess} the importance of { outflows} in the removal of dense gas around the protostar, one important quantity to consider is the outflow opening angle, as it can be used to estimate the volume of gas cleared in a protostellar core by the outflow.
To derive the outflow opening angle we follow a method based on that described by \citet{2011ApJ...743...91O,2020ApJ...896...11F}{ ; Dunham et al. in prep.} We first construct an outflow mask based on the $^{12}$CO integrated intensity maps with emission above 3$\sigma$, where we exclude velocity channels with cloud emission. We { calculate} the position angle for each pixel and then obtain the distribution of pixels at different angles within the outflow mask. We then fit this with a Gaussian to obtain the mean of the distribution ($\mu$), from which we obtain the outflow axis position angle, and the dispersion ($\sigma$). The outflow opening angle is then  defined as the full-width quarter maximum ($\Theta_{FWQM} = 3.3302 \sigma$) of the distribution. This definition is selected to match the opening angle derived by eye and to be consistent with past studies (e.g., \citealt{2011ApJ...743...91O}; Dunham et al. in prep.) 
The opening angles measured by this method are the projected outflow opening angles in the plane-of-sky and an inclination correction is needed. Without correction, the outflow opening angle for systems that are close to pole-on (i.e., small values of $i$) would be overestimated. We adopted a simple conical outflow model to correct the opening angles for the systems' inclination, as described by M.~Dunham et al.~(2023, in preparation). 

We were not able to obtain reliable estimates of the opening angle using this technique for the Class 0 HOPS-408, as its outflow is barely resolved by our observations, and the flat-spectrum sources HOPS-150 and HOPS-166, as their outflows are very clumpy or spotty which resulted in very poor Gaussian fits to the pixel angle distribution. For HOPS-194, the Gaussian fits to the $^{12}$CO pixel angle distribution is very poor, but the outflow cavity can clearly be traced by C$^{18}$O as shown in \autoref{fig:outflow_gallery}. Thus, for this source, we measured the outflow opening angle by eye using its C$^{18}$O moment 0 map. 

\autoref{fig:opening_angle} shows a plot of opening angle as a function of bolometric temperature ($T_{bol}$), which is usually considered a reasonable proxy for protostellar evolution (i.e., the more evolved protostars have higher $T_{bol}$ values, see \citet{1995ApJ...445..377C,1998ApJ...495..871L}). In \autoref{fig:opening_angle} we identify the protostellar evolutionary class of each source with one of three colors. It can be seen that while there is considerable scatter among protostars of the same evolutionary class, there is a clear trend where the outflow opening angle increases with the protostellar evolutionary stage. The increasing trend in the outflow opening angle is consistent with the $^{12}$CO maps shown in \autoref{fig:outflow_gallery}, and implies that outflows may play an important role in clearing the surrounding envelope material.      

\subsection{Momentum and energy opening angles }
\label{sec:EPopening_angle} 




\begin{figure*}[!htp]
\centering
\subfloat{%
  \includegraphics[width=0.48\textwidth]{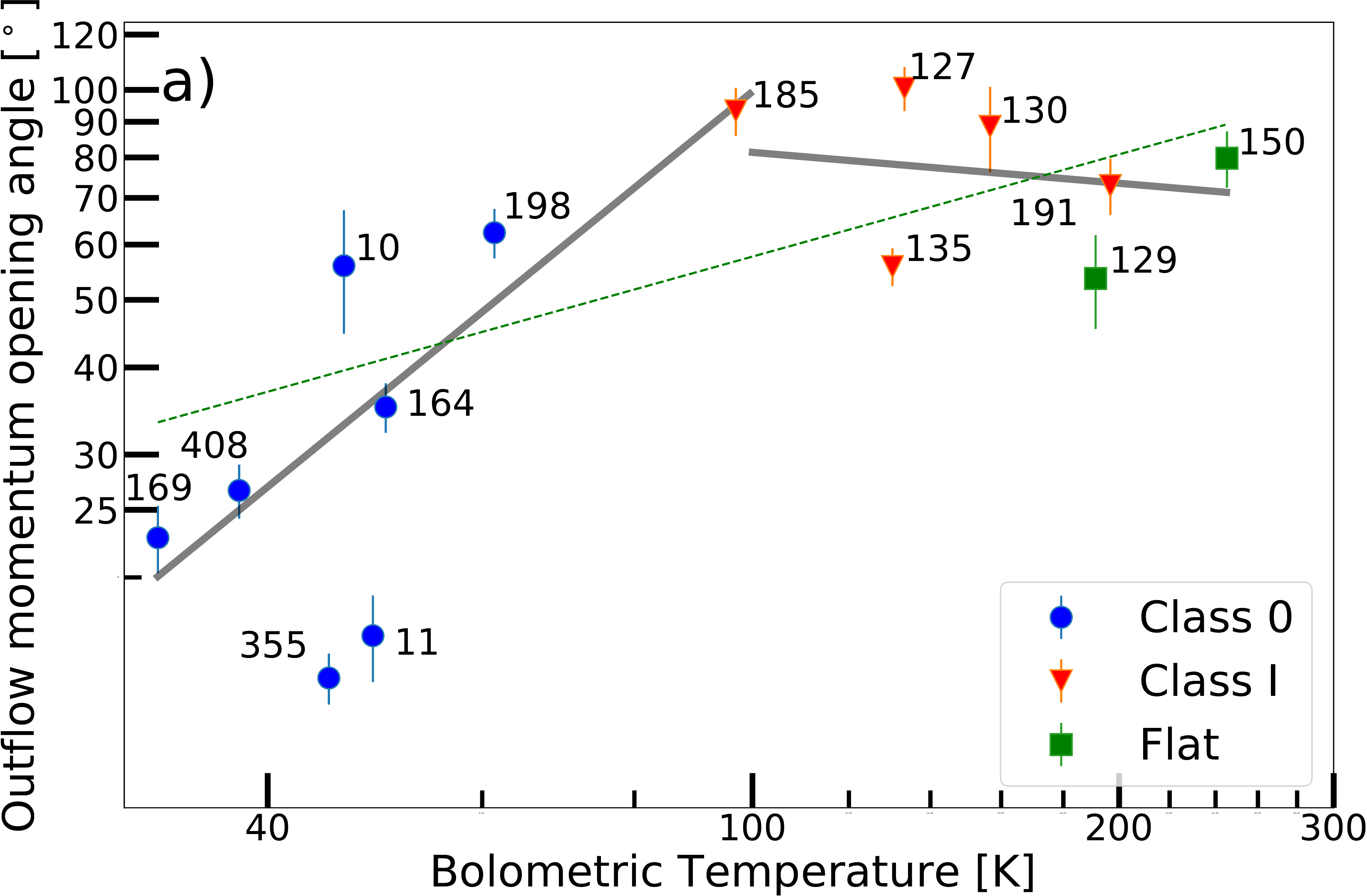}%
}\quad
\subfloat{%
  \includegraphics[width=0.48\textwidth]{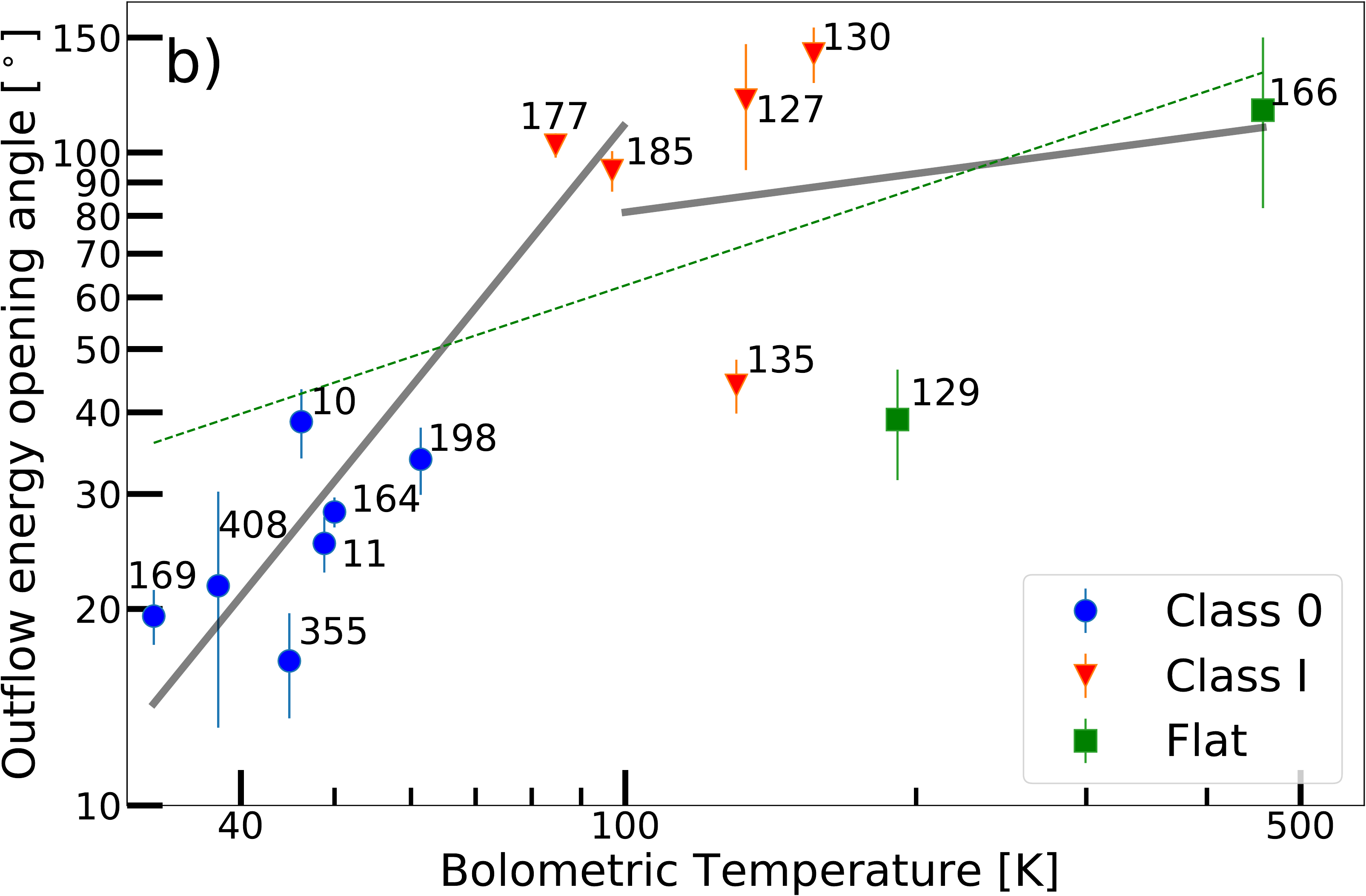}
}
\caption{Outflow momentum opening angle (left) and outflow energy opening angle (right) vs. $T_{bol}$ for the sources in our sample. In both panels, the green line represents a power-law fit to all the data using equation $y=10^c x^m$. { For the momentum opening angle, the fit to all the data gives $m = 0.44 \pm 0.18$ and $c = 0.79 \pm 0.36$. Similarly, for the energy opening angle $m = 0.49 \pm 0.17$ and $c = 0.81 \pm 0.38$.} In each panel, the two black lines represent broken power-law fits with a breakpoint at $T_{\rm bol} = 100\,K$. For the momentum opening angle, $m = 1.42 \pm 0.32$ and $c = -0.85 \pm 0.60$ for sources with  $T_{\rm bol} \le 100\,K$, and $m = -0.14 \pm 0.49$ and $c = 2.20 \pm 1.10$ for sources with $T_{\rm bol} > 100\,K$. For the energy opening angle, $m = 1.82 \pm 0.28 $ and $c = -1.59 \pm 0.52$ for sources with $T_{\rm bol} \le 100\,K$, and $m = 0.20 \pm 0.51$ and $c = 1.52 \pm 1.18$ for $T_{\rm bol} > 100\,K$. Note that HOPS-150 is taken out for the energy opening angle because it is contaminated by a high-velocity jet from a nearby source outside the field of view. The contamination is much weaker for the momentum map as compared to the energy map. 
Sources with ill-defined opening angles are not included in the plots. Opening angles have been corrected for the outflows' inclination.} 

\label{fig:PEopening_angle}
\end{figure*} 

\begin{figure*}[tbh]
\centering
\makebox[\textwidth]{\includegraphics[width=\textwidth]{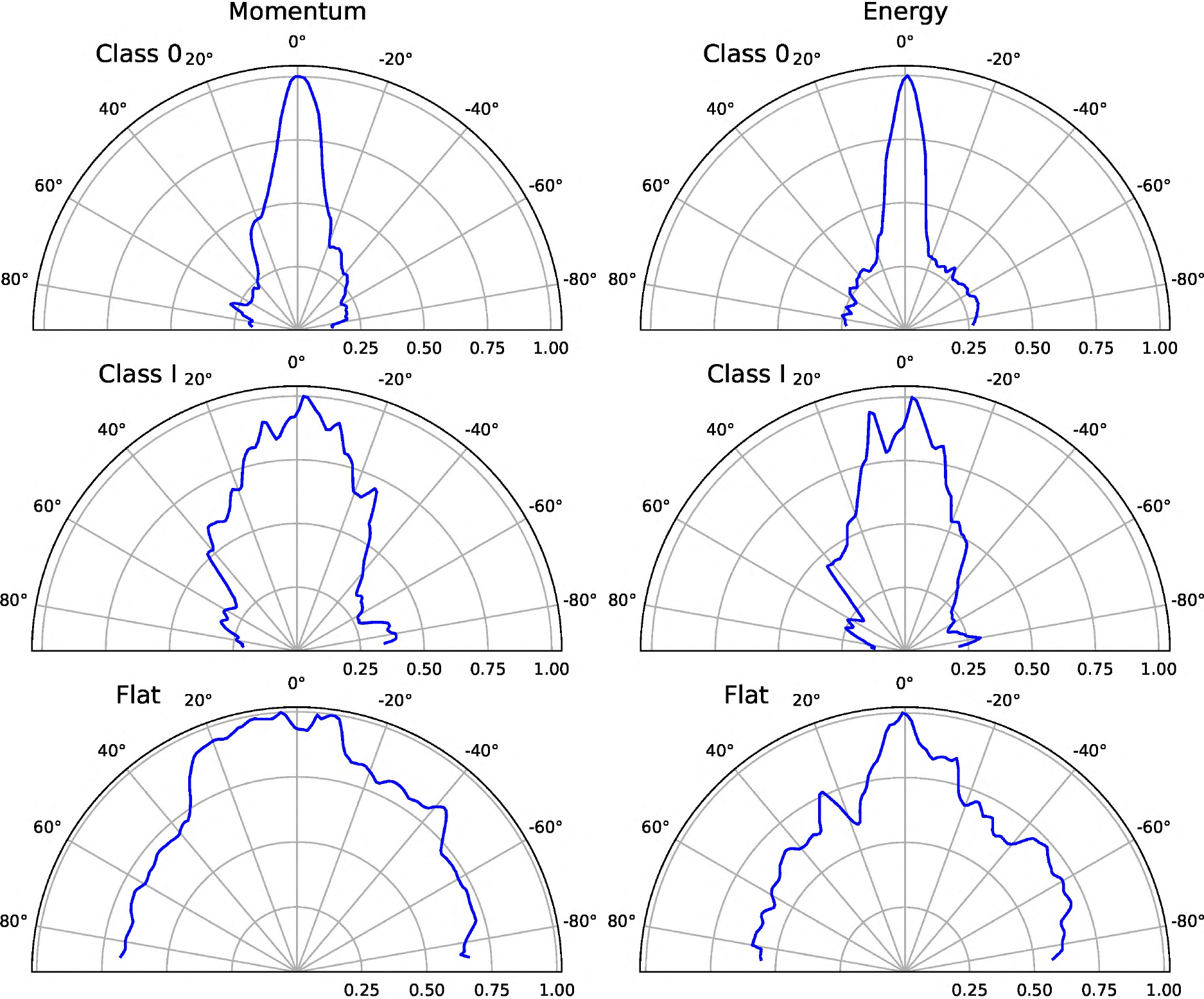}}
\caption{ Average normalized { momentum (left panels) and energy (right panels)} angular profile for protostars at different evolutionary stages: Class 0 (top), Class I (middle); and flat-spectrum (bottom). The 0$^\circ$ marks the direction of the outflow axis. { All profiles are measured within the region of our maps where the  sensitivity falls down to 40\% of the peak sensitivity at the center of the map  (i.e., 25\as0 radius, or about 10,000 au from the center). Emission perpendicular to the outflow direction (i.e., $\le -85^\circ $ and $\ge 85^\circ$) is not used to  estimate the average momentum and energy in order to avoid possible contamination from the circumstellar disk / inner envelope material. All angular profiles have been corrected for the outflows' inclination. 
} 
}
\label{fig:PE_polar}
\end{figure*}

To further understand how the outflow interacts with the surroundings, we { investigate} the angular distribution of the outflow momentum and energy. 
We start with estimating the outflow momentum and energy opening angle from the 2D momentum and energy map of each source (as shown in \autoref{fig:2D_MPE_dMdPdE} and \autoref{fig:HOPS10} to \autoref{fig:HOPS408} in the Appendix). We calculate the position angle of each pixel in the map, with respect to the source. We then determine the distribution of the average momentum/energy as a function of angle, and fit a Gaussian to estimate the momentum/energy opening angle. 
The opening angle is defined as the full-width quarter-maximum (FWQM) of the Gaussian fit. The measured momentum, energy opening angles are shown in \autoref{fig:PEopening_angle}. The outflow momentum and energy opening angles widen as a protostar evolves. To quantify the rate at which the opening angle widens we fitted a { power-law $y\,=\,10^c\,x^m$} to all the angular profiles shown above (See \autoref{fig:PEopening_angle}). { Note that several sources are not included  due to a poor Gaussian fit to their momentum/energy angle distribution.} 

In \autoref{fig:PE_polar} we show polar plots of the normalized average outflow momentum and energy for each evolutionary class. To construct these plots we normalized the momentum/energy profile (distribution) so that the peak value of each outflow has a value of one, and we then obtained the average profile for each protostellar evolutionary class. 
{ Note that the momentum/energy profiles include the high velocity jets as shown in \autoref{fig:dM_v}.} In \autoref{fig:PE_polar}, it is clearly seen that the average momentum and energy angular profiles change significantly between different evolutionary classes. For Class 0 sources the energy and momentum are much more collimated, compared to more evolved sources. 
The vast contrast in the angular profile of energy and momentum between Class 0 and Class I confirms that a significant widening of the outflow momentum and energy angular profile occurs { as the protostar evolves}. 


The significant widening of angular profiles shown \autoref{fig:PE_polar} { suggests} that the impact of protostellar outflows launched 
on the surrounding environment evolves rapidly at the earliest stage of star formation. In { Section \ref{sec:mass_rates_results} } we noted that while molecular outflows from  Class 0 sources have significantly higher momentum rate as compared to those from  Class I sources, the total mass loss rate for Class 0 molecular outflows is lower than the Class I outflows  (\autoref{fig:dM_dP_dE_r}).
These results are consistent with a picture in which 
the high velocity collimated Class 0 jets quickly { punch} a hole in the protostellar core. Later, in the Class I phase the entrained molecular outflow slows down and is spread out to a much larger volume. We thus expect the majority of the gas clearing occurs { during the} Class I stage.

\section{Discussion}
\label{sec:discussion}


\begin{figure*}[tbh]
\centering
\makebox[\textwidth]{\includegraphics[width=\textwidth]{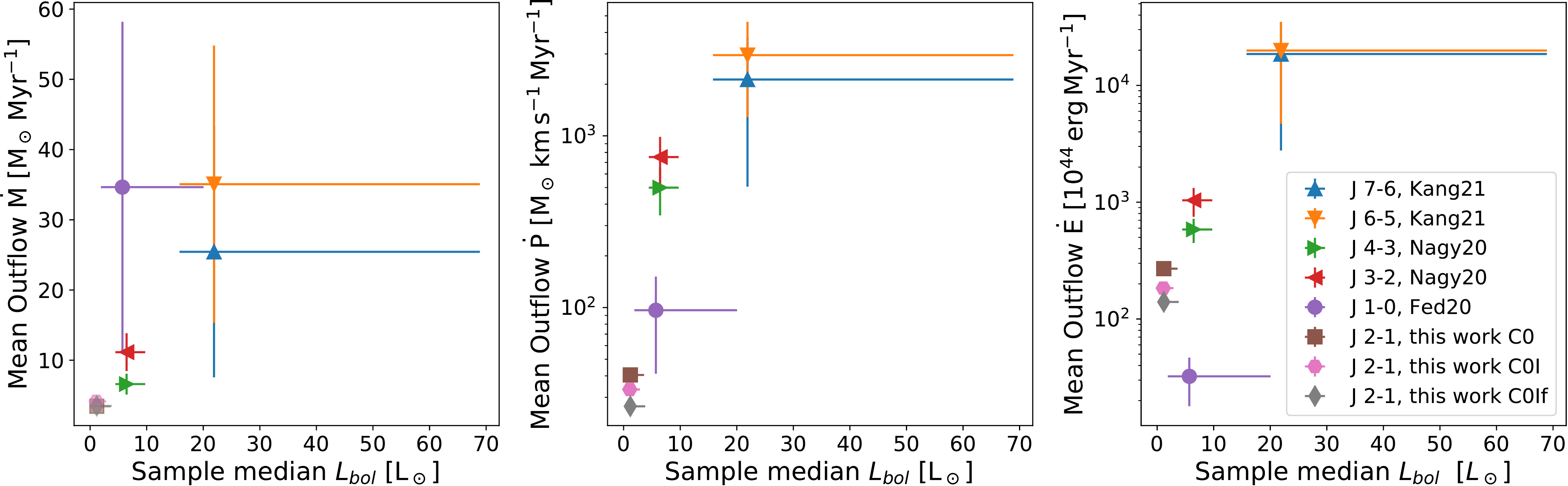}}
\caption{Comparison of  molecular outflow rates in Orion A. The left, middle, and right panels show the mean molecular outflow mass rate, momentum rate and energy rate of various samples against the median bolometric luminosity of the samples, respectively. Different colors represent different samples and CO rotational transitions used to measure the molecular outflow rates.   
 The sample by Nagy20 
includes 16 very young Class 0 protostars observed with APEX
 \citep{2020A&A...642A.137N}.
 The sample by Kang21  is composed  of 6 Class 0 and 3 Class I observed with APEX 
\citep{2021ApJS..255....2K}.
The Fed20 study includes 
45 protostars observed as part of the NRO-CARMA Orion survey \citep{2020ApJ...896...11F}. C0, C0I, C0If are the outflow rates from this study, corrected for the sample composition of Nagy20, Kang21 and Fed20, respectively, as described in the text.
The numbers in the legend next to the letter J represent the CO transition used in the survey.
}
\label{fig:compare_rate}
\end{figure*}

\subsection{ On selection effects: RM vs. R/$v_{max}$ methods
\label{rates_methods}
} 




Different methods used in measuring protostellar outflow rates have resulted in values varying by a factor of 5 to 6 \citep{2013A&A...556A..76V}. 
It is thus important to benchmark the different techniques, identifying all the assumptions and caveats. As described above, we used two methods to determine outflow rates in our sample: the Ring Method (RM) and the Pixel Flux-tracing Technique (PFT). We first compare our methods with the commonly used dynamical time approach, hereafter the $R_{lobe}/v_{\rm max}$ method, 
in which  dynamical time is estimated by using the simple equation $t_{dyn}=R_{lobe}/v_{max}$, where $R_{lobe}$ is the length of the outflow and $v_{max}$ is the maximum outflow velocity. Outflow mass-loss rate, momentum, and energy injection rate are estimated by dividing the total mass, momentum, and energy by $t_{dyn}$. This method thus ({ incorrectly}) assumes that the molecular gas inside the outflows moves from the source to the current position at a single velocity. In addition, as discussed below, this method highly depends on the sensitivity and extent of the observations. 

Using the $R_{lobe}/v_{\rm max}$ method  to estimate the  mass, momentum, and energy ejection rate of the
21 sources in our survey, we find estimates that are, on average,   { 80, 350, and 8} times higher than our RM results, respectively. The results highlight the problem for using a single dynamic time to estimate rates. 
The choice of a characteristic velocity ($v_{max}$), which typically is the maximum velocity relative to the system velocity where $^{12}$CO is detected at $3\sigma$, is dependent on the sensitivity of the observations.
Previous ALMA observations have { revealed} molecular outflow emission at much higher velocities than typical single-dish observations
\citep[e.g.,][]{2013ApJ...774...39A, 2016ApJ...823L..27H}. 
 Thus, it is not surprising that with our ALMA data we detect high values of  $v_{\rm max}$, which result in extremely { large} rate estimates when using the
 $R_{lobe}/v_{\rm max}$ method. This  clearly results in an overestimation of the outflow rates as only a small fraction of the outflow mass moves at the highest velocities. 
 For example, for HOPS-10 the $v_{\rm max}$ is close to 100\,km\,s$^{-1}$. Yet, only 0.9\% of the total molecular outflow mass is moving at this high speed.

 
 The $R_{lobe}/v_{\rm max}$ method also suffers from the fact that { it} depends on the measured outflow lobe length, which depends on the extent of the area observed around the outflow source and the sensitivity of the observations. In our case all our observations are restricted to 25'' (10,000 au) from the source, which in most cases is significantly smaller the the full extent of the outflow lobe. Hence, our observations would vastly underestimate $R_{lobe}$. We thus stress that this method should not be used to estimate dynamical times and rates as both $R_{lobe}$ and $v_{\rm max}$ highly depend  on the observations' parameters. In fact, two different (single-dish) studies with overlapping sources estimate the dynamical timescales of $^{12}$CO outflows that differ by a factor of 5 for the same sources \citep{2008ApJ...688..344T,2021ApJS..255....2K}. 

The unreliability of selecting outflow length and outflow characteristic velocity for the dynamical timescale estimates may have resulted in  conflicting results regarding outflow rates,  even in large outflow surveys. \citet{2020ApJ...896...11F} studied 45 protostellar outflows in the Orion A molecular cloud using a medium-resolution (8'') CO map. They applied the optical depth correction outlined by \citet{2014ApJ...783...29D} for the CO maps and they found that the molecular outflow rates are independent of evolutionary { stage}. This is in conflict with the Atacama Pathfinder EXperiment (APEX) $^{12}$CO $J = 3-2$ study of 49 sources by \citet{2017A&A...600A..99M} which found the outflow mass and momentum ejection rates decrease between Class 0 and Class I. { Similar} trends were also found by \citet{2015A&A...576A.109Y}.

\citet{2013A&A...556A..76V} compared seven different methods for obtaining 
outflow force (i.e., momentum rate). They found that estimates from these different methods can differ by as much as a factor of six for an individual source. Most of these methods rely on measuring the size of the outflow lobe or maximum outflow velocity. As discussed above these two parameters strongly depend on the observations' strategy and sensitivity. Of all the methods compared in that study, the Annulus Method developed by \citet{1996A&A...311..858B} on which we base our Ring Method, 
seems to be the one that depends the { least} on observing parameters (e.g., field of view or area covered by observations) and thus we prefer and recommend its use over the other methods discussed in \citet{2013A&A...556A..76V}. 

\subsection{Outflow Rates and Bolometric Luminosity}
\label{sec:outflow_rate_Lbol}

{ To} put our results in context with those of recent studies, we { compare our results to those of 
\citet[hereafter Nagy20]{2020A&A...642A.137N}, \citet[hereafter Kang21]{2021ApJS..255....2K}, and \citet[hereafter Fed20]{2020ApJ...896...11F} -- outflow studies of protostellar outflows from sources in the Orion molecular clouds.
{ To} ensure a fair comparison between our results and those of the other studies  (and remove any possible bias due to a differing percentage of sources at different evolutionary stages for the different studies), we re-estimate the average outflow rates for our sample by multiplying the average rate per evolutionary class we determined (Sec. \ref{sec:mass_rates_results}) by the percentage of sources in that evolutionary class in the sample of the particular study. { For example, in \citet{2020ApJ...896...11F}  sample,  51\% are Class 0 sources, 28\% are Class I, and 21\% are flat-spectrum sources. 
We then estimate the expected total molecular outflow mass rate we would derive if our sample had the same composition as that of the \citet{2020ApJ...896...11F} sample
to be:
51\% $\times$ 3.5 + 28\% $\times$ 5.6 + 21\% $\times$ 0.8 = 3.5\,M$_\odot$ Myr$^{-1}$, where 3.5, 5.6, and 0.8 M$_\odot$ Myr$^{-1}$ are the average mass-loss rates we derive from our sample for the Class 0, Class I and flat-spectrum sources, respectively.
In \autoref{fig:compare_rate}, we plot the average outflow rates for outflows driven by HOPS sources as determined   by Nagy20, Kang21, and Fed20, and we compare them to the average values of our study, corrected for the sample demographics of these other studies.} 
In the x-axis we plot the median bolometric luminosity for the sample in each study, with error bars showing the range between the first and third quartiles of $L_{bol}$ values in the sample. In the vertical axis we plot the sample average rates with error bars representing the standard deviations of the sample sources for Nagy20, Kang21 and our study. For Fed20, the vertical error-bars represent the range  of values obtained when including and excluding the low-velocity outflow components to estimate the rates.


{ {The most dominant factor that affects the outflow rate measurements is the selection of source bolometric luminosity range.} We see a very clear trend in \autoref{fig:compare_rate}, namely, studies with higher bolometric luminosity samples have significantly higher estimates of molecular outflow rates. The rate estimates can vary by a factor of 100 depending on the sample bolometric luminosity range. Molecular outflow rates  obtained from different $^{12}$CO transitions only varied by a factor of 2 for studies with the same sources (e.g.\  \citet{2021ApJS..255....2K} use $^{12}$CO transition 7-6 and 6-5).
The strong dependence on bolometric luminosity shows that the protostellar outflow feedback varies significantly depending on protostellar mass or accretion rate.  } 

These { increasing trends of the molecular outflow rates with bolometric luminosity} are consistent with the results shown by 
\citet{2015MNRAS.453..645M} for outflows from massive young stellar objects, with $L_{bol} > 10^3 L_{\sun}$. 
Previous studies have shown a clear correlation of the $^{12}$CO outflow force (i.e., outflow momentum rate) with the source's bolometric luminosity for a wide range of $L_{bol}$ values, from $\sim 10^{-2}$ to $ \sim 10^5 L_{\sun}$
\citep[e.g.,][]{1992A&A...261..274C,2016A&A...587A..17V,2015MNRAS.453..645M}.
From the left and right panels of \autoref{fig:compare_rate},
it appears that the increasing trend with $L_{bol}$ is also present 
in the outflow mass rate and the outflow energy  rate  for $L_{bol} < 10^2 { L_\odot}$, and not just for  { sources with} high values of $L_{bol}$ reported in \citet{2015MNRAS.453..645M}.
It is clear that in order to study (and detect) any possible evolutionary trends in the outflow rates one needs to {  rigorously control} the range in bolometric luminosity of the sample.

\subsection{Outflow mass loss as  evidence of core growth}

\begin{table*}
\setlength{\tabcolsep}{6pt} 
\caption{Total mass removed by the molecular outflow and its corresponding molecular outflow mass-loss rates at different evolutionary stages for different escape velocities }             
\label{table:core_growth}      
\centering                          
\begin{tabular}{c c c c c c c c c c c }        
\hline\hline                 
 & & $\overline{\dot{M}}_{out}$ & $\overline{\dot{M}}_{out}$ for & $\overline{\dot{M}}_{out}$ for & $M_{out}$ & $M_{out}$ for & $M_{out}$ for & Core  \\
Class & Duration  & for all gas &  $v_{esc} \ge 1\,$km$\,$s$^{-1}$ & $v_{esc} \ge 2\,$km$\,$s$^{-1}$& for all &  $v_{esc} \ge 1\,$ & $v_{esc} \ge 2\,$ & Mass\tablenotemark{a}  \\
& [Myr] &[M$_\odot$ Myr$^{-1}$] &  [M$_\odot$ Myr$^{-1}$] &  [M$_\odot$ Myr$^{-1}$] & gas [M$_\odot$] & km$\,$s$^{-1}$ [M$_\odot$] & km$\,$s$^{-1}$ [M$_\odot$] & [M$_\odot$] \\
\hline                        
0\tablenotemark{b} &0.13 - 0.26 & 3.5 & 3.2 & 2.6 & 0.5 - 0.9 & 0.4 - 0.8 & 0.3 - 0.7 & (0.95, 1.04) \\
\hline                                   
I  & 0.27 - 0.52 & 5.6 & 5.4 & 4.4 & 1.5 - 2.9 & 1.5 - 2.8 & 1.2 - 2.3 & (0.49, 0.49) \\
\hline   
flat\tablenotemark{c} & 0.27 - 0.52 & 0.8 & 0.7 & 0.5 & 0.2 - 0.4 & 0.2 - 0.4 & 0.1 - 0.2 & (0.46, 0.26) \\
\hline   

\end{tabular}
\tablenotetext{a}{ The first value is the average core mass from our ALMA C$^{18}$O data. The second value is the average core mass derived from the JCMT 850 $\mu$m dust continuum data \citep{2016ApJ...833...44L}. Both  core mass estimates are measured using a similar core size of $\sim$6000\,au.}
\tablenotetext{b}{HOPS-408 is excluded from these estimates as we were not able to obtain a reliable molecular outflow mass rate for this source.} 
\tablenotetext{c}{ HOPS-194 and HOPS-166 are excluded from the average as these two sources  have significantly higher core mass  derived from C$^{18}$O as compared to core masses derived from dust continuum. They also have the highest bolometric luminosity ($ > 10\,L_\odot$) in our sample.} 
\end{table*}

Our 
survey of 21 protostars has found that { by} the end of the { protostellar phase}, a typical outflow  can displace around 2.2 to { 4.2} M$_\odot$ { of core material}. However, this estimate includes all the gas that we identified as outflowing gas, and some of it, although perturbed and entrained by the protostellar wind, might still be gravitationally bound to the core. The bounded outflowing gas may eventually fall back towards the protostellar core forming a protostellar version of a ``Galactic fountain" \citep{2008A&A...484..743S}.  To determine how much core gas could be removed by an outflow, we need to estimate the core's escape velocity. For a 5 M$_\odot$ protostellar core, the corresponding escape velocity { ($v_{esc}$)} at a radius of 10,000\,au is $\sim 1$\,km s$^{-1}$. 
Alternatively, for a 2 M$_\odot$ protostellar envelope, the corresponding escape velocity ($v_{esc}$) at a radius of 1,000\,au is about $\sim 2$\,km s$^{-1}$. Each protostar has a different escape velocity depending on the mass of the system (protostellar + envelope). Unfortunately, we do not know the mass of each protostar in our sample. Hence, for simplicity, we adopted escape velocities of $1$\,km s$^{-1}$ and $2$\,km s$^{-1}$ for the whole sample.


{ We use the Ring Method to calculate the outflow mass loss rate (as in Sec.~\ref{sec:mass_rates_results})
for gas with a velocity greater than the escape velocity of $1$\,km s$^{-1}$ and $2$ \,km s$^{-1}$. Note that we assume the outflow gas moves mostly along the direction of the outflow axis, and we use the outflow inclination angle to obtain the gas velocities that we compared with the escape velocity. Some gas is expected to move perpendicular to the outflow direction due to, for example the thermal expansion of the gas in the post-shock zone behind the working surface
\citep{2001ApJ...557..443O,2007A&A...471..873D}. However this is expected to be much less massive than the gas that moves along the outflow direction.  
Hence, for simplicity we do not include gas moving perpendicular to the outflow direction. 


Adopting the protostellar ages for each phase from \citet{2015ApJS..220...11D},
see \autoref{table:core_growth},
we found for escape velocity of $1$\,km s$^{-1}$, the outflow will remove 0.4 - 0.8\,M$_\odot$ (using an average molecular outflow mass rate of $\overline{\dot{M}}_{out} = 3.2\,M_\odot$\,Myr$^{-1}$) of core material during the Class 0 phase , 1.5 - 2.8 M$_\odot$ (using $\overline{\dot{M}}_{out} = 5.4\,M_\odot$\,Myr$^{-1}$) during the Class 1 phase, and 0.2 - 0.4 M$_\odot$ (using $\overline{\dot{M}}_{out} = 0.7\,M_\odot$\,Myr$^{-1}$) in the flat-spectrum phase.}

{ Similarly, for cores with a higher escape velocity of $2$ \,km s$^{-1}$, the outflow will remove about 0.3 - 0.7\,M$_\odot$ of core material during the Class 0 phase, with an average mass loss rate of $\overline{\dot{M}}_{out} = 2.6\,M_\odot$\,Myr$^{-1}$, approximately 1.2 - 2.3\,M$_\odot$
of core material during the Class I phase, with an average mass loss rate of $\overline{\dot{M}}_{out} = 4.4\,M_\odot$\,Myr$^{-1}$, and roughly 0.1 - 0.2 M$_\odot$ of core material during the flat-spectrum phase with an average $\overline{\dot{M}}_{out} = 0.5\,M_\odot$\,Myr$^{-1}$. By the end of the protostellar phase, an outflow could remove a total of 2.1 - 4.0\,M$_\odot$ or 1.7 - 3.2\,M$_\odot$ of material from the protostellar core, assuming a core escape velocity of $1$ and $2$ \,km s$^{-1}$ respectively. The results are summarized in \autoref{table:core_growth}.} The derived values of the molecular outflow at 5 different distances from the source for velocity greater than the escape velocity of $1$\,km s$^{-1}$ and $2$ \,km s$^{-1}$, for all sources in our sample are listed in the Appendix (see \autoref{table:outflow_rate_v1_p1} to \autoref{table:outflow_rate_v2_p3}).


{ The averaged mass of the entrained outflow gas derived from our sample is comparable to or larger than the median of the protostellar core mass in the Orion A cloud \citep {2021PASJ...73..487T}
and the median of the cores mass in the Dragon infrared dark cloud \citep{2021ApJ...912..156K}. We also use our ALMA Cycle 6 C$^{18}$O data, as well as the JCMT 850 $\mu$m dust continuum core catalog identified by \citet{2016ApJ...833...44L}, to drive the mass of the cores harboring the protostars in our samples (see  \autoref{table:1} and \autoref{table:core_growth}). Both mass estimates were measured using a  similar core size of $\sim$6000\,au. Note that HOPS-408, HOPS-166 and HOPS-194 are excluded in the averaged outflow rate, mass estimates and the averaged core mass in \autoref{table:core_growth}. }

{ From \autoref{table:core_growth}, for escape velocity of 1\,km\,s$^{-1}$ we found that around 40\% to 80\% of the average core mass would be removed by the protostellar outflow by the end of Class 0. For Class I, the mass removed by the protostellar outflow is 3 to 6 times the average mass of a core in this stage. For flat-spectrum sources, we estimate that the outflow will remove
round 43\% to 87\% or 77\% to 154\% of the core mass, depending if we assume an average core mass derived from C$^{18}$O and dust continuum, respectively.} Protostellar cores { generally} do not live in isolated environments, but instead are embedded in larger filaments and clouds \citep{2014prpl.conf...27A,2021ApJ...908...92H}. { The high percentage of total core mass removed by outflows at each protostellar phase implies that the mass budget of the protostellar core must be continuously replenished by material from larger scales.} 




{ The biggest uncertainty in quantifying the mass-loss by protostellar outflows at each evolutionary stage is the time a protostellar system stays in each evolutionary class. The lifetime estimates we used here, reported by \citet{2015ApJS..220...11D}, are one of the most commonly used. \citet{2018A&A...618A.158K} use a different technique, a nuclear decay half-life model, to derive the duration of each evolutionary class. Converting the derived half-life to the average lifetime for each stage, the derived lifetime for Class 0, Class I, and flat-spectrum sources are $68 \, \pm \, 6$\,kyr, $130\, \pm \,10$\,kyr, $126 \, \pm \,12$\,kyr, respectively. They are a factor of 2 and 4 lower than the lower and upper bound of the lifetimes estimated by \citet{2015ApJS..220...11D}. \citet{2014prpl.conf..195D} and \citet{2022PASP..134d2001M} obtained protostellar lifetimes longer than the lifetimes from \citet{2018A&A...618A.158K}, but shorter than the values in \citet{2015ApJS..220...11D}. Even if we were to adopt the duration of the protostellar evolutionary classes from \citet{2018A&A...618A.158K} this would only lower the total core mass removed by protostellar outflows by a factor of 2; which is still a significant fraction of the total average core mass.
 } 

Recent studies have found a clear mass difference between protostellar and starless core in the Dragon infrared dark cloud (a.k.a., G28.37+0.07 or G28.37+0.06) \citep{2021ApJ...912..156K}. Adopting a typical temperature of 20\,K from the Herschel dust temperature map, the authors found that the medium value for protostellar core masses is 2.1\,M$_\odot$ and for starless cores is 0.37\,M$_\odot$. Protostellar cores are found to have similar densities but larger sizes as compared to starless cores. The mass difference can be attributed to the continuous mass accretion onto the protostellar cores from filaments, as predicted by simulations \citep{2019MNRAS.485.4509L}. 
{ The high protostellar outflow mass-loss from our ALMA survey poses serious challenges to the current concept of ``static" cores, as well-defined reservoir of mass, and the simplified core to star conversion ideas. Together with evidence of core growth predicted by simulations (e.g., \citealt{2019MNRAS.485.4509L}) and results from other observational studies (e.g., \citealt{2021ApJ...912..156K}), our results point to a more dynamic star formation at core scales.}


\subsection{Evolution of conventional outflow opening angle}

\begin{figure}
\centering
\includegraphics[width=\hsize]{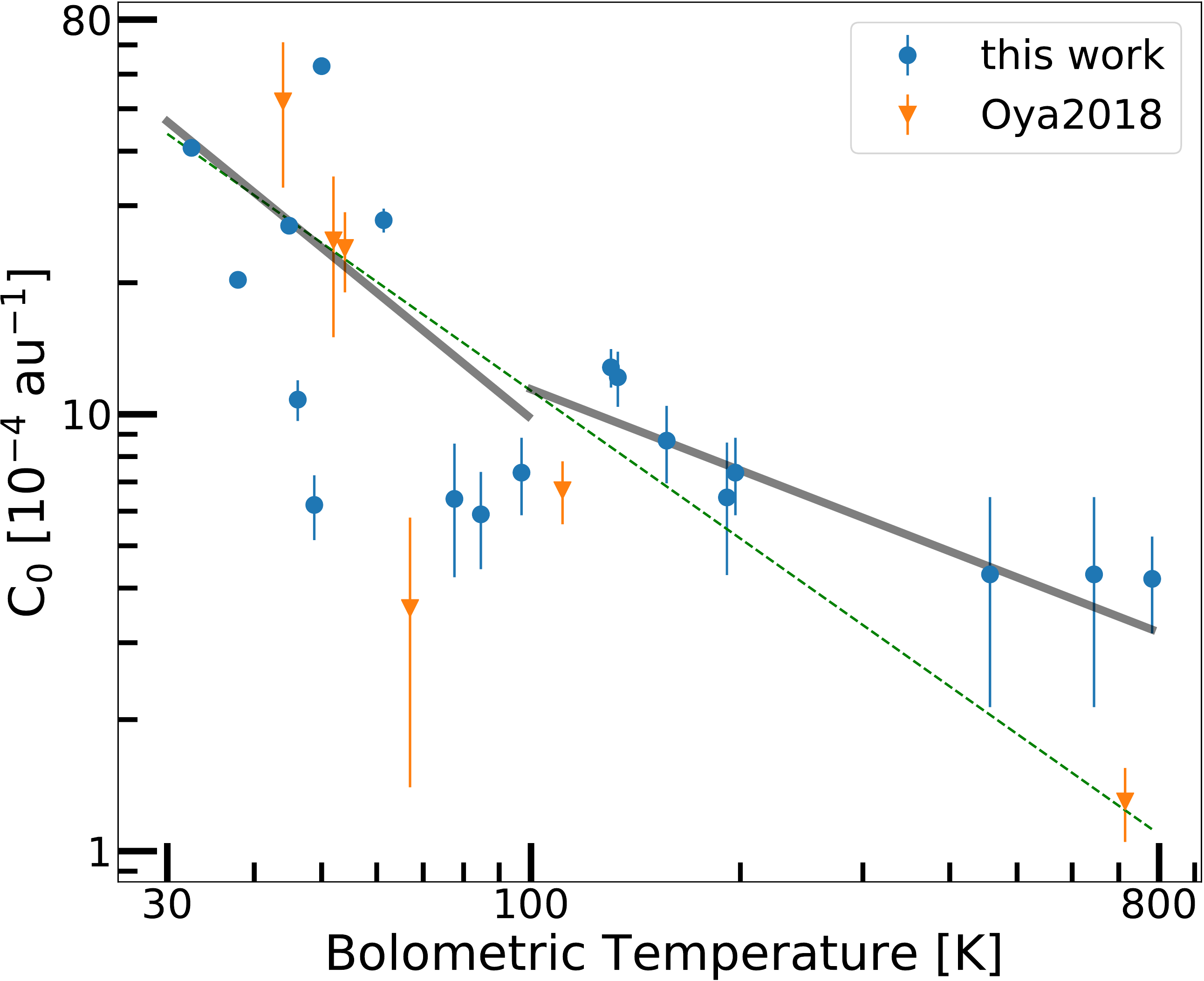}
\caption{Evolution of the Outflow curvature ($C_0$) from the wide-angle wind model. The orange points are literature values from \citet{2018ApJ...863...72O} which are also included in the fit. The thin dashed green line represents a power-law fit of all the data described by equation $y=10^c x^m$, with $m = -1.12 \pm 0.38$ and $c = 3.30 \pm 0.64$. The two thick grey lines represent the broken power-law fits with a break at $T_{\rm bol} = 100\,K$. For $T_{\rm bol} \le 100\,K$, $m = -1.30 \pm 0.72$ and $c = 3.59 \pm 1.20$; $T_{\rm bol} > 100\,K$, $m = -0.62 \pm 0.20$ and $c = 2.29 \pm 0.44$. Outflow curvature is an alternative way to quantify the outflow width. It is not affected by the source inclination angle and no inclination correction is needed. 
}
\label{fig:outflow_curvature}
\end{figure}

Protostellar outflows { play} an important role on core dispersal and setting the stellar mass \citep{2008ApJ...687..340M,2022MNRAS.515..167G}. The Spitzer Cores to Disks (c2d) legacy survey of 265 and 353 sources in { Ophiuchus} and Perseus has shown that all Class 0 protostars have associated cores observed by SCUBA instrument on James Clerk Maxwell Telescope. The number falls down to around 50\,\% for Class I and less than 3\% for more evolved Class II and Class III sources \citep{2008ApJ...683..822J}. A study of the Spectral Energy { Densities (SEDs)} of 330 Young Stellar Objects in the Herschel Orion Protostar Survey has shown that the envelope volume density  drops by a factor of 50 between Class 0 and flat-spectrum sources \citep{2016ApJS..224....5F}. Many simulations suggest that outflows are responsible for the dispersion of the protostellar core \citep{2013MNRAS.431.1719M,2014ApJ...784...61O,2017ApJ...847..104O}. Outflows from Class 0 sources are very collimated and therefore only impact a very limited volume, so if outflows are responsible for core dispersal, they need to become wider as they evolve. The widening of the outflow cavity found in multiple observations has suggested the impacted volume for outflow-core interactions increases over time \citep{2006ApJ...646.1070A,2014ApJ...783....6V,2017AJ....153..173H}. 

The recent { STARFORGE} simulations have shown that the widening of the outflow cavity allows protostellar outflows to disturb the accretion flows around the protostars, causing the gas to fragment into more low mass stars, setting the mass scale of stars and circumstellar disks \citep{2022MNRAS.515.4929G}. These { numerical} simulations also indicate that the peak of the IMF is sensitive to the momentum loading factor of the protostellar jets \citep{2022MNRAS.515.4929G}. Outflows are believed to be the main explanation for the low core-to-star efficiency of 30\% \citep{2007A&A...462L..17A}. Understanding how outflows interact with the protostellar core is one of the major { questions} in star formation, and the ``outflow opening angle" is one of the key parameters for understanding  the outflow-core interaction. 




We investigate the evolution of the outflow opening angles using our sample of 21 sources that includes seven sources of each of the protostellar classes (Class 0, Class I, and flat-spectrum). Performing a { least squared} fit to our data, as shown in \autoref{fig:opening_angle}, we find a correlation between the outflow opening angle ($\Theta_{\mathrm{FWQM}}$) and the bolometric temperature ($T_{bol}$) as:
 
\begin{gather}
\log \left(\frac{\Theta_{\mathrm{FWQM}}}{\operatorname{deg}}\right)=(1.5 \pm 0.2)+(0.18 \pm 0.09) \log \left(\frac{T_{\mathrm{bol}}}{\mathrm{K}}\right).
\end{gather} 
This is much shallower than the relationship derived from \citet{2006ApJ...646.1070A} based on the lower resolution 5\as0 (interferometer) maps of the $^{12}$CO outflows of 17 sources (combining results from their own survey and the literature): 
\begin{gather}
\log \left(\frac{\Theta_{\mathrm{FWQM}}}{\operatorname{deg}}\right)=(1.7 \pm 1.7 )+(0.38 \pm 0.96) \log \left(\frac{T_{\mathrm{bol}}}{\mathrm{K}}\right).
\end{gather} 
The difference in slope  might be caused by 
the fact that the 
\citet{2006ApJ...646.1070A}
study did not correct the outflow opening angle for the  outflow inclination. We conduct the same fit with our ALMA data using the opening angles without inclination correction in \autoref{table:opening_angles_noinc} and find the slope to be 0.38 $\pm$ 0.06 which is consistent with the results obtained by \citet{2006ApJ...646.1070A}.

A study of the outflow opening angle based on the Spitzer IRAC image of 31 YSOs by \citet{2014ApJ...783....6V} showed a clear increase in opening angle with age, with a break at 8000\,yr (which corresponds to a $T_{bol}$ of about $82\,$K), where the relationship then becomes approximately flat:


\begin{equation}
\log \left(\frac{\Theta_{\mathrm{FWQM}}}{\operatorname{deg}}\right)=
\begin{cases}
 2.05+0.76 \log \left(\frac{T_{\mathrm{bol}}}{\mathrm{K}}\right) 
& , \text {$T_{\mathrm{bol}}<82\mathrm{~K}$ }\\
2.03+0.05 \log \left(\frac{T_{\mathrm{bol}}}{\mathrm{K}}\right)
& , \text {$T_{\mathrm{bol}}>82\mathrm{~K}$.}
 \end{cases} 
\end{equation} 

The power-law index 0.76 derived by \citet{2014ApJ...783....6V} for younger YSOs is consistent with our high-resolution CO study with the power-law index of 1.24 for sources with $T_{\mathrm{bol}} \le 100$\,K.  For more evolved sources, the small power-law index of 0.05 in the \citet{2014ApJ...783....6V} suggests little or no widening of the outflow cavity after Class 0 ($T_{\mathrm{bol}} \ge 70$\,K). Our ALMA data also shows the widening of opening angles slows down for sources with $T_{\mathrm{bol}} \ge 100$\,K. \citet{2014ApJ...783....6V}'s result is consistent with the power-law index derived from our fit ($0.10 \pm 0.19$). 

Our results indicate that the widening of the outflow cavity slows down in Class I and flat-spectrum sources as shown in \autoref{fig:opening_angle}. For Class I and flat-spectrum sources, it is unclear whether or not the opening angle continues to (slightly) increase due to the large uncertainty in the fit.  
We conducted a Spearman rank-order test and a Pearson correlation test for the entire sample. We found the correlation coefficients are 0.49 and 0.44 with p-values of 0.04 and 0.07, respectively. As for the more evolved sources ($T_{\mathrm{bol}} \ge 100$\,K), we found a weak correlation with correlation coefficients of 0.40 and 0.20 and p-values of 0.29 and 0.60 for both tests. For the younger sources ($T_{\mathrm{bol}} \le 100$\,K), there is a strong positive relationship with correlation coefficients of 0.84 and 0.86 and p-values of 0.004 and 0.003, respectively. The statistical tests show a strong correlation between outflow opening angles and bolometric luminosity for the younger sources. The large p-values for the evolved sources are mostly due to the large scatter in the measurements. A larger sample is needed to determine whether the flattening in the opening angle-$T_{bol}$ relation for evolved sources ($T_{\mathrm{bol}} \ge 100$\,K) is real or due to a lack of fully mapped outflows at these later stages of protostellar evolution. 



{ The} flattening of slope for the outflow opening angles found by \citet{2014ApJ...783....6V}
is consistent with a recent study using Hubble Space Telescope (HST) Near-infrared (NIR) { scattered} light images of 30 protostars which shows no widening of the outflow cavities in protostars  \citep{2021ApJ...911..153H}. They also find evolved protostars with narrow cavities. The results from this HST study may partially be attributed to the infrared scattered light tracing a potentially narrower range of protostar ages with similar envelope optical depth (the sample is strongly dominated by Class I sources). High optical depths in the envelopes of younger more embedded sources (i.e., Class 0 sources) reduce the number of infrared photons illuminating the outflow cavity. Similarly, if the density of the outflow cavity is too low (e.g.\ flat-spectrum sources), it would also make it more difficult to scatter infrared photons. This { could} introduces a bias in the sample of the IR studies which results in the flattening or lack of trend found in the evolution of outflow opening angles.


The flattening of the outflow opening angle as protostars evolve might be partially attributed to the method used in inclination angle correction. The slope of the outflow opening angle versus bolometric temperature is shallower when an inclination angle correction is applied. Our inclination correction for the outflow opening angle is based on a simple conical outflow projection model. 
To test whether or not the flattening of the outflow opening angle is due to our inclination correction prescription,
we measured the outflow curvature, an alternative way to quantify the outflow width, by fitting the { wide-angle} wind model to the outflows in our sample. Opening angles derived by using the curvature of the parabolic outflow shell from the { wide-angle} wind model are not affected by inclination angle, and no inclinatio correction is needed \citep{2018ApJ...863...72O}. 
We modeled the Position-velocity (PV) diagrams perpendicular and parallel to each protostellar outflow and compare them with wide-angle wind models. The detailed modeling is described in \autoref{Appendix:B}, and an example is shown in \autoref{fig:inclination_angle}. The parameters of the best wide-angle wind models for each source are summarized in \autoref{table:wide_angle_wind}. 

In \autoref{fig:outflow_curvature}, we plot the evolution of outflow curvature (outflow curvature vs. outflow source bolometric temperature). The green line represents the best power-law fit through all the data points including the literature values from \citet{2018ApJ...863...72O}. The two thick grey lines represent the broken power-law fits with a break at $T_{\rm bol} = 100\,K$. For $T_{\rm bol} \le 100\,K$, $m = -1.30 \pm 0.72$ and $c = 3.59 \pm 1.20$; for $T_{\rm bol} > 100\,K$, $m = -0.62 \pm 0.20$ and $c = 2.29 \pm 0.44$.  We see a clear flattening of outflow curvature for more evolved sources, as seen in \autoref{fig:opening_angle}.
The agreement of outflow curvature from the wide-angle wind modeling, which is not affected by the inclination angle, and the outflow opening angle derived from integrated intensity maps suggests that the flattening trend is not an artifact introduced by our inclination correction.

{ In addition, different methods of measuring the protostellar outflow cavity can result in different opening angles. For example, \citet{2021ApJ...911..153H} used an edge detection technique combined with power-law fits to determine the opening angle and cavity volume. The cavity half-opening angle is defined at an envelope radius of 8000\,au. Adopting a different radius would change the values of the opening angles. While the absolute values of opening angles are dependent on the methods, samples using the same method should result in the same trend. Studies comparing different methods in determining the outflow cavity opening angles, similar to the study by \citet{2013A&A...556A..76V}  which compared outflow momentum rates using different techniques, are needed to establish the scaling relationships between each method. }
 

In Sec. \ref{sec:outflow_rate_Lbol} we showed that outflow rates depend on the bolometric luminosity of the source, and thus it is preferable that sources in a sample have a narrow range of $L_{bol}$ to detect evolutionary trends. Bolometric luminosity is a commonly used proxy for protostellar mass (e.g.\, \citet{2012A&A...545A..19S}).
{ It is still an open} question whether the opening angle { evolves} differently for low-mass and high-mass stars or sources with vastly different values of $L_{bol}$. { However,} \citet{2021ApJ...911..153H} { did} not found any clear trend of outflow opening angles versus bolometric luminosity (L$_{bol}$) over a sample range of 4 { orders} of magnitude in L$_{bol}$.

To investigate how the evolution of the outflow opening angle might depend on the protostellar mass, one place to look is by comparing our results to those obtained for low-luminosity objects (LLOs). The Canada–France–Hawaii Telescope K-band study of the opening angle using twelve (LLOs), with luminosity ranges $0.1 L_{\odot} \leqslant L_{\mathrm{int}} \leqslant 0.2 L_{\odot}$, shows a steeper { power-law} slope { than our study} \citep{2017AJ....153..173H}:

{
\begin{gather}
\log \left(\frac{\Theta}{\operatorname{deg}}\right)=(-1.42 \pm 0.39 )+(1.64 \pm 0.21) \log \left(\frac{T_{\mathrm{bol}}}{\mathrm{K}}\right).
\end{gather}
}
The bolometric luminosity is composed of both internal and external luminosity ($L_{bol} = L_\mathrm{int}+L_\mathrm{ext} $), where the external luminosity ($L_\mathrm{ext}$) is due to the heating of the envelope by the external, interstellar radiation field and is on the order of few tenths of L$_\odot$ \citep{2008ApJS..179..249D}. For our study, the range of $L_{bol}$ within the first and third quartile are $0.58 L_{\odot} \leqslant L_{\mathrm{bol}} \leqslant 3.82 L_{\odot}$. 

Low luminosity objects can be low mass protostars, extremely young protostars, or protostars in a quiescent phase of episodic accretion \citep{2014ApJ...783...29D,2012AJ....144..115S,2015A&A...579A..23J,2016ApJS..225...26K}. { Recent observations have shown ``two wind components'', a collimated jet and a wide-angle disk wind, co-exist in protostellar outflows (e.g., \citealt{2021ApJ...907L..41L}). }
At an early stage, due to the high column density of the dense envelope, only the collimated jet component can break out,  forming the jet-like outflows seen in Class 0 sources. As the envelope dissipates and accretes toward the central star-disk system over time, the wide-angle wind component starts to break out causing the outflow cavity to widen \citep[e.g.,][]{2006ApJ...646.1070A}. LLOs tend to drive much weaker outflows, and low luminosity can generally be attributed to a low accretion rate \citep{2017AJ....153..173H}. The envelope mass strongly correlates with bolometric luminosity $M_{env}\propto L_{bol}^{0.6}$ \citep{2013A&A...558A.125D}. Thus, the selection of LLOs systematically favors systems with less envelope mass. In systems with significantly lower envelope mass, the wide-angle wind component could break out more easily,  causing the outflow cavity to widen at a much earlier phase resulting in a much faster widening rate of the outflow cavity compared to higher luminosity sources. 
The larger power-law index found by \citet{2017AJ....153..173H} in their low-luminosity sample compared to the index found in our study (See \autoref{fig:opening_angle}) may be caused by the rapid widening of outflow cavity in LLOs. 

Alternatively, it is  possible that in these systems, the rapid widening of outflow cavities accelerates the removal of core gas resulting in a much lower envelope density, lower accretion rate and eventually lower bolometric luminosity. If that is the case, the rapid widening of outflow cavities in LLOs would be the cause rather than the result of gas dispersal in LLOs. 




\begin{figure*}[tbh]
\centering
\makebox[\textwidth]{\includegraphics[width=\textwidth]{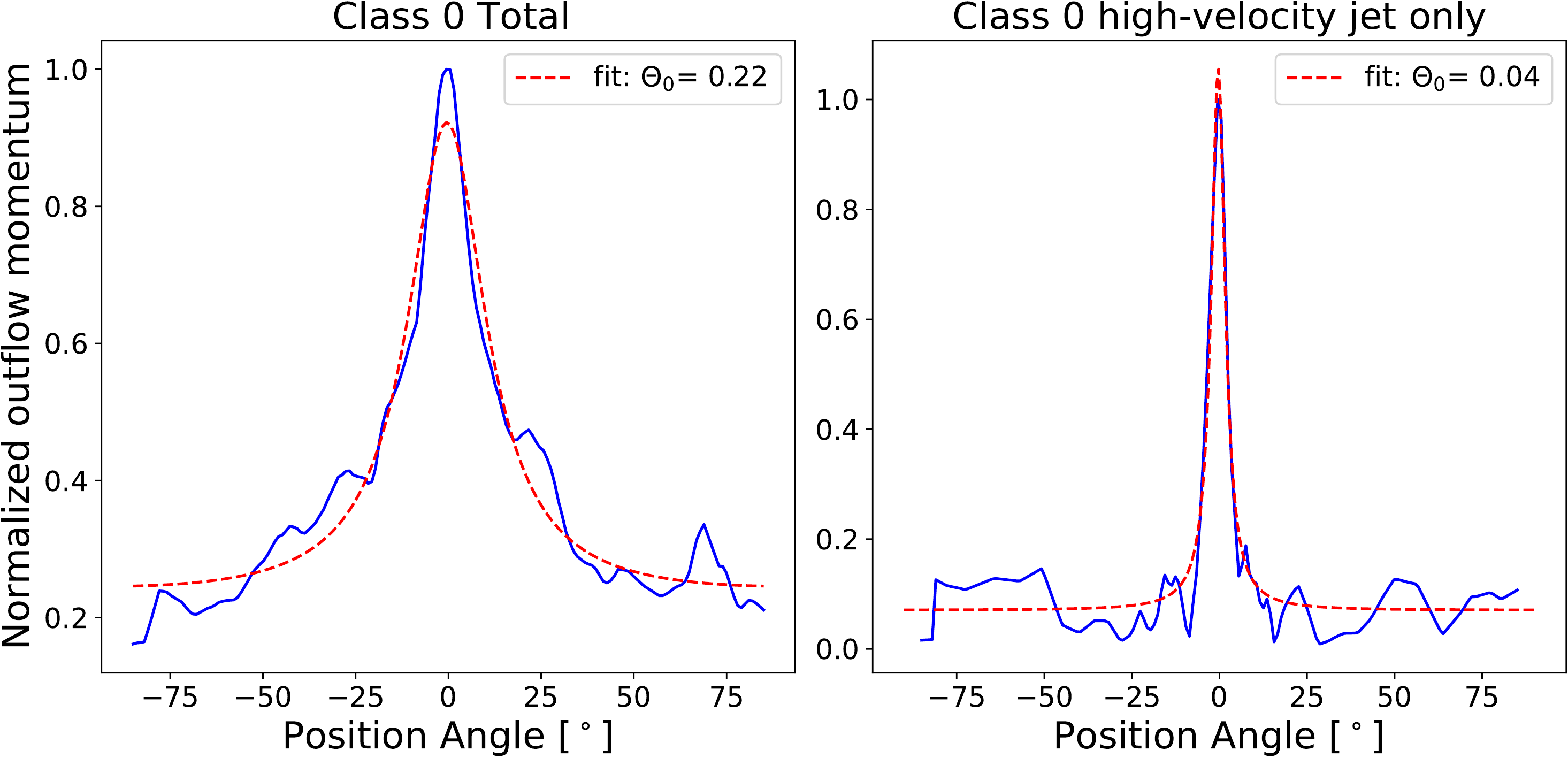}}
\caption{Normalized average outflow momentum angular profile for Class 0 protostars with no sign of outflow precession. The left figure shows the total (average) momentum profile (includes molecular outflow emission at all outflow velocities), and the right figure shows the average momentum profile for the high-velocity jets (HOPS-10 and HOPS-164) identified in \autoref{fig:dM_v}. 
The red dashed line shows the best fit using the theoretical model from Equation 16. All angular profiles have been corrected for the outflows' inclination. }

\label{fig:P_angular}
\end{figure*}

\subsection{Evolution of outflow momentum and energy opening angles}


The study of outflow cavity walls (as discussed above) has been usually conducted by measuring the outflow opening angle based on (integrated) intensity maps of CO or IR emission, which we will henceforth call it as the conventional outflow opening angle. The conventional outflow opening angle has been widely used as an indicator of how much material is impacted or cleared by the outflow and to determine whether or not the outflow-core interaction is responsible for the low-star formation efficiency \citep[e.g.][]{2006ApJ...646.1070A,2014ApJ...783....6V,2017AJ....153..173H,2021ApJ...911..153H}. While intensity maps likely trace the distribution of matter and the size of the outflow cavity, the velocity information, which is crucial for the outflow-core interaction, is missing from this picture. To capture the dynamics of outflow-core interaction and fold in the velocity information, we investigate two new quantities: the outflow momentum ($P_{out}$) opening angle, and the outflow kinetic energy ($E_{out}$) opening angle. Comparing these quantities for sources at different evolutionary stages allows us to quantitatively study how the angular distribution of outflow momentum and energy evolves. 


The least squared fits\footnote{These are not weighted fits. The error bars  in  \autoref{fig:PEopening_angle} show the uncertainty of the Gaussian fits to the angular profile of the outflows' momentum and energy, used to quantify the width of these quantities. The uncertainties    represent the range of possible values for the opening angles and  arise from the deviation of the measured angular profile from a Gaussian function. They are not due to random errors. Therefore, we do not use them as  weights for our power-law fits. } to the  momentum and energy opening angle as a function of $T_{bol}$ are: 
\begin{gather}
\log \left(\frac{\Theta_{\mathrm{P}}}{\operatorname{deg}}\right)=(0.79 \pm 0.35)+(0.49 \pm 0.16) \log \left(\frac{T_{\mathrm{bol}}}{\mathrm{K}}\right),
\end{gather}

\begin{gather}
\log \left(\frac{\Theta_{\mathrm{E}}}{\operatorname{deg}}\right)=(0.81 \pm 0.38)+(0.49 \pm 0.17) \log \left(\frac{T_{\mathrm{bol}}}{\mathrm{K}}\right).
\end{gather} 
Similar to the conventional opening angle, we find a clear trend in which the momentum and energy opening angles increase as protostars evolve. { The trend appears to flatten for more evolved sources as shown in black lines in \autoref{fig:PEopening_angle}. However, if we take out HOPS-166, the source with the highest bolometric luminosity and the highest C$^{18}$O core mass, both the momentum and energy opening angles would not show signs of flattening. More later-stage Class I and  flat-spectrum sources are needed to determine whether the $P_{out}$ and $E_{out}$ opening angles continue to increase or flatten at later protostellar evolutionary stages.} 


Comparing the momentum and energy opening angles with the conventional outflow opening angles shows that typically the outflow momentum and energy opening angles are around $10^\circ$ to $20^\circ$ smaller than the conventional opening angle, especially in Class 0 sources (See \autoref{table:opening_angles} in Appendix). 
The much narrower momentum and energy outflow opening angles in Class 0 sources show that $P_{out}$ and $E_{out}$  are concentrated in collimated jets at these early stages. As sources evolve, the outflow momentum and energy are transferred to the surrounding medium at a wider range of angles, resulting in a much steeper slope (\autoref{fig:PEopening_angle}) as compared to the conventional outflow opening angle measured from integrated intensity maps.


The most important consequence of narrower momentum and energy opening angles in Class 0 is that this indicates that the amount of gas entrainment is less in Class 0 sources as compared to Class I sources. The extremely collimated nature of outflows from Class 0 sources suggests these young outflows only impact a limited volume inside a protostellar core. Our results from Sec. \ref{sec:mass_rates_results} indicate that outflows from Class I show, on average, higher mass outflow rates than those from Class 0 sources (see \autoref{fig:dM_dP_dE_r}). This supports the idea that Class I outflows entrain more mass from the surrounding environment than younger sources.
This is consistent with \autoref{fig:PE_polar} 
which shows that the angular width of momentum and energy profile with respect to the outflow axis increases for more evolved sources. 

Studying the evolution of outflow opening angle or cavity size identified using integrated intensity maps in $^{12}$CO  \citep[e.g.,][]{2006ApJ...646.1070A}, or infrared scattered light \citep[e.g.,][]{2007ApJ...659.1404T,2008ApJ...675..427S,2021ApJ...911..153H} has been one of the main techniques used in understanding the feedback of outflows in halting accretion, dispersing the core and reducing the star formation rate in the past two decades. The outflow opening angle has been used to estimate the volume and mass affected by the outflow. To study the outflow-core interactions and determine feedback from outflows on core-to-star efficiency, one should also study the momentum or energy opening angles, which incorporate the velocity information. The evolution of the angular distribution of $P_{out}$ and $E_{out}$ is an important complement to the conventional outflow opening angles in determining how much core material is impacted by protostellar winds.

\subsection{Theoretical angular distribution of momentum}

The momentum angular distribution of winds launched by protostar-disk systems  has been studied theoretically, but the observational studies on this topic have been scarce. 
\citet{1999ApJ...526L.109M} indicate that for any hydromagnetic protostellar 
wind the  
 { normalized} angular distribution of the wind momentum can be described with the following equation:
\begin{gather}
 \zeta(\theta,\theta_0) =  \left[ln(\frac{2}{\theta_0})(sin^2\theta+\theta_0^2) \right]^{-1},
\end{gather}
where $\theta$ is the polar angle measured from the protostar's rotation axis, and { $\theta_0$ is the parameter which quantifies the angular width of the momentum profile.} This equation describes the momentum angular distribution  of the protostellar wind directly ejected from the central star-disk system. However, in observation we measure the momentum of the molecular outflow, which is the entrained gas. 
The outflow entrainmment mechanism (i.e., the wind-cloud interaction that produces the observed molecular outflow) is expected to be a momentum-conserving interaction, and therefore if the launched wind's axis does not change with time, then we would expect the molecular outflow momentum profile to be similar to that of the prostellar wind. 
For highly embedded, young protostars (i.e., Class 0 sources) there is still enough material throughout the envelope that the entrained material should trace well the wind momentum. Thus, the momentum angular profile of molecular outflows from Class 0 sources with no clear axis wandering (i.e., no change in outflow axis direction), should trace reasonably well the momentum angular distribution of the underlying protostellar wind. 

In \autoref{fig:P_angular} 
we  compare the observed molecular outflow momentum angular distribution with the distribution 
predicted by
\citet{1999ApJ...526L.109M}.
In the left panel of \autoref{fig:P_angular}
we show a fit to the { average} angular distribution of molecular outflow momentum from { Class 0 outflows}  with the generalized radial { hydromagnetic} wind profile (Equation 16).
When constructing this plot we exclude the outflows from HOPS-11 and HOPS-169 as they show signs of precession, and the outflow from HOPS-408 because it is barely resolved.} 
In the right panel of 
\autoref{fig:P_angular} we show only the  average outflow momentum profile for the high-velocity jet component  
from HOPS-10 and HOPS-164 
(the two non-precessing Class 0 outflows with clear jets, see \autoref{fig:dM_v}).
The average angular momentum profile is clearly more collimated in the 
 jet component
(with a $\theta_{0} = 0.04$)
compared to the profile obtained when considering all molecular outflow emissions at all velocities ($\theta_{0} \sim 0.2$).

\citet{1999ApJ...526L.109M} proposed that the momentum distribution in protostellar winds should be 
strongly collimated
with a  $\theta_{0} \le 0.05$ resulting in an extremely peaked profile with the majority of the momentum concentrated along the outflow axis. This is { consistent with our Class 0 high-velocity molecular jet momentum profile. However, 
 when including the low-velocity molecular component, the momentum profile is significantly wider.
 
 We may interpret these results in various ways. 
 One possible interpretation 
is that (non-precessing) jet-like prostellar winds  with can produce a much wider molecular outflow, even if the original wind has a very narrow momentum angular distribution and 
without the aide of another (much wider) wind component. This could be produced, for example, by the sideways ejection of material by a series of internal working surfaces \citep{1993A&A...276..539R}, turbulent mixing of the jet and a turbulent envelope \citep{2017ApJ...847..104O}, or the expanding shell driven by a pulsed jet in a stratified core \citep{2022A&A...664A.118R}.

Another possible interpretation is that if we assume that molecular outflows are solely produced by the direct (momentum-conversing) interaction between the underlying jet and surrounding envelope (and no other mechanism like those described above can produce a wider outflow), then the momentum angular distribution of the underlying jet should be  significantly wider (i.e., 
$\theta_{0}\sim 0.2$) than what was proposed by \citet{1999ApJ...526L.109M} and what is usually used in simulations \citep[e.g.,][]{2011ApJ...740..107C,2017ApJ...847..104O}. However, \citet{2014ApJ...784...61O} ran various simulations of the protostellar outflow evolution and gas entrainment in a core using different values of $\theta_{0}$ and found that even highly collimated outflows can entrain a significant amount of core gas and produce relatively wide molecular outflows. Yet, in that study the wide  outflow cavities are likely partially caused by  the  change in outflow axis seen in the simulations.  In  making the plots shown in \autoref{fig:P_angular} we only used outflows that exhibit a constant outflow axis direction, in order to exclude any possible molecular outflow widening due to a wandering jet 
\citep[as seen in the models by, e.g.,][]{1993ApJ...414..230M,2004MNRAS.347.1097R}.
Thus, comparing our results with those of simulation with jets with a time-varying axis direction is not a fair comparison.  

Yet another possible interpretation of our results is that there exist two wind components; a very collimated jet-like wind \citep[e.g.,][]{2000prpl.conf..789S}  and a wider (likely disk-driven) wind \citep[e.g.,][]{2000prpl.conf..759K}.
In such scenario, the (slower) wide-angle parts of the molecular outflow are driven by a disk wind, while the more collimated (high-velocity) component is driven by a jet-like wind. A thorough comparison of our sample of protostellar molecular outflows with  different  wind and entrainment models is needed in order to determine which of the scenarios described above is the most consistent with our data.

\section{Conclusions}
\label{sec:conclusion}


The paper presents the ALMA multi-line observations of protostellar environment around 21 protostars at different evolutionary stages. In short, the results can be summarized as the following:


\begin{enumerate}
\item We argue that the widely used method of estimating outflow rates using { a} single dynamical time, obtained by dividing a characteristic length by a characteristic velocity, is unreliable. 
To { better} estimate outflow properties and their impact on their surrounding, we devised the Pixel Flux-tracing Techniques (PFT) and modified the ``annulus method'' developed by \citet{1996A&A...311..858B} to measure outflow { mass, momentum, and energy} ejection rates. The PFT allows us to compute two-dimensional molecular outflow instantaneous rates maps for the first time. Instead of a single rate value obtained by previous methods, 
these maps contain spatial information of how the molecular outflow mass, momentum and energy rates vary within the outflow cavity.



\item On average, molecular outflows from  Class 0 sources have significantly higher momentum and energy ejection rates than those of more evolved protostars. However, molecular outflows from  Class I sources have, on average, a  60\% higher outflow mass rate compared to { their Class 0 counterpart}. This higher molecular outflow rate is partly due to the larger { molecular} outflow volume. 
Our results suggest that the dissipation of protostellar cores starts { when a Class 0 protostar} quickly opens up a hole with a high-velocity collimated jet. Then, in the Class I phase the { molecular} outflow cavity widens and a larger volume of core gas { is} impacted by the outflow. 

\item We find that by the end of the protostellar phase, a typical low-mass protostar will displace approximately $ 2$ to $4\,M_\odot$ of mass from the surrounding envelope. Similarly, we expect molecular outflows to inject a total momentum and energy of about $ 11$ to $22$\,M$_\odot$\,km s$^{-1}$ and about 4 to 8 $\times 10^{45}$\,erg into their natal cores by the end of the protostellar phase. Adopting the lifetimes for the different protostellar evolutionary phase from \citet{2015ApJS..220...11D},  our results indicate that the mass removed by outflows from Class 0, Class I and flat-spectrum sources are 40\% - 80\%, 300\% - 600\% and 40\% - 150\% of the (current) average estimated core mass, respectively. Thus, the { total mass removed by} outflows throughout the protostellar stage is comparable to or larger than the current protostellar core mass. This strongly suggests that the mass budget of  protostellar cores { is} continuously replenished by accretion from cloud gas at larger scales. { The high mass-loss from protostellar outflows poses serious challenges to the current concept of ``cores as the primary reservoir of stellar mass", and thus has strong implications for the origin of the IMF, and the simplified core to star conversion scenario.} 

\item The results from our study of 21 protostellar outflows at different evolutionary stages show that consistent with previous studies, the { molecular} outflow opening angle increases with the protostellar evolutionary phase, as traced by the bolometric temperature ($T_{bol}$). 
The rate at which the opening angle widens decline significantly for more evolved sources ($T_{bol} \ge 100\,K$), compared to younger (Class 0) sources. Yet, more outflows from more evolved protostars are needed to confirm the observed trend for Class I and flat-spectrum protostars.


\item To study the effect of outflow core interactions, we introduced two new quantities, the outflow momentum opening angle and the outflow energy opening angle, which quantify the angular profile of outflow momentum and energy. Our results show steeper slopes in the momentum and energy opening angles vs.~$T_{bol}$ plots compared to that of the traditional opening angle. This indicates that the outflow momentum and energy  are considerably more collimated (i.e., concentrated towards the outflow axis) in the Class 0 phase than at later evolutionary stages. 
The evolution of the angular distribution of the outflow momentum and energy is an important complement to the traditional opening angles in determining the volume fraction of the surrounding envelope impacted by molecular outflows.



\item Comparing the results of recent { molecular} outflow surveys in Orion with our results { indicates} that molecular outflow rates significantly depend on { the} protostellar bolometric luminosity. The strong dependence on bolometric luminosity { suggests} that  protostellar outflow feedback varies significantly depending on protostellar mass or accretion rate. 

\item 
We find that in Class 0 molecular outflows, the collimated high-velocity component has a narrow momentum angular distribution consistent with that expected for hydromagnetic winds. However, the average molecular outflow momentum angular distribution is significantly wider when including the molecular outflow emission at all velocities. We argue that this could be interepreted in various ways. One possibility is that highly collimated winds may be able to produce molecular outflows that are much wider than the driving jet. On the other hand, it may be that the original (driving) wind is wider than the collimated jets generally assumed in simulations. Yet another possibility is that two co-existing wind components (one jet-like and the other a wide-angle wind) entrain the surrounding core gas to produce the observed molecular outflow. Further comparison between molecular outflow observations and wind entrainment models are needed to determine which of these scenarios describes best the data.


\end{enumerate}



\begin{acknowledgements}
      CHH acknowledges support from the National Radio Astronomy Observatory (NRAO) Student Observing Support (SOS). HGA and CHH acknowledge support from NSF award AST-1714710. ZYL is supported in part by NASA 80NSSC20K0533 and NSF AST-1910106. AS gratefully acknowledges support by the Fondecyt Regular (project code 1220610), and ANID BASAL projects ACE210002 and FB210003. STM received funding from NSF AST grant 210827 and NASA ADAP grants 80NSSC19K0591 and 80NSSC18K1564. SSRO acknowledges funding support from NSF Career 1748571. This paper makes use of the following ALMA data: ADS/JAO.ALMA \#2016.1.01338S ALMA is a partnership of ESO (representing its member states), NSF (USA) and NINS (Japan), together with NRC (Canada) and NSC and ASIAA (Taiwan) and KASI (Republic of Korea), in cooperation with the Republic of Chile. The Joint ALMA Observatory is operated by ESO, AUI/NRAO and NAOJ. The National Radio Astronomy Observatory is a facility of the National Science Foundation operated under cooperative agreement by Associated Universities, Inc. 
\end{acknowledgements}

\begin{appendix}

\section{Cloud Subtraction}
\label{sec:Appendix_A}
Total Power array data is sensitive to large-scale emission and is crucial to recover the outflow mass. However, while it recovers the large-scale outflow emission it also picks up the cloud emission. Even though we tried to avoid including cloud emission by not including velocity channels in the outflow maps that were close to the cloud velocity, there were several sources in which emission from the parent cloud (or extended ``contaminating'' CO emission at different velocities), could not be avoided even at velocities more than 1 km\,s$^{-1}$ away from the parent cloud velocity. Therefore, we developed cloud subtraction routines for the Ring Method  and the Pixel Flux-tracing Technique as shown in \autoref{fig:RM_background_sub} and \autoref{fig:PFT_background_sub} respectively.

The upper left panel of \autoref{fig:RM_background_sub} shows the $^{12}$CO integrated intensity map of the red-shifted lobe of the HOPS-135 outflow, which clearly shows large-scale cloud emission. To avoid contamination from cloud emission in the estimation of outflow parameters using the RM, we first estimate the averaged background per pixel inside the ring mask, but outside the outflow mask (as shown in the upper right panel of \autoref{fig:RM_background_sub}).
For tracers with cloud contamination, we subtracted the background cloud spectrum from the outflow spectrum, as shown in the lower panels of \autoref{fig:RM_background_sub}. 
We then use the background subtracted data to construct the mass-spectrum inside the rings to estimate the molecular outflow rates.

As for the Pixel Flux-tracing Technique (PFT), the mass spectrum is calculated for each individual pixel (instead of using rings). Hence, we adopted a slightly different method for the background subtraction as shown in \autoref{fig:PFT_background_sub}. We first obtained the averaged profile for the quantity of interests as a function of distance from the central protostar perpendicular to the outflow direction. To avoid subtracting outflow emission near the protostar, we set the background value in the inner region ($\le 1050\,$au) to a constant as shown in \autoref{fig:PFT_background_sub} upper plot. We rotate the radial profile to create a 2D background map (see lower middle panel of \autoref{fig:PFT_background_sub}). Then we subtracted out the background to get rid of the large-scale cloud emission
(see lower right panel of \autoref{fig:PFT_background_sub}).

\begin{figure*}[tbh]
\centering
\makebox[\textwidth]{\includegraphics[width=\textwidth]{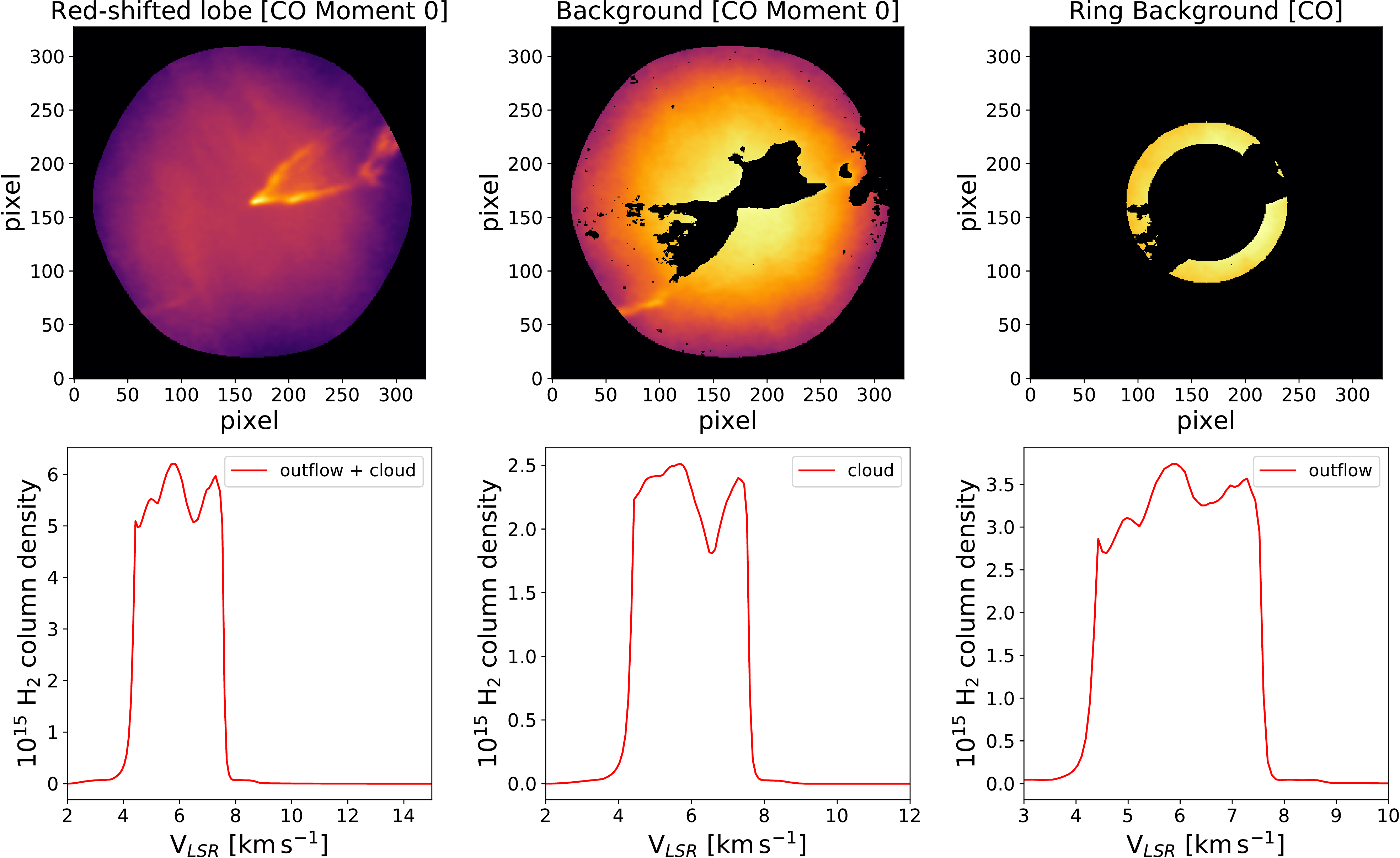}}
\caption{Subtraction of cloud emission using the Ring Method  for HOPS-135. The red-shifted outflow lobe sits on-top of cloud shown in $^{12}$CO Momentum 0 (upper left panel). We measure the background emission in a ring outside the outflow mask as shown in the upper right panel. For each rings at different distances from source, we measured its corresponding background.
For tracers with cloud contamination, { we subtracted the background cloud spectrum from the spectrum obtained over the area covered by the outflow (lobe) mask, as shown in the bottom 3 panels.} In the bottom right panel the `outflow' label represents the background subtracted column density spectrum.
}
\label{fig:RM_background_sub}
\end{figure*} 

\begin{figure*}[tbh]
\centering
\makebox[\textwidth]{\includegraphics[width=\textwidth]{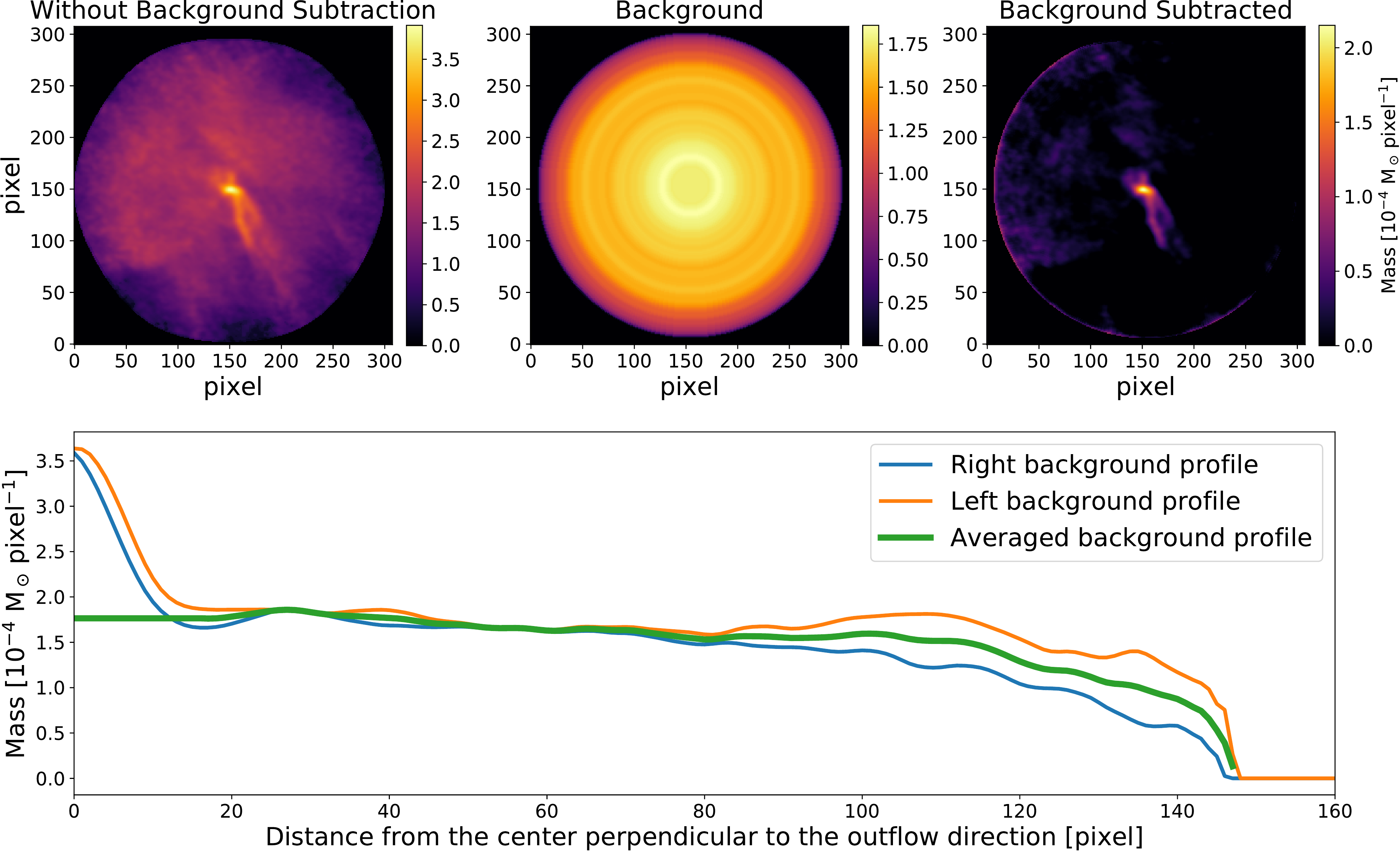}}
\caption{Cloud subtraction method for the PFT. In this example, we show (in the upper left panel) the uncorrected outflow mass map for HOPS-10, This clearly shows background cloud emission not associated with the outflow. 
We measure the background emission as a function of distance from the central protostar (perpendicular to the outflow direction), as shown in the upper panel. The orange and blue lines represent the left side and the right side of the background emission profile from the protostar. We set the central region ($\le 1050\,$au) of the profile to constant to avoid subtracting outflow emission near the protostar. We used the radial profile to generate a two-dimensional background map (upper middle panel). We then subtract the background map to remove the cloud contamination, resulting in the background-subtracted map (upper right panel). 
}
\label{fig:PFT_background_sub}
\end{figure*}

\section{Outflow inclination properties and wide angle wind modelling}
\label{Appendix:B}
In this section, we compared the inclination angle derived from the wide-angle wind modeling to the inclination angle derived from the disk major to minor axis ratio. The inclination angles derived from both methods are shown in \autoref{table:inclination_angle}. 

We obtain the outflow inclination angle by fitting a wide-angle wind model \citep{1996ApJ...472..211L,2000ApJ...542..925L}. The wide-angle wind model has been used by others to estimate the inclination angle (and other properties) of observed molecular outflows 
\citep[e.g.,][]{2010ApJ...717...58H,2013ApJ...774...39A,2017ApJ...834..178Y}. In this model, the entrained material can be described by a parabolic shell with a  expansion velocity analogous to the Hubble law. 
Many of the outflows from  Class I 
sources (and some from flat-spectrum sources) show parabolic-like cavities similar to those expected from the wide-angle wind model. On the other hand,
this wide-angle wind model may not be appropriate for modeling  molecular outflows from Class 0 sources, as they are typically very collimated, and show a  jet-like morphology (specially at high velocities). However,  some of these Class 0 outflows also exhibit wider outflow cavities at low velocities (e.g.\ HOPS-10, HOPS-198). Hence, modeling these outflow cavity walls with a wide-angle wind model can provide an estimate of the outflow inclination angle. 

The wide-angle wind model of  \citet{2000ApJ...542..925L} can be described by 3 independent parameters: $c_0$, $v_0$, and $i$. Considering z is the outflow direction, and R is the radial distance from the outflow center, then a cylindrical symmetric parabolic shell surface can be described by:
\begin{gather}
 z = c_0 R^2,
\end{gather}
and the corresponding velocity follows:
\begin{gather}
 v_z = v_0 z,\\
 v_r = v_0 r,\\
 v_{\phi} = 0.
\end{gather}

\autoref{fig:inclination_angle} outlines the basic steps for fitting the wide-angle wind model to constrain the outflow inclination. We start with creating an outflow mask from the $^{12}$CO integrated intensity (moment 0) map. The outflow mask is used to isolate the outflow from the surroundings and fit the outflow lobe morphology  to the model. Then we applied the { wide-angle} wind model to constrain the pair solutions for $c_0$ and $i$ for each lobe (red-shifted and blue-shifted lobe). The parabola should fit the edge of the outflow mask. For every 5 degrees of inclination angle, from 0 degrees to 180 degrees, we search for the best solutions for $c_0$. In total, we obtained 36 pairs of solutions (for parameters $c_0$ and $i$).

Then for each pair of $c_0$ and $i$ solution, we run different { wide-angle} wind models with a different characteristic velocity ($v_0$) using 0.3, 0.5, 1.0, 1.5, 2.0, 2.5, 3.0, 3.5, 4.0, 4.5, 5.0, 5.5, and 6.0 km s$^{-1}$. In total, we produced 468 models to compare with the $^{12}$CO position-velocity (PV) diagram along the outflow axis. The central panel in \autoref{fig:inclination_angle} shows the residual plot with red (value of 1) representing the data, and blue (value of -1) representing the wide-angle wind model. We eliminate solutions with large residuals in the PV diagram. After reducing the number of solutions, we used the remaining solutions (with parameters $c_0$, $i$, and $v_0$) to compare with the PV perpendicular to the outflow. For the wide-angle wind model, the PV diagram perpendicular to the outflow axis is an ellipse. By matching the ellipse with the observational data, we can constrain the inclination angle of the outflow axis (with respect to the line-of-sight) as shown for HOPS-130 in \autoref{fig:inclination_angle}. The resulting estimated inclinations are shown in  \autoref{table:inclination_angle} with 0 degrees being pole-on and 90 degrees being an edge-on system. The parameters for the best wide-angle wind model for each source are shown in \autoref{table:wide_angle_wind}. Note that we divide the $c_0$ and $v_0$ by the source distance of 400$\,$pc to convert the values in the unit of au instead of arcseconds.


In \autoref{fig:inclination_diff} we plot the distribution of the difference in inclination values derived using the two different methods: the outflow wide-angle wind fitting method and the disk continuum method. For the disk continuum method, the inclination angle can be derived by using the disk major to minor axis ratio and assuming the protostellar outflows are perpendicular to the disk plane. While there is a large dispersion in \autoref{fig:inclination_diff}, the distribution peaks at $10\sim20^\circ$ and shows that in general both methods are consistent with each other. For the vast majority of sources, the estimated $i$ from the two methods are different within $35\arcdeg$, and only three sources (i.e., 14\% of the sample) show differences in the estimate of $i$ of $45\arcdeg$ or more. The small difference between the two methods might be partially explained by the assumption of a thin disk for the disk minor to major axis ratio method. Due to the disk thickness, the method cannot fully recover edge-on disk systems (small outflow inclination angles). Even so, the inclination angle derived from the high-resolution disk continuum is a likely more accurate estimate than that derived using the wide-angle wind model because it involves significantly fewer assumptions. Moreover, it is likely that not all molecular outflows are entrained by a wide-angle wind  
(the main assumptions when using the wide-angle wind model). In particular, many of our Class 0 outflows are jet-like and deviate from the parabolic shell structure expected in outflows driven by the wide-angle winds.

We also compare our results to the disk inclination angle derived through modeling of SED of protostars by \citet{2016ApJS..224....5F}. We find that the distribution of the derived inclination angle difference is completely random. This strongly suggests that the disk inclination angle from SED modeling is not reliable, and should not be used in determining disk and outflow inclination angles. Alternatively, machine learning approaches may be used to predict both the plane-of-sky orientation and line-of-sight outflow inclination from CO spectral data \citep{2022ApJ...941...81X}.

\begin{figure*}
\centering
\includegraphics[width=\hsize]{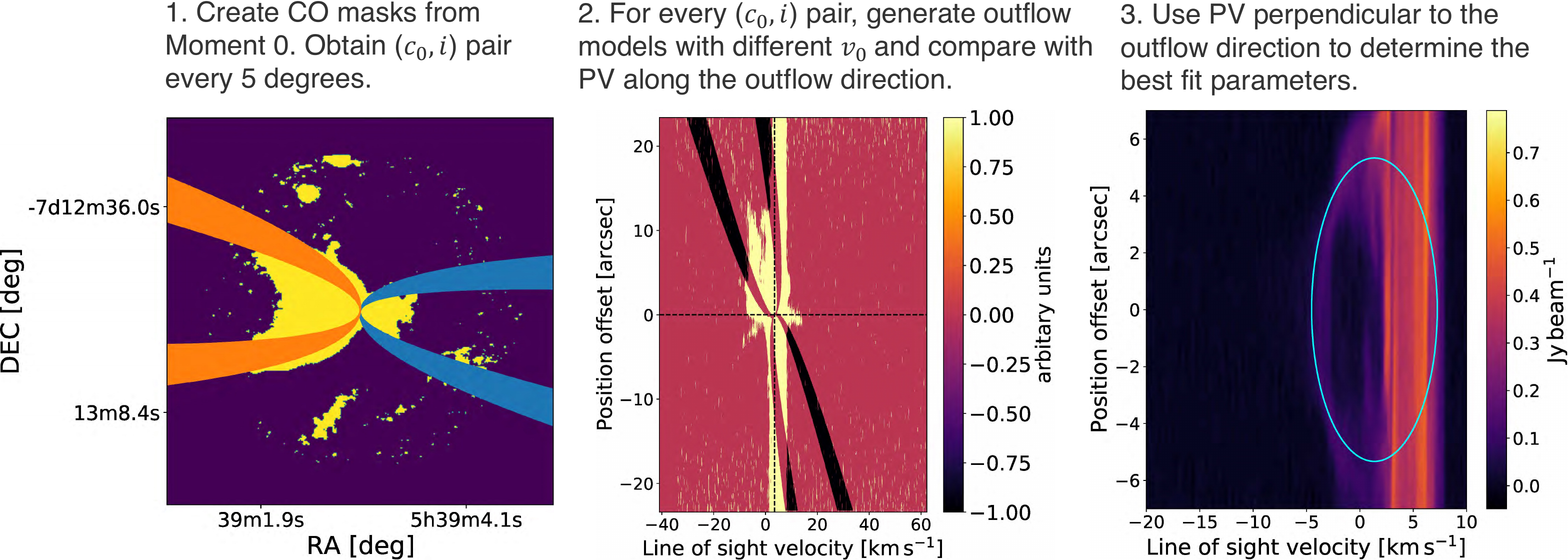}
\caption{Method for constraining the outflow inclination angle by fitting a wide-angle wind model. The example shown here uses the outflow from HOPS-130. 
}
\label{fig:inclination_angle}
\end{figure*}

\begin{table}[H]
\setlength{\tabcolsep}{6pt} 
\caption{Outflow inclination angles}             
\label{table:inclination_angle}      
\centering                          
\begin{tabular}{c c c c c c }        
\hline\hline                 
Source  & Outflow inclination   & Disk & Disk  & Disk  & Outflow inclination \\    
& from wide-angle wind  & major axis & minor axis & minor to major  & derived from   \\ 
& model fit [deg]&   [arcsec]\tablenotemark{a} &    [arcsec]\tablenotemark{a}&  axis ratio &  disk [deg]  \\ 
\hline                        
HOPS-10  & 40 & 0\as14 & 0\as07 & 0.50 & 60 \\
\hline                                   
HOPS-11  & 45 & 0\as13 & 0\as12 & 0.92 & 23 \\ 
\hline
HOPS-13  & 50& 0\as19 & 0\as09 & 0.47 & 62 \\ 
\hline
HOPS-127  & 25 & 0\as05 & 0\as03 & 0.60 & 53 \\ 
\hline
HOPS-129  & 20& 0\as10 & 0\as06 & 0.60 & 53 \\ 
\hline
HOPS-130  & 25& 0\as46 & 0\as15 & 0.33 & 71 \\ 
\hline
HOPS-134  & 45& 0\as06 & 0\as05 & 0.83 & 34 \\ 
\hline
HOPS-135  & 35& 0\as11 & 0\as06 & 0.55 & 57 \\ 
\hline
HOPS-150B  & 25& 0\as03 & 0\as07 & 0.78 & 39  \\ 
\hline
HOPS-157  & 20 & 0\as53 & 0\as46 & 0.87 & 30 \\ 
\hline
HOPS-164  & 55 (blue)/25 (red)& 0\as22 & 0\as14 & 0.64 & 50 \\ 
\hline
HOPS-166  & 20& 0\as16 & 0\as15 & 0.94 & 20 \\ 
\hline
HOPS-169  & 70& 0\as17 & 0\as13 & 0.77 & 40 \\ 
\hline
HOPS-177  & 30& 0\as19 & 0\as03 & 0.16 & 81\tablenotemark{c} \\ 
\hline
HOPS-185  & 30\tablenotemark{b} & 0\as27 & 0\as10 & 0.37 & 68 \\ 
\hline
HOPS-191  & 30& 0\as05 & 0\as02 & 0.41 & 66 \\ 
\hline
HOPS-194  & 20& 0\as19 & 0\as17 & 0.89 & 27 \\ 
\hline
HOPS-198  & 25\tablenotemark{b} & 0\as41 & 0\as07 & 0.17 & 80 \\ 
\hline
HOPS-200  & NA& 0\as46 & 0\as16 & 0.35 & 70 \\ 
\hline
HOPS-355  & 45& 0\as08 & 0\as07 & 0.88 & 29 \\ 
\hline
HOPS-408  & 70& 0\as28 & 0\as24 & 0.88 & 59 \\ 
\hline

\end{tabular}

\begin{flushleft}
\tablenotetext{a}{Deconvolved sizes from ALMA  0.87 mm continuum data published by  \citet{2020ApJ...890..130T}.
}
\tablenotetext{b}{Suspicious results. The derived inclination angle is very different compared to the expected value from the molecular outflow morphology (\autoref{fig:outflow_gallery}).} 
\tablenotetext{c}{As described in \ref{sec:inclination_correction}, we believe this estimate of the inclination angle is wrong because the continuum image for this source has a poor signal-to-noise. We instead use the mean inclination angle from a random uniform distribution of outflow orientations on the sky ($57.3\arcdeg$) for this source.} 
\end{flushleft}
\end{table}

\begin{figure*}[tbh]
\centering
\makebox[\textwidth]{\includegraphics[width=\textwidth]{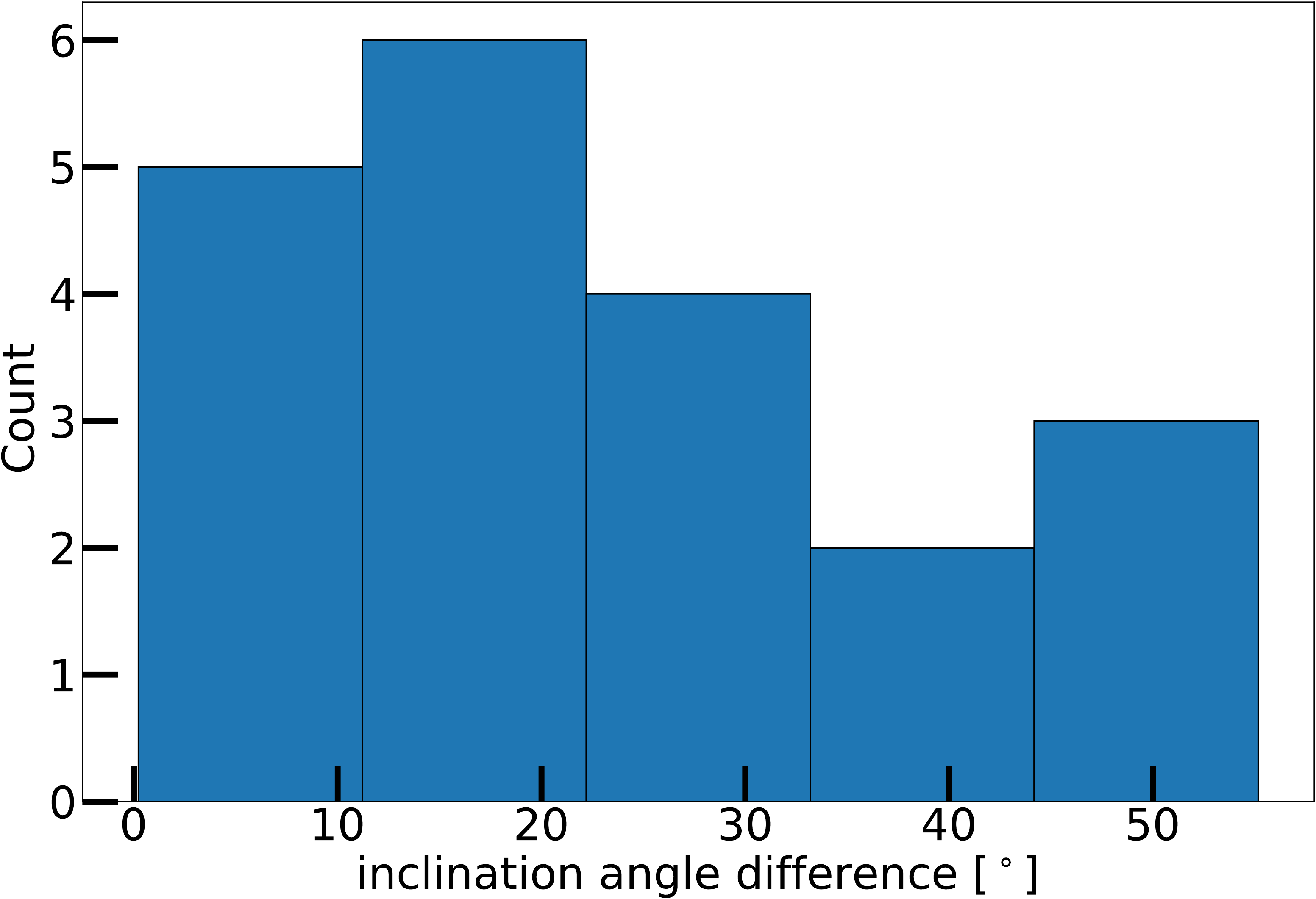}}
\caption{Histogram of inclination angle differences between those derived from the wide-angle wind model fit to the inclinations derived from the disk minor to major axis ratio.
}
\label{fig:inclination_diff}
\end{figure*}

\begin{table}[H]
\setlength{\tabcolsep}{6pt} 
\caption{ Parameters for the best outflow wide-angle wind model for each source}             
\label{table:wide_angle_wind}      
\centering                          
\begin{tabular}{c c c c }        
\hline\hline                 
Source  & $i$   & $c_0$\tablenotemark{a} & $v_0$\tablenotemark{b} \\    
& [deg] & [$10^{-4}$ au$^{-1}$] & [$10^{-4}$ km\,s$^{-1}$] \\ 
\hline                        
HOPS-10  & 40 & 40.3 (blue) / 10.8 (red) &	37.5  \\
\hline                                   
HOPS-11  & 45 & 6.2 &	12.5 \\ 
\hline
HOPS-13  & 50& 25.0\tablenotemark{c} &	7.5 \\
\hline
HOPS-127  & 25 & 13.9 (blue) / 10.4 (red) &	25.0 \\
\hline
HOPS-129  & 20& 4.3 (blue) / 8.6 (red) & 12.5 (blue) / 7.5 (red) \\
\hline
HOPS-130  & 25& 7.0 (blue) / 10.4 (red) & 12.5 \\
\hline
HOPS-134  & 45& 4.2 &	12.5 \\
\hline
HOPS-135  & 35& 12.8 &	12.5 \\
\hline
HOPS-150B  & 25& 52.2\tablenotemark{c} &	7.5 \\
\hline
HOPS-157  & 20& 64.5 &	37.5 \\
\hline
HOPS-164  & 55 (blue)/25 (red)& 45.2 (blue) / 80.0 (red) &	25.0 (blue) / 7.5 (red) \\
\hline
HOPS-166  & 20& 4.3 &	7.5 \\
\hline
HOPS-169  & 70& 40.7 &	12.5 \\
\hline
HOPS-177  & 30& 5.9 &	7.5 \\
\hline
HOPS-185  & 30\tablenotemark{d} & 5.9 (blue) / 8.8 (red) &	25.0 \\
\hline
HOPS-191  & 30& 8.8 (blue) / 5.9 (red) & 12.5  \\
\hline
HOPS-194  & 20&  4.3 &	12.5 (blue) / 7.5 (red) \\
\hline
HOPS-198  & 25\tablenotemark{d} &  10.4 (blue) / 45.2 (red)	& 25.0 \\
\hline
HOPS-200  & NA&  NA & NA\\
\hline
HOPS-355  & 45&  27.0 &	12.5 \\
\hline
HOPS-408  & 70&  20.3 &	25.0 \\
\hline

\end{tabular}

\begin{flushleft}
\tablenotetext{a}{The uncertainty for $c_0$ is 0.74/$sin(i) \times 10^{-4}$ au$^{-1}$. }
\tablenotetext{b}{The values are either 7.5, 12.5, 25.0 or 37.5 $\,10^{-4}$km\,s$^{-1}$ because these are the values for the grid parameter search. The uncertainty for $v_0$ is half the interval between the values evaluated by the model. For 7.5$\times \,10^{-4}$ km\,s$^{-1}$, the uncertainty is $\pm 3.8 \times 10^{-4}$ km\,s$^{-1}$, for all other values the the uncertainty is $\pm 6.3 \times 10^{-4}$ km\,s$^{-1}$.}
\tablenotetext{c}{Unreliable results, poorly constrained.}
\tablenotetext{d}{Suspicious results. The derived inclination angle is very different compared to the expected value from the molecular outflow morphology (\autoref{fig:outflow_gallery}).} 
 
\end{flushleft}
\end{table}

\section{Molecular Outflow Mass, Momentum, and Energy velocity range information}
In this section we show the velocity range used for constructing the molecular outflow mass, momentum, and energy spectrum. An example of outflow mass spectrum constructed from CO, $^{13}$CO, and C$^{18}$O is shown in \autoref{fig:mass_spectrum}. 


\begin{table}[H]
\setlength{\tabcolsep}{6pt} 
\caption{Velocity ranges for constructing Outflow Mass, Momentum, and Energy Spectra}             
\label{table:velocity_range}      
\centering                          
\begin{tabular}{c | c c c | c c c c c }        
\hline\hline                 
&  & Blue-shifted &  &  & Red-shifted &  &   \\
\hline
Source  & $^{12}$CO & $^{13}$CO &  C$^{18}$O  & C$^{18}$O & $^{13}$CO & $^{12}$CO   \\    
&($km\,s^{-1}$) &($km\,s^{-1}$) & ($km\,s^{-1}$) & ($km\,s^{-1}$) & ($km\,s^{-1}$) & ($km\,s^{-1}$)  \\ 
\hline                        
HOPS-10  & $ v \le7.52$ & $7.52\le v \le7.76$ & NA & NA & $ 9.58 \le v \le 9.74$ & $9.74 \le v$   \\
\hline                                   
HOPS-11  & $ v \le5.88$ & $5.88\le v \le6.68$ & $6.68\le v \le7.10$ & $8.50\le v \le10.33$ & $ 10.33 \le v \le 11.12$ & $11.12 \le v$ \\
\hline
HOPS-13  & $ v \le 5.02$ & $ 5.02 \le v \le 5.58$ & $5.58 \le v \le 5.66$ & NA & NA & NA \\
\hline
HOPS-127  & $ v \le 3.27$ & $ 3.27\le v \le 3.75$ & NA & $ 5.02 \le v \le 5.65$ & $ 5.65 \le v \le  6.84$ & $6.84 \le v$ \\
\hline
HOPS-129  & $ v \le 2.52$& $2.52 \le v \le 2.84$ & $2.84 \le v \le 2.92$ & NA & NA & NA   \\
  & $2.92 \le v \le 3.07$ &  &  &  &  &  \\
\hline
HOPS-130  & $ v \le 2.45$ & NA & NA & NA & NA & NA  \\
\hline
HOPS-134  & NA & NA & NA & NA & $ 6.93 \le v \le 7.80 $ & $7.80 \le v$   \\
\hline
HOPS-135  & $ v \le 3.20$ & $3.20 \le v \le 3.68$ & $3.68 \le v \le 4.39$ & $4.95 \le v \le 5.20$ & $5.20  \le v \le  6.30$ & $6.30 \le v$  \\
\hline
HOPS-150  & $ v \le 3.28$ & $3.28 \le v \le 3.62$ & $3.62 \le v \le 3.68$ & NA & NA & NA \\
\hline
HOPS-157  & $ v \le 3.64$ & $4.64 \le v \le 4.75$ & NA & NA & $6.64  \le v \le 9.28 $ & $9.28 \le v$ \\
\hline
HOPS-164  & $ v \le 4.61$ & $4.61 \le v \le 4.92$ & $4.92 \le v \le 5.09$ & NA & $6.58  \le v \le  6.83$ & $6.83 \le v$ \\
\hline
HOPS-166  & $ v \le 5.85$ & NA & NA & NA & $ 10.05 \le v \le 12.05 $ & $12.05 \le v$ \\
\hline
HOPS-169  & $ v \le 2.95$ & $2.95 \le v \le 5.96$ & $5.96 \le v \le 6.67$ & $7.47 \le v \le 8.42$ & $9.58  \le v \le 9.62 $ & $9.62 \le v$ \\
\hline
HOPS-177  & $ v \le 6.78$ & $6.78 \le v \le 8.37$ & NA & NA & $ 9,40 \le v \le 9.96 $ & $9.96 \le v$ \\
\hline
HOPS-185  & $ v \le 6.38$ & $6.38 \le v \le 6.61$ & $6.61 \le v \le 6.94$ & NA & NA & $9.87 \le v$ \\
\hline
HOPS-191  & $ v \le 7.64$ & NA & NA & NA & NA & $10.5 \le v$  \\
\hline
HOPS-194  & NA & NA & NA & NA & NA & $ 9.85 \le v$ \\
\hline
HOPS-198  & $ v \le 4.72$ & $4.72 \le v \le 4.79$ & $4.79 \le v \le 4.88$ & NA & NA & $5.99 \le v \le 6.54$ \\
  &  & & &  &  & $10.51 \le v$ \\
\hline
HOPS-200  & $ v \le 6.59$ & $6.59 \le v \le 6.66$ & $6.66 \le v \le 7.75$ & NA & NA & NA \\
\hline
HOPS-355  & $ v \le 4.89$ & $6.05 \le v \le 6.14$ & $6.14 \le v \le 6.40$ & NA & NA & $10.05 \le v$  \\
\hline
HOPS-408 & $ v \le 2.38$ & $2.38 \le v \le 2.83$ & NA & $4.21 \le v \le 6.15$ & $ 6.15 \le v \le  6.90$ & $6.90 \le v$  \\
\hline
\end{tabular}
\end{table}

\pagebreak

\section{Summarize the sources excluded for each analysis}
\label{Appendix:Sources_analysis} In this section we summarize the sources excluded for each analysis and give explanations. The results are summarized in \autoref{table:analysis_sources}.

\begin{table*}
\setlength{\tabcolsep}{6pt} 
\caption{Sources excluded for each analysis}             
\label{table:analysis_sources}      
\centering                          
\begin{tabular}{c | c | c | c | c  }        
\hline\hline                 
Main Analysis & Analysis item & Section/ Figure & Sources excluded & Reason \\
\hline                        
 &  &   & HOPS-408 & Outflow is barely resolved.\\
\cline{4-5}
Mass, & Radial profile & \autoref{fig:dM_dP_dE_r} & HOPS-166, HOPS-194 & Significantly higher bolometric luminosity\\
Momentum,  & & &  & and core mass compared to other sources.\\
\cline{2-5}
Energy&  & \autoref{fig:compare_rate},  & HOPS-408  & Outflow is barely resolved.\\
\cline{4-5}
rate & Class average & \autoref{table:MPEdMdPdE_results}, \autoref{table:core_growth}& HOPS-166, HOPS-194 & Significantly higher bolometric luminosity\\
& & &  & and core mass compared to other sources.\\
\hline
 &  &  & HOPS-11 blue-shifted lobe & Outflow lobe is strongly asymmetric. \\
 \cline{4-5} 
 &&& HOPS-13, HOPS-200& Very messy outflow.\\
  \cline{4-5}
 &Outflow Momentum &\autoref{fig:PE_polar}& HOPS-134 blue-shifted lobe& No detection.\\
  \cline{4-5} 
&angular profile && HOPS-135 blue-shifted lobe& Low S/N ratio.\\
  \cline{4-5} 
  &  & & HOPS-177 red-shifted lobe& Low S/N ratio. \\
\cline{2-5}         
  &  &  &HOPS-11 blue-shifted lobe & Outflow lobe is strongly asymmetric.\\
\cline{4-5}
Angular &&& HOPS-13, HOPS-200& Very messy outflow.\\
  \cline{4-5}
Profile& && HOPS-130 red-shifted lobe& Compact emission, hard to separate out.\\
  \cline{4-5} 
 &  && HOPS-134 blue-shifted lobe& No detection.\\
  \cline{4-5}
   & Outflow Energy& \autoref{fig:PE_polar}& HOPS-135 blue-shifted lobe& Low S/N ratio.\\
  \cline{4-5}
  &angular profile && HOPS-150 & Contaminated by nearby source. \\
  \cline{4-5}
&&& HOPS-157 red-shifted lobe& Compact emission, hard to separate out.\\
  \cline{4-5} 
&&& HOPS-177 red-shifted lobe& Low S/N ratio.\\
  \cline{4-5} 
& && HOPS-191 red-shifted lobe& Low S/N ratio. \\
 \cline{4-5} 
& && HOPS-194 & Low S/N ratio. \\
\hline      
 &   &  & HOPS-408 & Outflow is barely resolved.\\
 \cline{4-5}
 &  &  & HOPS-150, HOPS-166 & Opening angle is ill defined.\\
 \cline{4-5}
 & Conventional  & \autoref{fig:opening_angle} &  & Poor Gaussian fits to the $^{12}$CO data. \\
 & Opening Angle & & HOPS-194 &  Outflow opening angle measured from \\
& & & &   C$^{18}$O moment 0 map by eye.\\
\cline{2-5}
Opening  &   &  & HOPS-13, HOPS-134 & \\
Angle & Momentum &\autoref{fig:PEopening_angle}&HOPS-157, { HOPS-166},  &  Poor Gaussian fits to the Momentum Maps.\\
 & Opening Angle && HOPS-177, HOPS-194,  &\\
 & & & HOPS-200 &\\
\cline{2-5}
  &   &  & HOPS-13, HOPS-157 & \\
& Energy &\autoref{fig:PEopening_angle}& HOPS-134, HOPS-191  &  Poor Gaussian fits to the Energy Maps.\\
 & Opening Angle && HOPS-194, HOPS-200 &\\
\cline{4-5}
 &  && HOPS-150 & Contaminated by nearby source.\\
\cline{2-5}
& Outflow &\autoref{fig:outflow_curvature}& HOPS-13, HOPS-150  &  Poorly constrained.\\
 & Curvature && HOPS-200\\
\hline
\end{tabular}
\end{table*}

\pagebreak

\section{Molecular Outflow rates}
In this section, we report all the measurements of outflow mass ($\dot{M}$), momentum ($\dot{P}$), and energy ($\dot{E}$) ejection rate for both the red-shifted (R) and blue-shifted (B) lobe at different radius. We present the outflow rate using all outflow gas in \autoref{table:outflow_rate_vall_p1}, \autoref{table:outflow_rate_vall_p2}, and \autoref{table:outflow_rate_vall_p3}. For gas with escape velocity $v_{out} > 1$ km s$^{-1}$, the outflow rates are reported in \autoref{table:outflow_rate_v1_p1}, \autoref{table:outflow_rate_v1_p2}, and \autoref{table:outflow_rate_v1_p3}. Similarly, for gas with escape velocity $v_{out} > 2$ km s$^{-1}$, the outflow rates are reported in \autoref{table:outflow_rate_v2_p1}, \autoref{table:outflow_rate_v2_p2}, and \autoref{table:outflow_rate_v2_p3}.


\begin{table}[H]
\setlength{\tabcolsep}{5.8pt} 
\caption{Molecular outflow rates using all outflow gas, within $2380-3740$ au and $3749-5100$ au away from source }             
\label{table:outflow_rate_vall_p1}      
\centering                          
\begin{tabular}{c | c c c c c c |c c c c c c  }        
\hline\hline                 
 &  &  2380 & $-$ & 3740 & AU & &  &  3740 & $-$ & 5100 & AU &  \\
\hline
Source  & $\dot{M_B}$\tablenotemark{a} & $\dot{M_R}$\tablenotemark{a} &  $\dot{P_R}$\tablenotemark{b}  &  $\dot{P_B}$\tablenotemark{b} &  $\dot{E_R}$\tablenotemark{c} &  $\dot{E_B}$\tablenotemark{c} & $\dot{M_B}$\tablenotemark{a} & $\dot{M_R}$\tablenotemark{a} &  $\dot{P_R}$\tablenotemark{b}  &  $\dot{P_B}$\tablenotemark{b} &  $\dot{E_R}$\tablenotemark{c} &  $\dot{E_B}$\tablenotemark{c} \\    
\hline                        
\hline 
HOPS10  &  1.46  &  3.11  &  5.14  &  71.73  &  5.78E+41  &  4.75E+43  &  1.79  &  2.93  &  10.54  &  68.68  &  1.79E+42  &  4.94E+43 \\
\hline
HOPS10 &  1.46  &  2.67  &  5.14  &  30.38  &  5.78E+41  &  8.30E+42  &  1.79  &  2.48  &  10.54  &  25.07  &  1.79E+42  &  6.49E+42 \\
no jet && &&&&&\\
\hline
HOPS11  &  1.70  &  4.30  &  7.66  &  26.69  &  8.19E+41  &  4.08E+42  &  2.29  &  5.43  &  11.06  &  39.39  &  1.24E+42  &  7.64E+42 \\
\hline
HOPS11  &  1.62  &  3.94  &  5.73  &  17.14  &  3.58E+41  &  1.51E+42  &  2.04  &  4.99  &  6.51  &  24.93  &  3.52E+41  &  2.74E+42 \\
no jet & & &&&&&\\
\hline
HOPS164  &  1.76  &  2.95  &  10.23  &  20.82  &  2.61E+42  &  1.00E+43  &  1.34  &  3.17  &  9.01  &  28.17  &  2.78E+42  &  1.44E+43 \\
\hline
HOPS164 &  1.63  &  2.75  &  5.01  &  8.18  &  2.22E+41  &  6.44E+41  &  1.22  &  2.81  &  3.90  &  7.49  &  1.87E+41  &  4.58E+41 \\
no jet & & &&&&&\\
\hline
HOPS169  &  3.79  &  2.30  &  24.99  &  15.18  &  3.09E+42  &  2.02E+42  &  3.11  &  2.13  &  17.75  &  18.24  &  1.82E+42  &  2.44E+42 \\
\hline
HOPS198  &  0.46  &  3.01  &  5.31  &  22.22  &  8.27E+41  &  4.62E+42  &  0.31  &  1.56  &  3.85  &  28.58  &  6.08E+41  &  1.25E+43 \\
\hline
HOPS355  &  0.22  &  0.05  &  0.50  &  0.36  &  2.13E+40  &  2.76E+40  &  0.17  &  0.06  &  0.43  &  0.46  &  2.16E+40  &  3.87E+40 \\
\hline
HOPS408  & NA & NA & NA & NA &  NA  &  NA  & NA & NA & NA & NA &  NA  &  NA \\
\hline
HOPS127  &  0.34  &  2.37  &  2.39  &  6.59  &  2.81E+41  &  2.62E+41  &  0.23  &  2.42  &  1.47  &  6.82  &  1.55E+41  &  3.25E+41 \\
\hline
HOPS130  &  0.71  & NA &  10.02  & NA &  1.76E+42  &  NA  &  0.48  & NA &  6.70  & NA &  1.21E+42  &  NA \\
\hline
HOP135  &  0.19  &  13.37  &  0.39  &  43.09  &  1.99E+40  &  1.67E+42  &  0.19  &  11.52  &  0.28  &  36.11  &  1.14E+40  &  1.38E+42 \\
\hline
HOPS157  &  0.93  & NA &  2.76  & NA &  1.93E+41  &  NA  &  1.27  & NA &  4.02  & NA &  3.38E+41  &  NA \\
\hline
HOPS177  &  7.12  &  0.74  &  20.14  &  1.05  &  6.43E+41  &  1.92E+40  &  5.77  &  0.79  &  16.98  &  1.28  &  5.68E+41  &  3.01E+40 \\
\hline
HOPS185  &  8.28  &  0.12  &  30.95  &  0.85  &  2.05E+42  &  6.29E+40  &  4.62  &  0.16  &  16.25  &  1.14  &  6.85E+41  &  8.25E+40 \\
\hline
HOPS191  &  0.61  &  0.06  &  2.55  &  0.60  &  1.47E+41  &  7.33E+40  &  0.83  &  0.02  &  3.93  &  0.21  &  2.62E+41  &  2.02E+40 \\
\hline
HOPS129  &  0.90  &  0.04  &  1.08  &  0.02  &  1.58E+40  &  1.63E+38  &  0.70  &  0.04  &  0.97  &  0.03  &  1.80E+40  &  1.87E+38 \\
\hline
HOPS134  & NA &  0.39  & NA &  1.71  &  NA  &  1.32E+41  & NA &  0.29  & NA &  0.94  &  NA  &  6.31E+40 \\
\hline
HOPS13  &  1.69  & NA &  6.34  & NA &  3.58E+41  &  NA  &  1.81  & NA &  7.98  & NA &  5.20E+41  &  NA \\
\hline
HOPS150  &  0.31  & NA &  0.28  & NA &  6.68E+39  &  NA  &  0.15  & NA &  0.23  & NA &  6.97E+39  &  NA \\
\hline
HOPS166  &  0.24  &  2.51  &  1.36  &  8.02  &  9.98E+40  &  3.93E+41  &  0.36  &  1.74  &  1.78  &  6.63  &  9.81E+40  &  3.90E+41 \\
\hline
HOPS194  & NA &  0.16  & NA &  0.50  &  NA  &  1.94E+40  & NA &  0.17  & NA &  0.50  &  NA  &  1.68E+40 \\
\hline
HOPS200  & NA & NA & NA & NA &  NA  &  NA  & NA & NA & NA & NA &  NA  &  NA \\
\hline

\end{tabular}
\begin{flushleft}
\tablenotetext{a}{In units of M$_\odot$ Myr$^{-1}$.}\tablenotetext{b}{In units of $M_\odot\,km\,s^{-1}$ Myr$^{-1}$.}\tablenotemark{c}{In units of erg Myr$^{-1}$.}



\end{flushleft}
\end{table}

\begin{table}[H]
\setlength{\tabcolsep}{5.3pt} 
\caption{Molecular outflow rates using all outflow gas, within $5100-6460$ au and $6460-7820$ au away from source}             
\label{table:7}      
\centering                          
\begin{tabular}{c | c c c c c c |c c c c c c  }        
\hline\hline                 
&  & 5100  & $-$ & 6460 & AU & &  & 6460  & $-$ & 7820 & AU &  \\
\hline
Source  & $\dot{M_B}$\tablenotemark{a} & $\dot{M_R}$\tablenotemark{a} &  $\dot{P_R}$\tablenotemark{b}  &  $\dot{P_B}$\tablenotemark{b} &  $\dot{E_R}$\tablenotemark{c} &  $\dot{E_B}$\tablenotemark{c} & $\dot{M_B}$\tablenotemark{a} & $\dot{M_R}$\tablenotemark{a} &  $\dot{P_R}$\tablenotemark{b}  &  $\dot{P_B}$\tablenotemark{b} &  $\dot{E_R}$\tablenotemark{c} &  $\dot{E_B}$\tablenotemark{c} \\    
\hline                        
\hline 
HOPS10  &  1.96  &  3.00  &  14.87  &  78.66  &  3.34E+42  &  6.40E+43  &  1.77  &  2.82  &  13.43  &  95.12  &  3.63E+42  &  7.86E+43 \\
\hline
HOPS10 &  1.96  &  2.41  &  14.87  &  19.39  &  3.34E+42  &  3.86E+42  &  1.77  &  2.08  &  13.43  &  20.96  &  3.63E+42  &  4.12E+42 \\
no jet & & &&&&&\\
\hline
HOPS11  &  2.24  &  4.95  &  10.04  &  36.55  &  1.01E+42  &  6.73E+42  &  2.05  &  3.73  &  7.74  &  28.12  &  5.51E+41  &  4.85E+42 \\
\hline
HOPS11  &  2.05  &  4.76  &  6.63  &  29.34  &  3.59E+41  &  3.86E+42  &  2.05  &  3.22  &  7.74  &  15.31  &  5.51E+41  &  1.27E+42 \\
no jet & & &&&&&\\
\hline
HOPS164  &  0.50  &  3.22  &  5.61  &  29.14  &  2.58E+42  &  1.72E+43  & NA &  2.28  &  0.02  &  20.79  &  4.92E+39  &  1.29E+43 \\
\hline
HOPS164  &  0.40  &  2.85  &  1.07  &  6.55  &  3.46E+40  &  1.93E+41  & NA &  2.03  &  0.01  &  4.88  &  2.28E+38  &  1.57E+41 \\
no jet && &&&&&\\
\hline
HOPS169  &  2.54  &  1.71  &  16.10  &  16.66  &  1.73E+42  &  2.28E+42  &  1.48  &  1.52  &  10.20  &  14.62  &  1.02E+42  &  1.74E+42 \\
\hline
HOPS198  &  0.15  &  1.24  &  1.50  &  16.93  &  1.87E+41  &  6.50E+42  &  0.09  &  0.33  &  0.74  &  2.05  &  7.41E+40  &  2.85E+41 \\
\hline
HOPS355  &  0.14  &  0.05  &  0.36  &  0.41  &  1.54E+40  &  3.67E+40  &  0.15  &  0.04  &  0.42  &  0.31  &  1.65E+40  &  2.64E+40 \\
\hline
HOPS408  & NA & NA & NA & NA &  NA  &  NA  & NA & NA & NA & NA &  NA  &  NA \\
\hline
HOPS127  &  0.10  &  2.35  &  0.42  &  6.76  &  2.91E+40  &  4.17E+41  &  0.05  &  2.23  &  0.19  &  5.55  &  1.01E+40  &  1.97E+41 \\
\hline
HOPS130  &  0.21  & NA &  2.35  & NA &  3.59E+41  &  NA  &  0.10  & NA &  0.76  & NA &  8.06E+40  &  NA \\
\hline
HOP135  &  0.11  &  2.35  &  0.12  &  7.21  &  2.86E+39  &  2.73E+41  & NA & NA & NA & NA &  NA  &  NA \\
\hline
HOPS157  &  0.31  & NA &  0.45  & NA &  9.73E+39  &  NA  & NA & NA & NA & NA &  NA  &  NA \\
\hline
HOPS177  &  5.31  &  1.03  &  16.03  &  1.67  &  5.65E+41  &  4.12E+40  &  3.88  &  1.25  &  11.95  &  1.96  &  4.45E+41  &  4.99E+40 \\
\hline
HOPS185  &  5.30  &  0.18  &  19.20  &  1.25  &  8.30E+41  &  8.84E+40  &  5.18  &  0.06  &  18.55  &  0.42  &  7.12E+41  &  3.36E+40 \\
\hline
HOPS191  &  0.80  &  0.01  &  4.45  &  0.13  &  3.53E+41  &  1.17E+40  &  0.36  &  0.01  &  1.50  &  0.05  &  9.17E+40  &  4.47E+39 \\
\hline
HOPS129  &  0.37  &  0.04  &  0.55  &  0.03  &  1.14E+40  &  1.88E+38  &  0.25  &  0.02  &  0.40  &  0.01  &  9.21E+39  &  8.83E+37 \\
\hline
HOPS134  & NA &  0.19  & NA &  0.37  &  NA  &  7.63E+39  & NA &  0.10  & NA &  0.22  &  NA  &  5.16E+39 \\
\hline
HOPS13  &  0.18  & NA &  0.74  & NA &  3.95E+40  &  NA  & NA & NA & NA & NA &  NA  &  NA \\
\hline
HOPS150  &  0.05  & NA &  0.03  & NA &  2.70E+38  &  NA  &  0.06  & NA &  0.07  & NA &  1.66E+39  &  NA \\
\hline
HOPS166  &  0.33  &  1.46  &  1.62  &  5.62  &  8.71E+40  &  3.35E+41  &  0.32  &  0.76  &  1.64  &  2.89  &  8.96E+40  &  1.61E+41 \\
\hline
HOPS194  & NA &  0.13  & NA &  0.37  &  NA  &  1.13E+40  & NA &  0.10  & NA &  0.28  &  NA  &  8.95E+39 \\
\hline
HOPS200  & NA & NA & NA & NA &  NA  &  NA  & NA & NA & NA & NA &  NA  &  NA \\
\hline

\end{tabular}
\begin{flushleft}
\tablenotetext{a}{In units of M$_\odot$ Myr$^{-1}$.}\tablenotetext{b}{In units of $M_\odot\,km\,s^{-1}$ Myr$^{-1}$.}\tablenotemark{c}{In units of erg Myr$^{-1}$.}
\label{table:outflow_rate_vall_p2}  
\end{flushleft}
\end{table}

\begin{table}[H]
\setlength{\tabcolsep}{5.3pt} 
\caption{Molecular outflow rates using all outflow gas, within $7820-9180$ au away from source}             
\label{table:outflow_rate_vall_p3}      
\centering                          
\begin{tabular}{c | c c c c c c   }        
\hline\hline                 
 &  &  7820 & $-$ & 9180 & AU     \\
\hline
Source  & $\dot{M_B}$\tablenotemark{a} & $\dot{M_R}$\tablenotemark{a} &  $\dot{P_R}$\tablenotemark{b}  &  $\dot{P_B}$\tablenotemark{b} &  $\dot{E_R}$\tablenotemark{c} &  $\dot{E_B}$\tablenotemark{c}    \\    
\hline                        
\hline
HOPS10  &  0.79  &  1.90  &  4.05  &  76.61  &  5.59E+41  &  6.43E+43 \\
\hline
HOPS10  &  0.79  &  1.29  &  4.05  &  15.65  &  5.59E+41  &  3.03E+42 \\
no jet & \\
\hline
HOPS11  &  1.63  &  2.30  &  6.90  &  21.40  &  4.60E+41  &  3.63E+42 \\
\hline
HOPS11 &  1.63  &  1.32  &  6.90  &  4.48  &  4.60E+41  &  1.88E+41 \\
no jet & \\
\hline
HOPS164  & NA &  1.77  & NA &  19.95  &  NA  &  1.41E+43 \\
\hline
HOPS164  & NA &  1.55  & NA &  3.89  &  NA  &  1.46E+41 \\
no jet & \\
\hline
HOPS169  &  1.09  &  1.38  &  8.87  &  11.86  &  8.80E+41  &  1.22E+42 \\
\hline
HOPS198  &  0.10  & NA &  0.88  & NA &  9.77E+40  &  NA \\
\hline
HOPS355  &  0.13  &  0.03  &  0.34  &  0.19  &  1.31E+40  &  1.22E+40 \\
\hline
HOPS408  & NA & NA & NA & NA &  NA  &  NA \\
\hline
HOPS127  &  0.10  &  2.00  &  0.33  &  5.47  &  1.73E+40  &  2.04E+41 \\
\hline
HOPS130  & NA & NA & NA & NA &  NA  &  NA \\
\hline
HOP135  & NA & NA & NA & NA &  NA  &  NA \\
\hline
HOPS157  & NA & NA & NA & NA &  NA  &  NA \\
\hline
HOPS177  &  1.31  &  1.14  &  4.13  &  1.76  &  1.61E+41  &  4.43E+40 \\
\hline
HOPS185  & NA & NA & NA & NA &  NA  &  NA \\
\hline
HOPS191  & NA & NA & NA & NA &  NA  &  NA \\
\hline
HOPS129  &  0.13  & NA &  0.28  & NA &  8.94E+39  &  4.91E+35 \\
\hline
HOPS134  & NA &  0.08  & NA &  0.16  &  NA  &  4.10E+39 \\
\hline
HOPS13  & NA & NA & NA & NA &  NA  &  NA \\
\hline
HOPS150  &  0.06  & NA &  0.08  & NA &  1.88E+39  &  NA \\
\hline
HOPS166  &  0.14  &  0.49  &  0.66  &  2.35  &  3.31E+40  &  1.58E+41 \\
\hline
HOPS194  & NA &  0.05  & NA &  0.13  &  NA  &  4.10E+39 \\
\hline
HOPS200  & NA & NA & NA & NA &  NA  &  NA \\
\hline

\end{tabular}
\tablenotetext{a}{In units of M$_\odot$ Myr$^{-1}$.}\tablenotetext{b}{In units of $M_\odot\,km\,s^{-1}$ Myr$^{-1}$.}\tablenotemark{c}{In units of erg Myr$^{-1}$.}

\end{table}

\begin{table}[H]
\setlength{\tabcolsep}{5.8pt} 
\caption{Molecular outflow rates for gas with $v_{out} > 1$ km s$^{-1}$, within $2380-3740$ au and $3749-5100$ au away from source}             
\label{table:outflow_rate_v1_p1}      
\centering                          
\begin{tabular}{c | c c c c c c |c c c c c c  }        
\hline\hline                 
 &  &  2380 & $-$ & 3740 & AU & &  &  3740 & $-$ & 5100 & AU &  \\
\hline
Source  & $\dot{M_B}$\tablenotemark{a} & $\dot{M_R}$\tablenotemark{a} &  $\dot{P_R}$\tablenotemark{b}  &  $\dot{P_B}$\tablenotemark{b} &  $\dot{E_R}$\tablenotemark{c} &  $\dot{E_B}$\tablenotemark{c} & $\dot{M_B}$\tablenotemark{a} & $\dot{M_R}$\tablenotemark{a} &  $\dot{P_R}$\tablenotemark{b}  &  $\dot{P_B}$\tablenotemark{b} &  $\dot{E_R}$\tablenotemark{c} &  $\dot{E_B}$\tablenotemark{c} \\    
\hline                        
\hline 
HOPS10  &  1.13  &  3.11  &  4.93  &  72.53  &  6.46e+41  &  4.86E+43  &  1.43E+00  &  2.93  &  10.44  &  69.29  &  1.95E+42  &  5.02E+43 \\
\hline
HOPS10 &  1.13  &  3.11  &  4.93  &  72.53  &  6.46e+41  &  4.86E+43  &  1.43E+00  &  2.93  &  10.44  &  69.29  &  1.95E+42  &  5.02E+43 \\
no jet & & &&&&&\\
\hline
HOPS11  &  1.49  &  4.03  &  7.51  &  26.96  &  8.32e+41  &  4.42E+42  &  2.02E+00  &  5.19  &  10.93  &  39.85  &  1.29E+42  &  8.11E+42 \\
\hline
HOPS11  &  1.49  &  4.03  &  7.51  &  26.96  &  8.32e+41  &  4.42E+42  &  2.02E+00  &  5.19  &  10.93  &  39.85  &  1.29E+42  &  8.11E+42 \\
no jet & & &&&&&\\
\hline
HOPS164  &  1.76  &  2.95  &  10.23  &  20.82  &  2.61e+42  &  1.00E+43  &  1.34E+00  &  3.17  &  9.01  &  28.17  &  2.78E+42  &  1.44E+43 \\
\hline
HOPS164  &  1.76  &  2.95  &  10.23  &  20.82  &  2.61e+42  &  1.00E+43  &  1.34E+00  &  3.17  &  9.01  &  28.17  &  2.78E+42  &  1.44E+43 \\
no jet & & &&&&&\\
\hline
HOPS169  &  3.35  &  1.83  &  24.73  &  14.95  &  3.11e+42  &  2.07E+42  &  2.59E+00  &  1.67  &  17.43  &  18.02  &  1.84E+42  &  2.49E+42 \\
\hline
HOPS198  &  0.50  &  3.04  &  8.71  &  30.68  &  5.6e+42  &  2.53E+43  &  3.88E-01  &  1.59  &  9.90  &  33.63  &  7.09E+42  &  2.28E+43 \\
\hline
HOPS355  &  0.13  &  0.05  &  0.49  &  0.41  &  3.15e+40  &  4.95E+40  &  1.03E-01  &  0.06  &  0.42  &  0.55  &  2.82E+40  &  8.30E+40 \\
\hline
HOPS408  & NA & NA & NA & NA &  0  &  NA  &  NA  & NA & NA & NA &  NA  &  NA \\
\hline
HOPS127  &  0.35  &  2.05  &  2.50  &  6.53  &  3.31e+41  &  3.49E+41  &  2.29E-01  &  2.14  &  1.62  &  6.78  &  2.21E+41  &  4.12E+41 \\
\hline
HOPS130  &  0.71  & NA &  10.76  & NA &  2.42e+42  &  NA  &  4.90E-01  & NA &  7.45  & NA &  1.83E+42  &  NA \\
\hline
HOP135  &  0.09  &  12.89  &  0.44  &  43.14  &  8.42e+40  &  1.99E+42  &  6.40E-02  &  10.97  &  0.28  &  35.97  &  5.82E+40  &  1.57E+42 \\
\hline
HOPS157  &  0.59  & NA &  2.54  & NA &  2.12e+41  &  NA  &  8.42E-01  & NA &  3.74  & NA &  3.62E+41  &  NA \\
\hline
HOPS177  &  7.13  &  0.55  &  20.51  &  1.42  &  8.64e+41  &  4.20E+41  &  5.78E+00  &  0.60  &  17.38  &  1.67  &  8.22E+41  &  4.34E+41 \\
\hline
HOPS185  &  8.28  &  0.13  &  31.43  &  2.27  &  2.48e+42  &  1.61E+42  &  4.62E+00  &  0.18  &  16.68  &  3.21  &  1.08E+42  &  2.33E+42 \\
\hline
HOPS191  &  0.63  &  0.07  &  3.61  &  1.47  &  9.73e+41  &  9.09E+41  &  8.54E-01  &  0.03  &  5.34  &  0.48  &  1.37E+42  &  2.75E+41 \\
\hline
HOPS129  &  0.55  & NA &  0.93  & NA &  7.78e+40  &  NA  &  4.82E-01  & NA &  0.97  & NA &  9.78E+40  &  NA \\
\hline
HOPS134  & NA &  0.40  & NA &  2.22  &  0  &  3.53E+41  &  NA  &  0.30  & NA &  1.43  &  NA  &  2.92E+41 \\
\hline
HOPS13  &  1.68  & NA &  6.33  & NA &  3.58e+41  &  NA  &  1.82E+00  & NA &  8.13  & NA &  5.59E+41  &  NA \\
\hline
HOPS150  &  0.08  & NA &  0.51  & NA &  1.44e+41  &  NA  &  7.36E-02  & NA &  0.51  & NA &  1.39E+41  &  NA \\
\hline
HOPS166  &  0.25  &  2.52  &  1.46  &  8.17  &  1.33e+41  &  4.23E+41  &  3.64E-01  &  1.75  &  1.92  &  6.76  &  1.47E+41  &  4.16E+41 \\
\hline
HOPS194  & NA &  0.21  & NA &  1.82  &  0  &  4.41E+41  &  NA  &  0.22  & NA &  1.94  &  NA  &  4.73E+41 \\
\hline
HOPS200  & NA & NA & NA & NA &  0  &  NA  &  NA  & NA & NA & NA &  NA  &  NA \\
\hline

\end{tabular}
\begin{flushleft}
\tablenotetext{a}{In units of M$_\odot$ Myr$^{-1}$.}\tablenotetext{b}{In units of $M_\odot\,km\,s^{-1}$ Myr$^{-1}$.}\tablenotemark{c}{In units of erg Myr$^{-1}$.}


\end{flushleft}
\end{table}

\begin{table}[H]
\setlength{\tabcolsep}{5.8pt} 
\caption{Molecular outflow rates for gas with $v_{out} > 1$ km s$^{-1}$, within $5100-6460$ au and $6460-7820$ au away from source}             
\label{table:outflow_rate_v1_p2}      
\centering                          
\begin{tabular}{c | c c c c c c |c c c c c c  }        
\hline\hline                 
 &  & 5100  & $-$ & 6460 & AU & &  & 6460  & $-$ & 7820 & AU &  \\
\hline
Source  & $\dot{M_B}$\tablenotemark{a} & $\dot{M_R}$\tablenotemark{a} &  $\dot{P_R}$\tablenotemark{b}  &  $\dot{P_B}$\tablenotemark{b} &  $\dot{E_R}$\tablenotemark{c} &  $\dot{E_B}$\tablenotemark{c} & $\dot{M_B}$\tablenotemark{a} & $\dot{M_R}$\tablenotemark{a} &  $\dot{P_R}$\tablenotemark{b}  &  $\dot{P_B}$\tablenotemark{b} &  $\dot{E_R}$\tablenotemark{c} &  $\dot{E_B}$\tablenotemark{c} \\    
\hline                        
\hline 
HOPS10  &  1.60  &  3.01  &  14.76  &  79.83  &  3.52E+42  &  6.55E+43  &  1.43  &  2.83  &  13.44  &  96.41  &  3.88E+42  &  8.03E+43 \\
\hline
HOPS10  &  1.60  &  3.01  &  14.76  &  79.83  &  3.52E+42  &  6.55E+43  &  1.43  &  2.83  &  13.44  &  96.41  &  3.88E+42  &  8.03E+43 \\
no jet & & &&&&&\\
\hline
HOPS11  &  1.95  &  4.79  &  9.92  &  37.04  &  1.07E+42  &  7.18E+42  &  1.66  &  3.62  &  7.73  &  28.53  &  6.77E+41  &  5.21E+42 \\
\hline
HOPS11  &  1.95  &  4.79  &  9.92  &  37.04  &  1.07E+42  &  7.18E+42  &  1.66  &  3.62  &  7.73  &  28.53  &  6.77E+41  &  5.21E+42 \\
no jet & & &&&&&\\
\hline
HOPS164  &  0.50  &  3.22  &  5.61  &  29.14  &  2.58E+42  &  1.72E+43  & NA &  2.28  &  0.02  &  20.79  &  4.92E+39  &  1.29E+43 \\
\hline
HOPS164  &  0.50  &  3.22  &  5.61  &  29.14  &  2.58E+42  &  1.72E+43  & NA &  2.28  &  0.02  &  20.79  &  4.92E+39  &  1.29E+43 \\
no jet & & &&&&&\\
\hline
HOPS169  &  2.25  &  1.49  &  15.94  &  16.63  &  1.75E+42  &  2.35E+42  &  1.40  &  1.47  &  10.22  &  14.80  &  1.05E+42  &  1.86E+42 \\
\hline
HOPS198  &  0.16  &  1.28  &  3.38  &  21.88  &  3.17E+42  &  1.64E+43  &  0.09  &  0.34  &  1.82  &  3.32  &  1.97E+42  &  3.40E+42 \\
\hline
HOPS355  &  0.10  &  0.05  &  0.38  &  0.48  &  2.95E+40  &  6.94E+40  &  0.13  &  0.04  &  0.47  &  0.41  &  3.99E+40  &  6.89E+40 \\
\hline
HOPS408  & NA & NA & NA & NA &  NA  &  NA  & NA & NA & NA & NA &  NA  &  NA \\
\hline
HOPS127  &  0.10  &  1.94  &  0.52  &  6.62  &  6.98E+40  &  5.03E+41  &  0.05  &  1.82  &  0.26  &  5.44  &  4.02E+40  &  2.90E+41 \\
\hline
HOPS130  &  0.22  & NA &  3.14  & NA &  1.04E+42  &  NA  &  0.11  & NA &  1.81  & NA &  9.79E+41  &  NA \\
\hline
HOP135  &  0.03  &  2.22  &  0.09  &  7.23  &  1.61E+40  &  3.61E+41  & NA & NA & NA & NA &  NA  &  NA \\
\hline
HOPS157  &  0.20  & NA &  0.39  & NA &  1.99E+40  &  NA  & NA & NA & NA & NA &  NA  &  NA \\
\hline
HOPS177  &  5.31  &  0.78  &  16.46  &  2.00  &  8.29E+41  &  4.27E+41  &  3.89  &  0.94  &  12.30  &  2.51  &  6.50E+41  &  6.58E+41 \\
\hline
HOPS185  &  5.31  &  0.21  &  19.63  &  4.35  &  1.19E+42  &  3.43E+42  &  5.18  &  0.09  &  18.87  &  3.09  &  1.00E+42  &  2.95E+42 \\
\hline
HOPS191  &  0.82  &  0.02  &  5.53  &  0.49  &  1.19E+42  &  3.93E+41  &  0.37  &  0.01  &  2.27  &  0.16  &  6.96E+41  &  1.01E+41 \\
\hline
HOPS129  &  0.25  & NA &  0.57  & NA &  6.62E+40  &  NA  &  0.18  & NA &  0.54  & NA &  1.03E+41  &  NA \\
\hline
HOPS134  & NA &  0.21  & NA &  0.84  &  NA  &  2.20E+41  & NA &  0.11  & NA &  0.55  &  NA  &  1.65E+41 \\
\hline
HOPS13  &  0.18  & NA &  0.74  & NA &  3.95E+40  &  NA  & NA & NA & NA & NA &  NA  &  NA \\
\hline
HOPS150  &  0.04  & NA &  1.04  & NA &  4.03E+41  &  NA  &  0.02  & NA &  0.17  & NA &  5.09E+40  &  NA \\
\hline
HOPS166  &  0.34  &  1.47  &  1.87  &  5.83  &  1.46E+41  &  3.75E+41  &  0.32  &  0.76  &  1.74  &  2.92  &  1.24E+41  &  1.68E+41 \\
\hline
HOPS194  & NA &  0.19  & NA &  1.92  &  NA  &  5.09E+41  & NA &  0.14  & NA &  1.59  &  NA  &  4.33E+41 \\
\hline
HOPS200  & NA & NA & NA & NA &  NA  &  NA  & NA & NA & NA & NA &  NA  &  NA \\
\hline

\end{tabular}
\begin{flushleft}
\tablenotetext{a}{In units of M$_\odot$ Myr$^{-1}$.}\tablenotetext{b}{In units of $M_\odot\,km\,s^{-1}$ Myr$^{-1}$.}\tablenotemark{c}{In units of erg Myr$^{-1}$.}


\end{flushleft}
\end{table}

\begin{table}[H]
\setlength{\tabcolsep}{5.8pt} 
\caption{Molecular outflow rates for gas with $v_{out} > 1$ km s$^{-1}$, within $7820-9180$ au away from source}             
\label{table:outflow_rate_v1_p3}      
\centering                          
\begin{tabular}{c | c c c c c c   }        
\hline\hline                 
 &  &  7820 & $-$ & 9180 & AU     \\
\hline
Source  & $\dot{M_B}$\tablenotemark{a} & $\dot{M_R}$\tablenotemark{a} &  $\dot{P_R}$\tablenotemark{b}  &  $\dot{P_B}$\tablenotemark{b} &  $\dot{E_R}$\tablenotemark{c} &  $\dot{E_B}$\tablenotemark{c}    \\    

\hline                        
\hline 
HOPS10  &  0.66  &  1.91  &  5.65  &  78.07  &  1.55E+42  &  6.62E+43 \\
\hline
HOPS10  &  0.66  &  1.91  &  5.65  &  78.07  &  1.55E+42  &  6.62E+43 \\
no jet & \\
\hline
HOPS11  &  1.45  &  2.28  &  7.00  &  22.14  &  5.64E+41  &  4.13E+42 \\
\hline
HOPS11  &  1.45  &  2.28  &  7.00  &  22.14  &  5.64E+41  &  4.13E+42 \\
no jet & \\
\hline
HOPS164  & NA &  1.77  & NA &  19.95  &  NA  &  1.41E+43 \\
\hline
HOPS164  & NA &  1.77  & NA &  19.95  &  NA  &  1.41E+43 \\
no jet & \\
\hline
HOPS169  &  1.08  &  1.37  &  9.01  &  12.16  &  9.49E+41  &  1.40E+42 \\
\hline
HOPS198  &  0.11  & NA &  3.18  & NA &  4.21E+42  &  NA \\
\hline
HOPS355  &  0.11  &  0.04  &  0.39  &  0.36  &  3.50E+40  &  9.02E+40 \\
\hline
HOPS408  & NA & NA & NA & NA &  NA  &  NA \\
\hline
HOPS127  &  0.10  &  1.71  &  0.46  &  5.56  &  7.15E+40  &  3.61E+41 \\
\hline
HOPS130  & NA & NA & NA & NA &  NA  &  NA \\
\hline
HOP135  & NA & NA & NA & NA &  NA  &  NA \\
\hline
HOPS157  & NA & NA & NA & NA &  NA  &  NA \\
\hline
HOPS177  &  1.31  &  0.86  &  4.35  &  2.48  &  2.93E+41  &  7.40E+41 \\
\hline
HOPS185  & NA & NA & NA & NA &  NA  &  NA \\
\hline
HOPS191  & NA & NA & NA & NA &  NA  &  NA \\
\hline
HOPS129  &  0.12  & NA &  0.36  & NA &  5.06E+40  &  NA \\
\hline
HOPS134  & NA &  0.08  & NA &  0.54  &  NA  &  1.89E+41 \\
\hline
HOPS13  & NA & NA & NA & NA &  NA  &  NA \\
\hline
HOPS150  &  0.03  & NA &  0.22  & NA &  6.30E+40  &  NA \\
\hline
HOPS166  &  0.14  &  0.49  &  0.73  &  2.40  &  5.67E+40  &  1.68E+41 \\
\hline
HOPS194  & NA &  0.08  & NA &  1.08  &  NA  &  3.10E+41 \\
\hline
HOPS200  & NA & NA & NA & NA &  NA  &  NA \\
\hline
\end{tabular}
\tablenotetext{a}{In units of M$_\odot$ Myr$^{-1}$.}\tablenotetext{b}{In units of $M_\odot\,km\,s^{-1}$ Myr$^{-1}$.}\tablenotemark{c}{In units of erg Myr$^{-1}$.}


\end{table}

\begin{table}[H]
\setlength{\tabcolsep}{5.8pt} 
\caption{Molecular outflow rates for gas with $v_{out} > 2$ km s$^{-1}$, within $2380-3740$ au and $3749-5100$ au away from source}             
\label{table:outflow_rate_v2_p1}      
\centering                          
\begin{tabular}{c | c c c c c c |c c c c c c  }        
\hline\hline                 
 &  &  2380 & $-$ & 3740 & AU & &  &  3740 & $-$ & 5100 & AU &  \\
\hline
Source  & $\dot{M_B}$\tablenotemark{a} & $\dot{M_R}$\tablenotemark{a} &  $\dot{P_R}$\tablenotemark{b}  &  $\dot{P_B}$\tablenotemark{b} &  $\dot{E_R}$\tablenotemark{c} &  $\dot{E_B}$\tablenotemark{c} & $\dot{M_B}$\tablenotemark{a} & $\dot{M_R}$\tablenotemark{a} &  $\dot{P_R}$\tablenotemark{b}  &  $\dot{P_B}$\tablenotemark{b} &  $\dot{E_R}$\tablenotemark{c} &  $\dot{E_B}$\tablenotemark{c} \\    
\hline                        
\hline 
HOPS10  &  0.39  &  3.11  &  4.02  &  72.53  &  6.35E+41  &  4.86E+43  &  0.67  &  2.93  &  9.52  &  69.29  &  1.94E+42  &  5.02E+43 \\
\hline
HOPS10  &  0.39  &  3.11  &  4.02  &  72.53  &  6.35E+41  &  4.86E+43  &  0.67  &  2.93  &  9.52  &  69.29  &  1.94E+42  &  5.02E+43 \\
no jet && &&&&& \\
\hline
HOPS11  &  1.00  &  3.18  &  6.78  &  25.76  &  8.21E+41  &  4.40E+42  &  1.35  &  4.25  &  9.94  &  38.55  &  1.27E+42  &  8.09E+42 \\
\hline
HOPS11  &  1.00  &  3.18  &  6.78  &  25.76  &  8.21E+41  &  4.40E+42  &  1.35  &  4.25  &  9.94  &  38.55  &  1.27E+42  &  8.09E+42 \\
no jet && &&&&&\\
\hline
HOPS164  &  1.13  &  1.56  &  9.26  &  18.50  &  2.59E+42  &  1.00E+43  &  0.95  &  1.70  &  8.39  &  25.75  &  2.77E+42  &  1.43E+43 \\
\hline
HOPS164 &  1.13  &  1.56  &  9.26  &  18.50  &  2.59E+42  &  1.00E+43  &  0.95  &  1.70  &  8.39  &  25.75  &  2.77E+42  &  1.43E+43 \\
no jet & &&&&&\\
\hline
HOPS169  &  2.77  &  1.25  &  23.87  &  14.08  &  3.09E+42  &  2.05E+42  &  2.26  &  1.58  &  16.94  &  17.86  &  1.83E+42  &  2.48E+42 \\
\hline
HOPS198  &  0.50  &  3.04  &  8.71  &  30.68  &  5.60E+42  &  2.53E+43  &  0.39  &  1.59  &  9.90  &  33.63  &  7.09E+42  &  2.28E+43 \\
\hline
HOPS355  &  0.13  &  0.05  &  0.49  &  0.41  &  3.15E+40  &  4.95E+40  &  0.10  &  0.06  &  0.42  &  0.55  &  2.82E+40  &  8.30E+40 \\
\hline
HOPS408  & NA & NA & NA & NA &  NA  &  NA  & NA & NA & NA & NA &  NA  &  NA \\
\hline
HOPS127  &  0.26  &  1.42  &  2.36  &  5.59  &  3.29E+41  &  3.35E+41  &  0.17  &  1.33  &  1.52  &  5.58  &  2.20E+41  &  3.94E+41 \\
\hline
HOPS130  &  0.71  & NA &  10.76  & NA &  2.42E+42  &  NA  &  0.49  & NA &  7.45  & NA &  1.83E+42  &  NA \\
\hline
HOP135  &  0.06  &  9.81  &  0.39  &  38.52  &  8.35E+40  &  1.92E+42  &  0.03  &  8.11  &  0.24  &  31.68  &  5.77E+40  &  1.51E+42 \\
\hline
HOPS157  &  0.33  & NA &  2.21  & NA &  2.08E+41  &  NA  &  0.39  & NA &  3.16  & NA &  3.54E+41  &  NA \\
\hline
HOPS177  &  5.46  &  0.11  &  17.88  &  0.85  &  8.22E+41  &  4.13E+41  &  4.49  &  0.16  &  15.37  &  1.10  &  7.90E+41  &  4.26E+41 \\
\hline
HOPS185  &  7.79  &  0.13  &  30.51  &  2.27  &  2.47E+42  &  1.61E+42  &  4.45  &  0.18  &  16.35  &  3.21  &  1.07E+42  &  2.33E+42 \\
\hline
HOPS191  &  0.45  &  0.07  &  3.34  &  1.47  &  9.69E+41  &  9.09E+41  &  0.65  &  0.03  &  5.02  &  0.48  &  1.36E+42  &  2.75E+41 \\
\hline
HOPS129  &  0.08  & NA &  0.34  & NA &  7.03E+40  &  NA  &  0.12  & NA &  0.51  & NA &  9.19E+40  &  NA \\
\hline
HOPS134  & NA &  0.22  & NA &  1.98  &  NA  &  3.50E+41  & NA &  0.12  & NA &  1.16  &  NA  &  2.88E+41 \\
\hline
HOPS13  &  1.56  & NA &  6.17  & NA &  3.55E+41  &  NA  &  1.76  & NA &  8.04  & NA &  5.58E+41  &  NA \\
\hline
HOPS150  &  0.05  & NA &  0.46  & NA &  1.43E+41  &  NA  &  0.06  & NA &  0.48  & NA &  1.38E+41  &  NA \\
\hline
HOPS166  &  0.25  &  1.73  &  1.46  &  6.85  &  1.33E+41  &  4.01E+41  &  0.36  &  1.20  &  1.92  &  5.86  &  1.47E+41  &  4.01E+41 \\
\hline
HOPS194  & NA &  0.16  & NA &  1.74  &  NA  &  4.39E+41  & NA &  0.18  & NA &  1.86  &  NA  &  4.72E+41 \\
\hline
HOPS200  & NA & NA & NA & NA &  NA  &  NA  & NA & NA & NA & NA &  NA  &  NA \\
\hline

\end{tabular}
\begin{flushleft}
\tablenotetext{a}{In units of M$_\odot$ Myr$^{-1}$.}\tablenotetext{b}{In units of $M_\odot\,km\,s^{-1}$ Myr$^{-1}$.}\tablenotemark{c}{In units of erg Myr$^{-1}$.}


\end{flushleft}
\end{table}

\begin{table}[H]
\setlength{\tabcolsep}{5.8pt} 
\caption{Molecular outflow rates for gas with $v_{out} > 1$ km s$^{-1}$, within $5100-6460$ au and $6460-7820$ au away from source}             
\label{table:outflow_rate_v2_p2}      
\centering                          
\begin{tabular}{c | c c c c c c |c c c c c c  }        
\hline\hline                 
 &  & 5100  & $-$ & 6460 & AU & &  & 6460  & $-$ & 7820 & AU &  \\
\hline
Source  & $\dot{M_B}$\tablenotemark{a} & $\dot{M_R}$\tablenotemark{a} &  $\dot{P_R}$\tablenotemark{b}  &  $\dot{P_B}$\tablenotemark{b} &  $\dot{E_R}$\tablenotemark{c} &  $\dot{E_B}$\tablenotemark{c} & $\dot{M_B}$\tablenotemark{a} & $\dot{M_R}$\tablenotemark{a} &  $\dot{P_R}$\tablenotemark{b}  &  $\dot{P_B}$\tablenotemark{b} &  $\dot{E_R}$\tablenotemark{c} &  $\dot{E_B}$\tablenotemark{c} \\    
\hline                        
\hline 
HOPS10  &  0.81  &  3.01  &  13.79  &  79.83  &  3.51E+42  &  6.55E+43  &  0.67  &  2.83  &  12.50  &  96.41  &  3.87E+42  &  8.03E+43 \\
\hline
HOPS10  &  0.81  &  3.01  &  13.79  &  79.83  &  3.51E+42  &  6.55E+43  &  0.67  &  2.83  &  12.50  &  96.41  &  3.87E+42  &  8.03E+43 \\
no jet & & &&&&&\\
\hline
HOPS11  &  1.31  &  4.21  &  8.98  &  36.26  &  1.05E+42  &  7.17E+42  &  1.18  &  3.32  &  7.03  &  28.11  &  6.67E+41  &  5.20E+42 \\
\hline
HOPS11  &  1.31  &  4.21  &  8.98  &  36.26  &  1.05E+42  &  7.17E+42  &  1.18  &  3.32  &  7.03  &  28.11  &  6.67E+41  &  5.20E+42 \\
no jet & & &&&&&\\
\hline
HOPS164  &  0.38  &  1.73  &  5.43  &  26.69  &  2.58E+42  &  1.72E+43  & NA &  1.27  &  0.02  &  19.13  &  4.90E+39  &  1.28E+43 \\
\hline
HOPS164  &  0.38  &  1.73  &  5.43  &  26.69  &  2.58E+42  &  1.72E+43  & NA &  1.27  &  0.02  &  19.13  &  4.90E+39  &  1.28E+43 \\
no jet & & &&&&&\\
\hline
HOPS169  &  1.94  &  1.48  &  15.51  &  16.61  &  1.75E+42  &  2.35E+42  &  1.24  &  1.46  &  9.97  &  14.78  &  1.04E+42  &  1.86E+42 \\
\hline
HOPS198  &  0.16  &  1.28  &  3.38  &  21.88  &  3.17E+42  &  1.64E+43  &  0.09  &  0.34  &  1.82  &  3.32  &  1.97E+42  &  3.40E+42 \\
\hline
HOPS355  &  0.10  &  0.05  &  0.38  &  0.48  &  2.95E+40  &  6.94E+40  &  0.13  &  0.04  &  0.47  &  0.41  &  3.99E+40  &  6.89E+40 \\
\hline
HOPS408  & NA & NA & NA & NA &  NA  &  NA  & NA & NA & NA & NA &  NA  &  NA \\
\hline
HOPS127  &  0.06  &  1.21  &  0.46  &  5.57  &  6.89E+40  &  4.87E+41  &  0.03  &  1.21  &  0.21  &  4.55  &  3.95E+40  &  2.76E+41 \\
\hline
HOPS130  &  0.22  & NA &  3.14  & NA &  1.04E+42  &  NA  &  0.11  & NA &  1.81  & NA &  9.79E+41  &  NA \\
\hline
HOP135  &  0.01  &  1.60  &  0.06  &  6.29  &  1.59E+40  &  3.46E+41  & NA & NA & NA & NA &  NA  &  NA \\
\hline
HOPS157  &  0.04  & NA &  0.18  & NA &  1.70E+40  &  NA  & NA & NA & NA & NA &  NA  &  NA \\
\hline
HOPS177  &  4.13  &  0.19  &  14.62  &  1.24  &  7.99E+41  &  4.17E+41  &  3.00  &  0.20  &  10.94  &  1.59  &  6.29E+41  &  6.46E+41 \\
\hline
HOPS185  &  5.18  &  0.21  &  19.39  &  4.35  &  1.18E+42  &  3.43E+42  &  5.05  &  0.09  &  18.63  &  3.09  &  9.98E+41  &  2.95E+42 \\
\hline
HOPS191  &  0.65  &  0.02  &  5.27  &  0.49  &  1.19E+42  &  3.93E+41  &  0.26  &  0.01  &  2.12  &  0.16  &  6.94E+41  &  1.01E+41 \\
\hline
HOPS129  &  0.08  & NA &  0.36  & NA &  6.34E+40  &  NA  &  0.07  & NA &  0.39  & NA &  1.01E+41  &  NA \\
\hline
HOPS134  & NA &  0.08  & NA &  0.63  &  NA  &  2.17E+41  & NA &  0.06  & NA &  0.47  &  NA  &  1.64E+41 \\
\hline
HOPS13  &  0.18  & NA &  0.74  & NA &  3.95E+40  &  NA  & NA & NA & NA & NA &  NA  &  NA \\
\hline
HOPS150  &  0.04  & NA &  1.03  & NA &  4.03E+41  &  NA  &  0.01  & NA &  0.16  & NA &  5.06E+40  &  NA \\
\hline
HOPS166  &  0.34  &  1.00  &  1.87  &  5.06  &  1.46E+41  &  3.63E+41  &  0.32  &  0.51  &  1.74  &  2.49  &  1.24E+41  &  1.61E+41 \\
\hline
HOPS194  & NA &  0.15  & NA &  1.85  &  NA  &  5.08E+41  & NA &  0.12  & NA &  1.55  &  NA  &  4.32E+41 \\
\hline
HOPS200  & NA & NA & NA & NA &  NA  &  NA  & NA & NA & NA & NA &  NA  &  NA \\
\hline

\end{tabular}
\begin{flushleft}
\tablenotetext{a}{In units of M$_\odot$ Myr$^{-1}$.}\tablenotetext{b}{In units of $M_\odot\,km\,s^{-1}$ Myr$^{-1}$.}\tablenotemark{c}{In units of erg Myr$^{-1}$.}


\end{flushleft}
\end{table}

\begin{table}[H]
\setlength{\tabcolsep}{5.8pt} 
\caption{Molecular outflow rates for gas with $v_{out} > 1$ km s$^{-1}$, within $7820-9180$ au away from source}             
\label{table:outflow_rate_v2_p3}      
\centering                          
\begin{tabular}{c | c c c c c c   }        
\hline\hline                 
 &  &  7820 & $-$ & 9180 & AU     \\
\hline
Source  & $\dot{M_B}$\tablenotemark{a} & $\dot{M_R}$\tablenotemark{a} &  $\dot{P_R}$\tablenotemark{b}  &  $\dot{P_B}$\tablenotemark{b} &  $\dot{E_R}$\tablenotemark{c} &  $\dot{E_B}$\tablenotemark{c}    \\    

\hline                        
\hline 
HOPS10  &  0.32  &  1.91  &  5.23  &  78.07  &  1.54E+42  &  6.62E+43 \\
\hline
HOPS10  &  0.32  &  1.91  &  5.23  &  78.07  &  1.54E+42  &  6.62E+43 \\
no jet & \\
\hline
HOPS11  &  1.18  &  2.18  &  6.59  &  22.01  &  5.58E+41  &  4.13E+42 \\
\hline
HOPS11  &  1.18  &  2.18  &  6.59  &  22.01  &  5.58E+41  &  4.13E+42 \\
no jet & \\
\hline
HOPS164  & NA &  1.00  & NA &  18.66  &  NA  &  1.40E+43 \\
\hline
HOPS164  & NA &  1.00  & NA &  18.66  &  NA  &  1.40E+43 \\
no jet & \\
\hline
HOPS169  &  1.02  &  1.37  &  8.91  &  12.15  &  9.47E+41  &  1.40E+42 \\
\hline
HOPS198  &  0.11  & NA &  3.18  & NA &  4.21E+42  &  NA \\
\hline
HOPS355  &  0.11  &  0.04  &  0.39  &  0.36  &  3.50E+40  &  9.02E+40 \\
\hline
HOPS408  & NA & NA & NA & NA &  NA  &  NA \\
\hline
HOPS127  &  0.07  &  1.29  &  0.40  &  4.91  &  7.05E+40  &  3.50E+41 \\
\hline
HOPS130  & NA & NA & NA & NA &  NA  &  NA \\
\hline
HOP135  & NA & NA & NA & NA &  NA  &  NA \\
\hline
HOPS157  & NA & NA & NA & NA &  NA  &  NA \\
\hline
HOPS177  &  0.96  &  0.17  &  3.81  &  1.64  &  2.84E+41  &  7.29E+41 \\
\hline
HOPS185  & NA & NA & NA & NA &  NA  &  NA \\
\hline
HOPS191  & NA & NA & NA & NA &  NA  &  NA \\
\hline
HOPS129  &  0.05  & NA &  0.28  & NA &  4.96E+40  &  NA \\
\hline
HOPS134  & NA &  0.05  & NA &  0.49  &  NA  &  1.88E+41 \\
\hline
HOPS13  & NA & NA & NA & NA &  NA  &  NA \\
\hline
HOPS150  &  0.02  & NA &  0.20  & NA &  6.27E+40  &  NA \\
\hline
HOPS166  &  0.14  &  0.39  &  0.73  &  2.23  &  5.67E+40  &  1.65E+41 \\
\hline
HOPS194  & NA &  0.07  & NA &  1.06  &  NA  &  3.10E+41 \\
\hline
HOPS200  & NA & NA & NA & NA &  NA  &  NA \\
\hline

\end{tabular}
\tablenotetext{a}{In units of M$_\odot$ Myr$^{-1}$.}\tablenotetext{b}{In units of $M_\odot\,km\,s^{-1}$ Myr$^{-1}$.}\tablenotemark{c}{In units of erg Myr$^{-1}$.}


\end{table}

\pagebreak

\section{Two-dimensional Mass, Momentum, Energy, Mass rate, Momentum rate (Force), and Energy rate (mechanical luminosity) maps for all outflows}
In this section, we presents the two-dimensional mass, momentum, energy, mass rate, momentum rate (force), and energy rate (mechanical luminosity) maps for all outflows. All the rate maps are computed from our Pixel Flux-tracing Technique (PFT). Instead of a single rate value obtained by previous methods, the PFT allows us to compute two-dimensional molecular outflow instantaneous rates maps for the first time. 



\begin{figure*}[tbh]
\centering
\makebox[\textwidth]{\includegraphics[width=\textwidth]{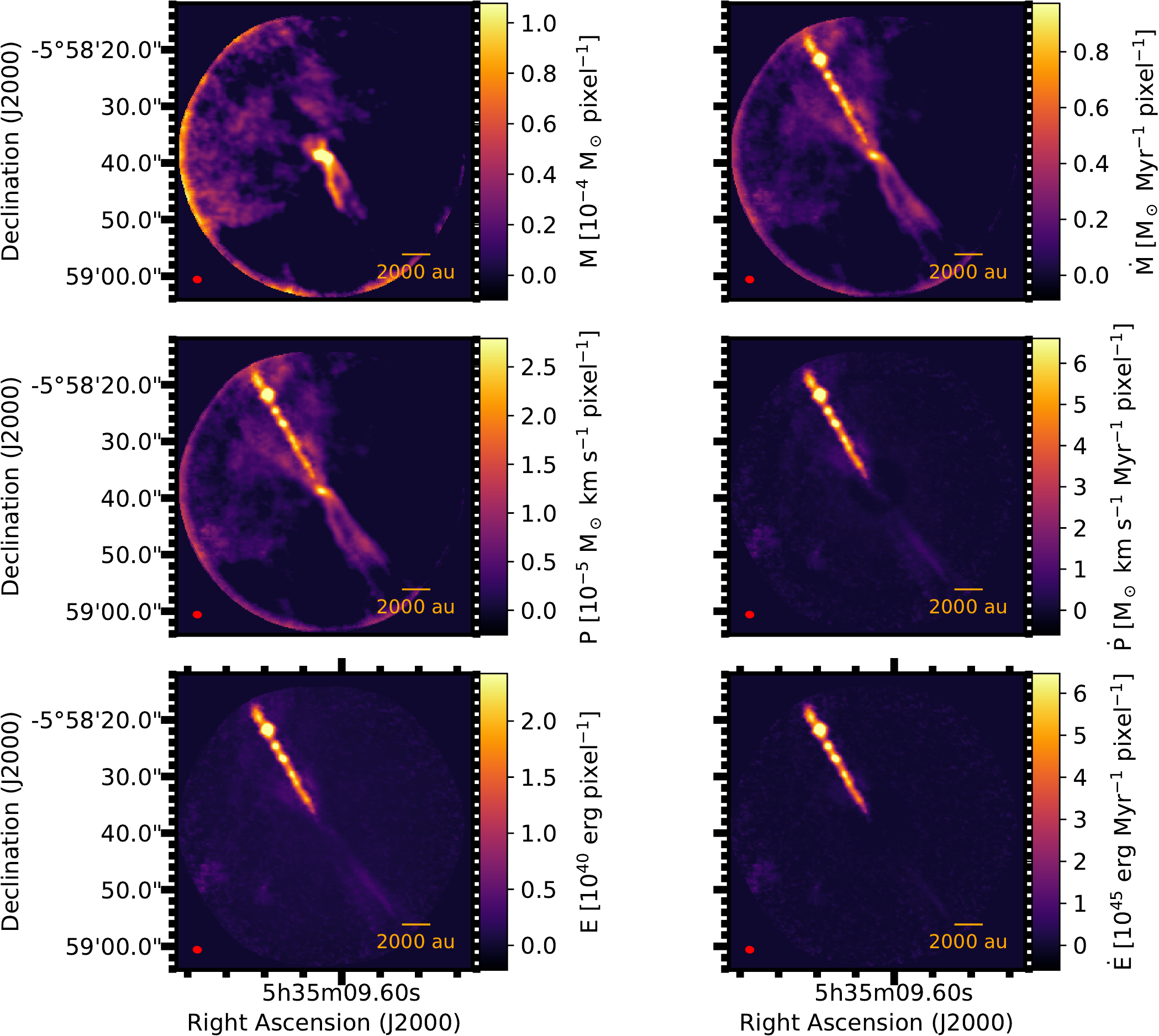}}
\caption{HOPS-10 molecular outflow mass, momentum, and energy  maps (panels in left column), and the corresponding  rate maps (panels in right column). {  The red ellipse in the bottom left of each panel represents the synthesized beam.  Note that the pixel size is the same for all maps and is 0\as17 $\times$  0\as17.}
}
\label{fig:HOPS10}
\end{figure*}


\begin{figure*}[tbh]
\centering
\makebox[\textwidth]{\includegraphics[width=\textwidth]{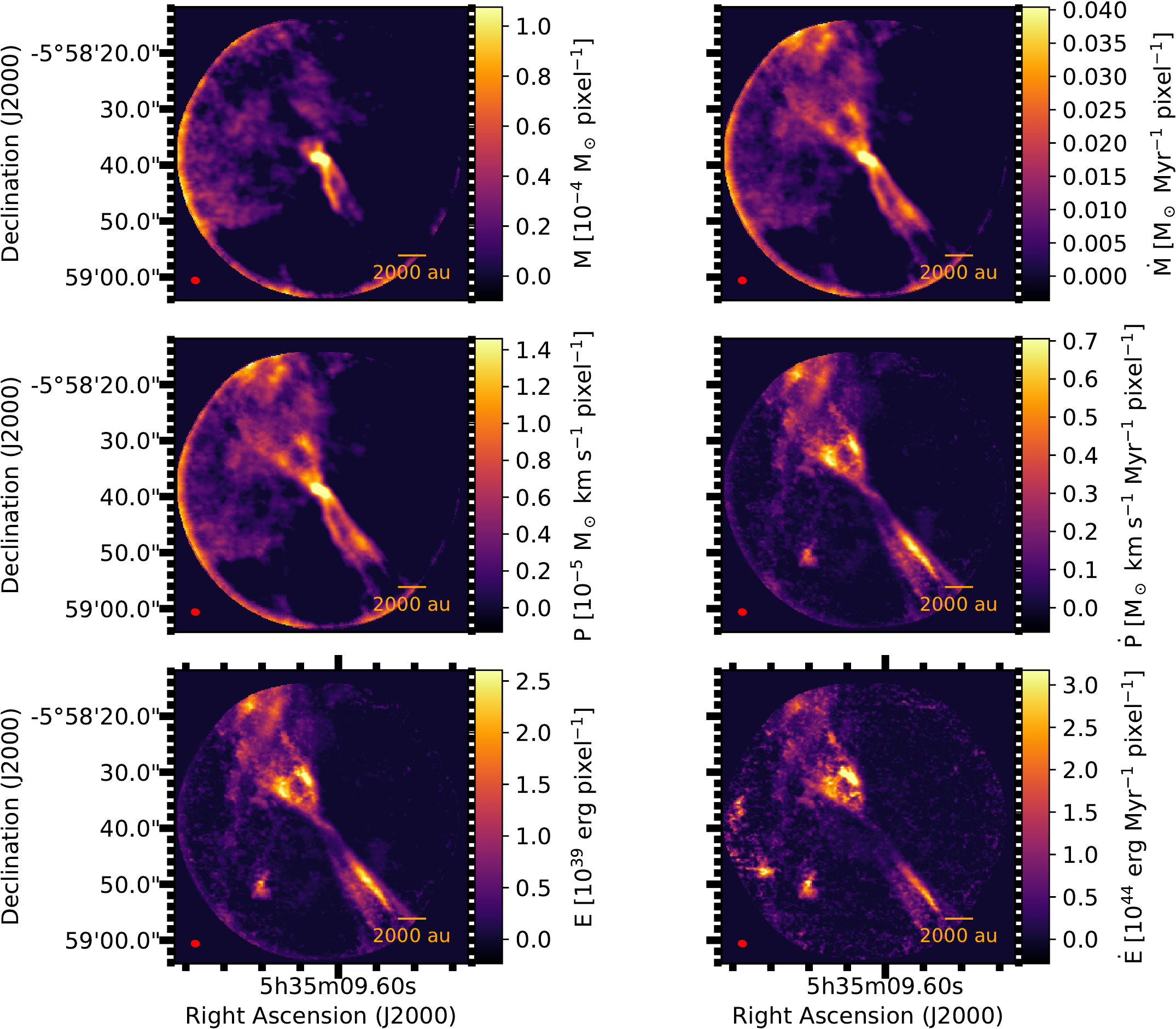}}
\caption {HOPS-10 outflow mass, momentum, energy maps, and the corresponding instantaneous rate maps. Note for this figure set, we exclude the high-velocity jet components. { The pixel size is 0\as17 $\times$  0\as17. The red ellipses in the bottom left of plots represent the synthesized beam.} 
}
\label{fig23}
\end{figure*}

\begin{figure*}[tbh]
\centering
\makebox[\textwidth]{\includegraphics[width=\textwidth]{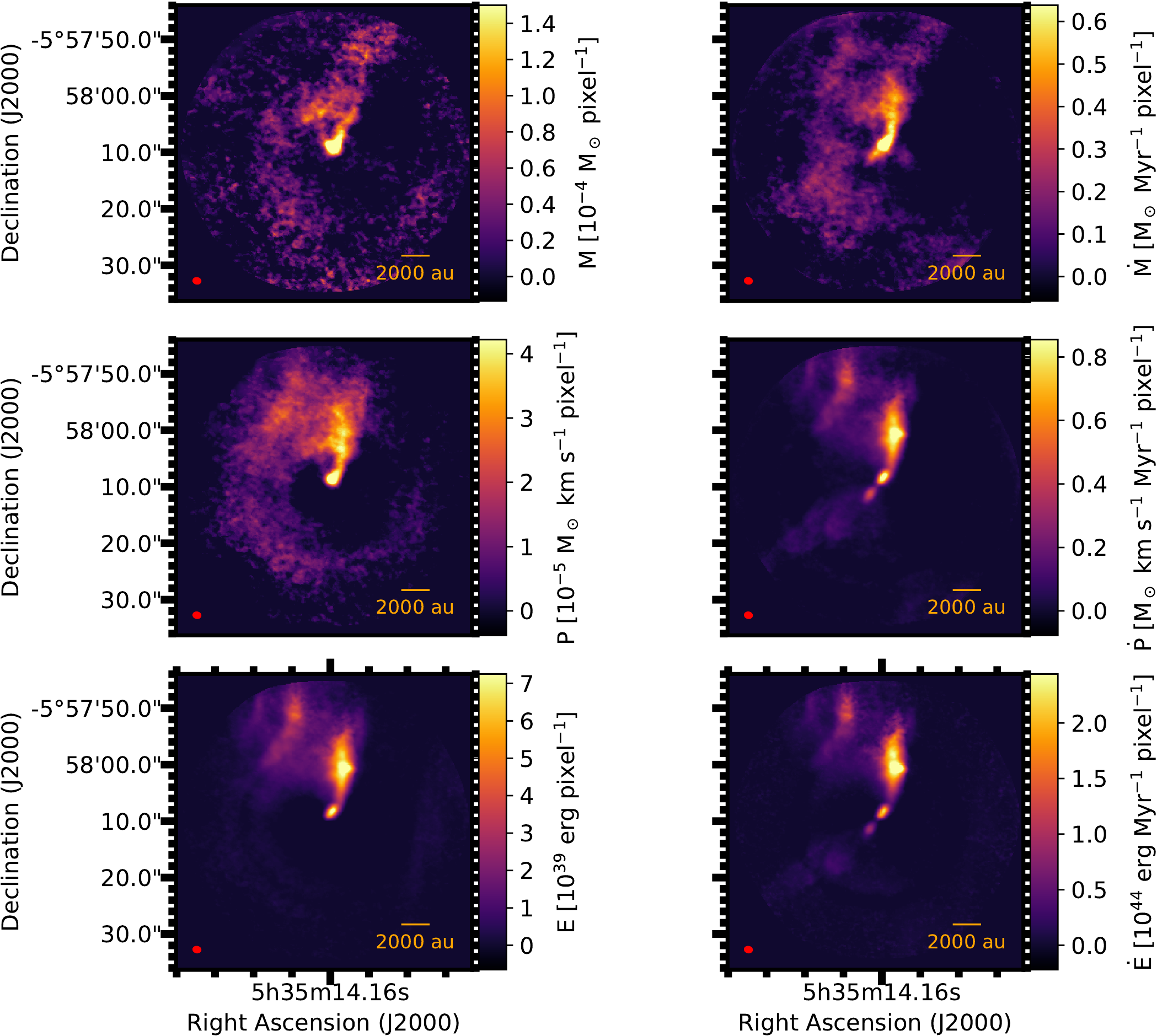}}
\caption {HOPS-11 outflow mass, momentum, energy  maps, and the corresponding instantaneous rate maps. 
{ Note that the pixel size is 0\as17 $\times$  0\as17. The red ellipses in the bottom left of plots represent the synthesized beam.}}
\label{fig24}
\end{figure*}

\begin{figure*}[tbh]
\centering
\makebox[\textwidth]{\includegraphics[width=\textwidth]{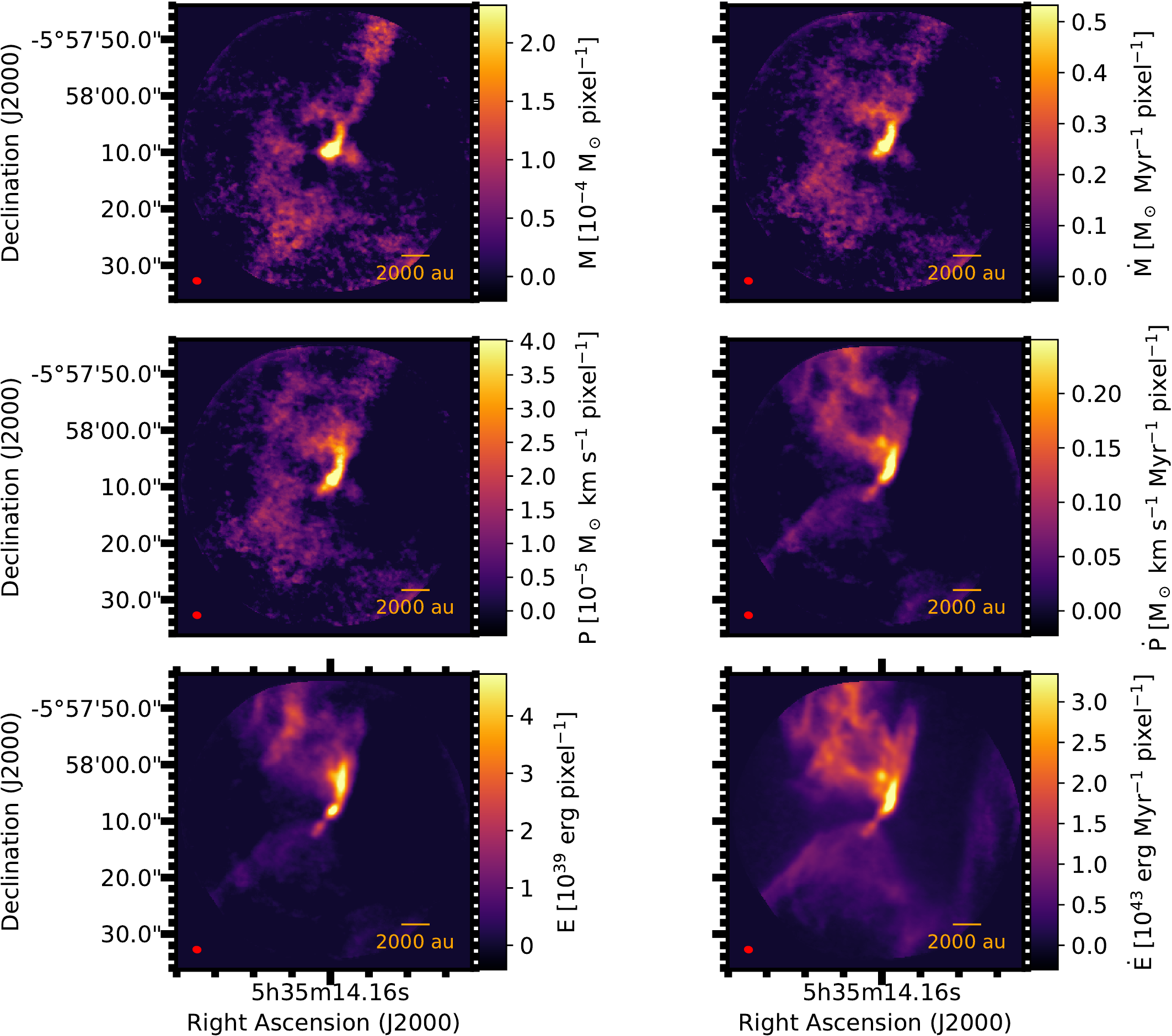}}
\caption {HOPS-11 outflow mass, momentum, energy  maps, and the corresponding instantaneous rate maps. Note for this figure set, we exclude the high-velocity jet components. 
{ Note that the pixel size is 0\as17 $\times$  0\as17. The red ellipses in the bottom left of plots represent the synthesized beam.}}
\label{fig25}
\end{figure*}

\begin{figure*}[tbh]
\centering
\makebox[\textwidth]{\includegraphics[width=\textwidth]{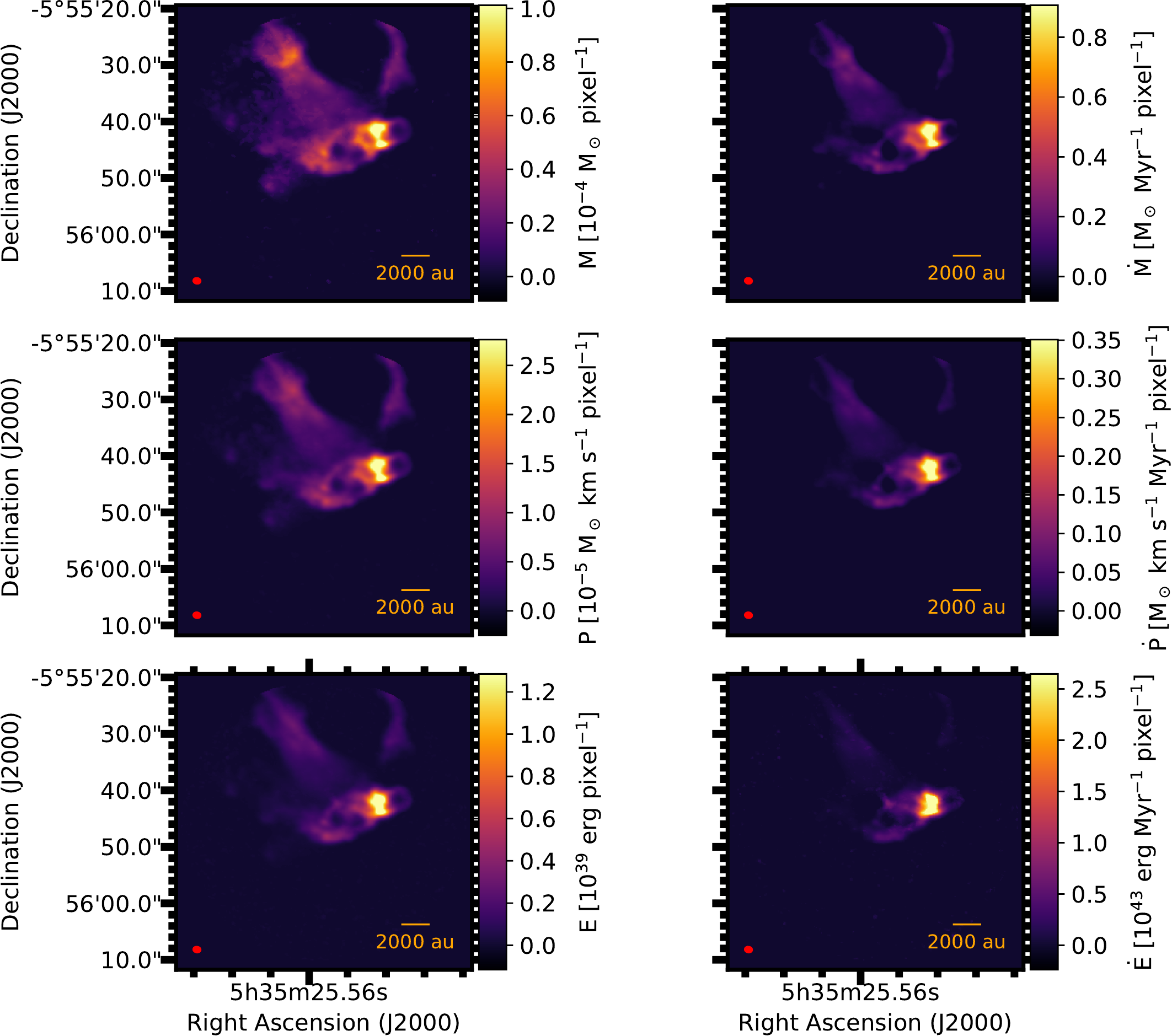}}
\caption {HOPS-13 outflow mass, momentum, energy  maps, and the corresponding instantaneous rate maps. 
{ Note that the pixel size is 0\as17 $\times$  0\as17. The red ellipses in the bottom left of plots represent the synthesized beam.}}
\label{fig26}
\end{figure*}

\begin{figure*}[tbh]
\centering
\makebox[\textwidth]{\includegraphics[width=\textwidth]{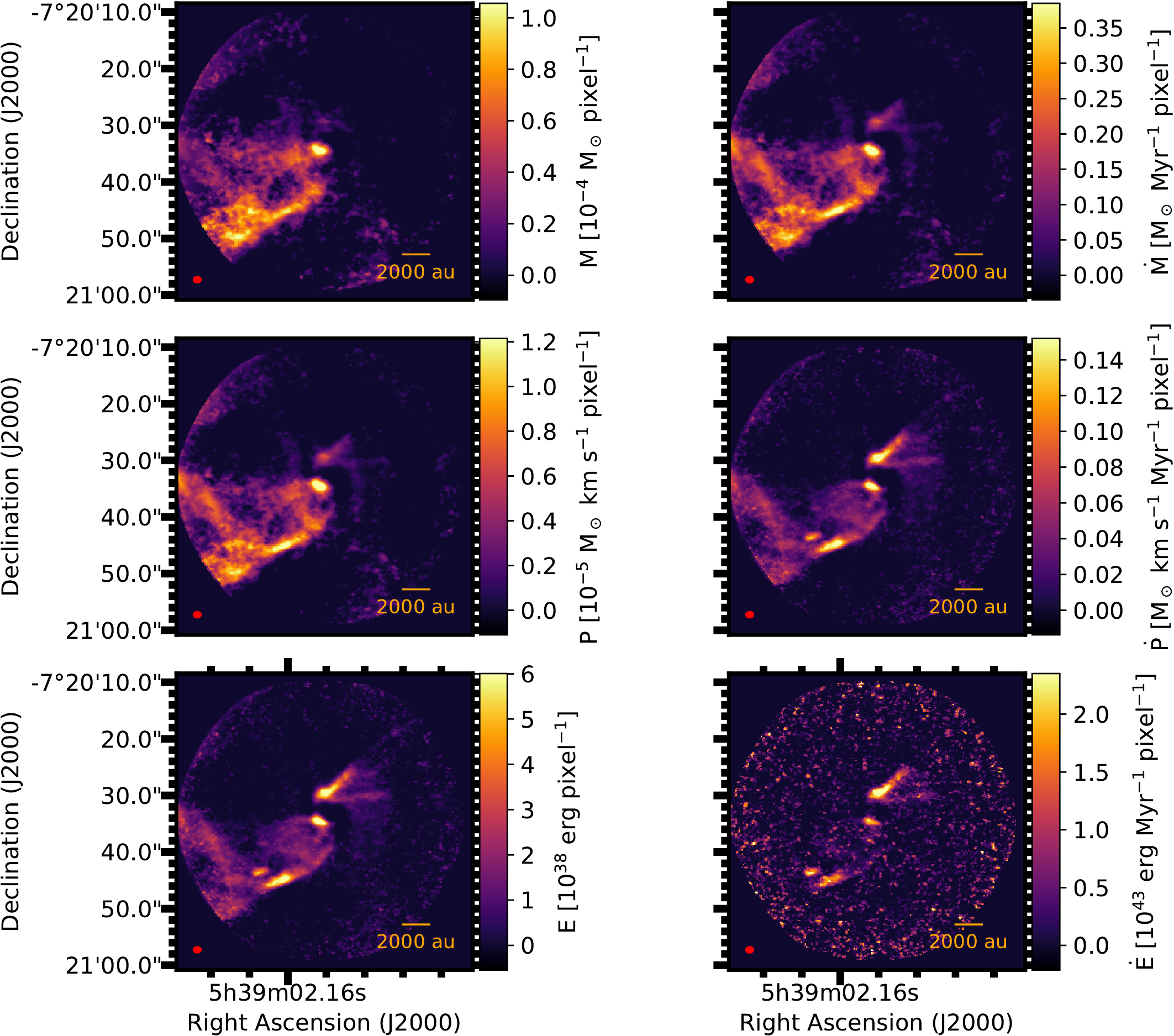}}
\caption {HOPS-127 outflow mass, momentum, energy  maps, and the corresponding instantaneous rate maps. 
{ Note that the pixel size is 0\as17 $\times$  0\as17. The red ellipses in the bottom left of plots represent the synthesized beam.}}
\label{fig27}
\end{figure*}

\begin{figure*}[tbh]
\centering
\makebox[\textwidth]{\includegraphics[width=\textwidth]{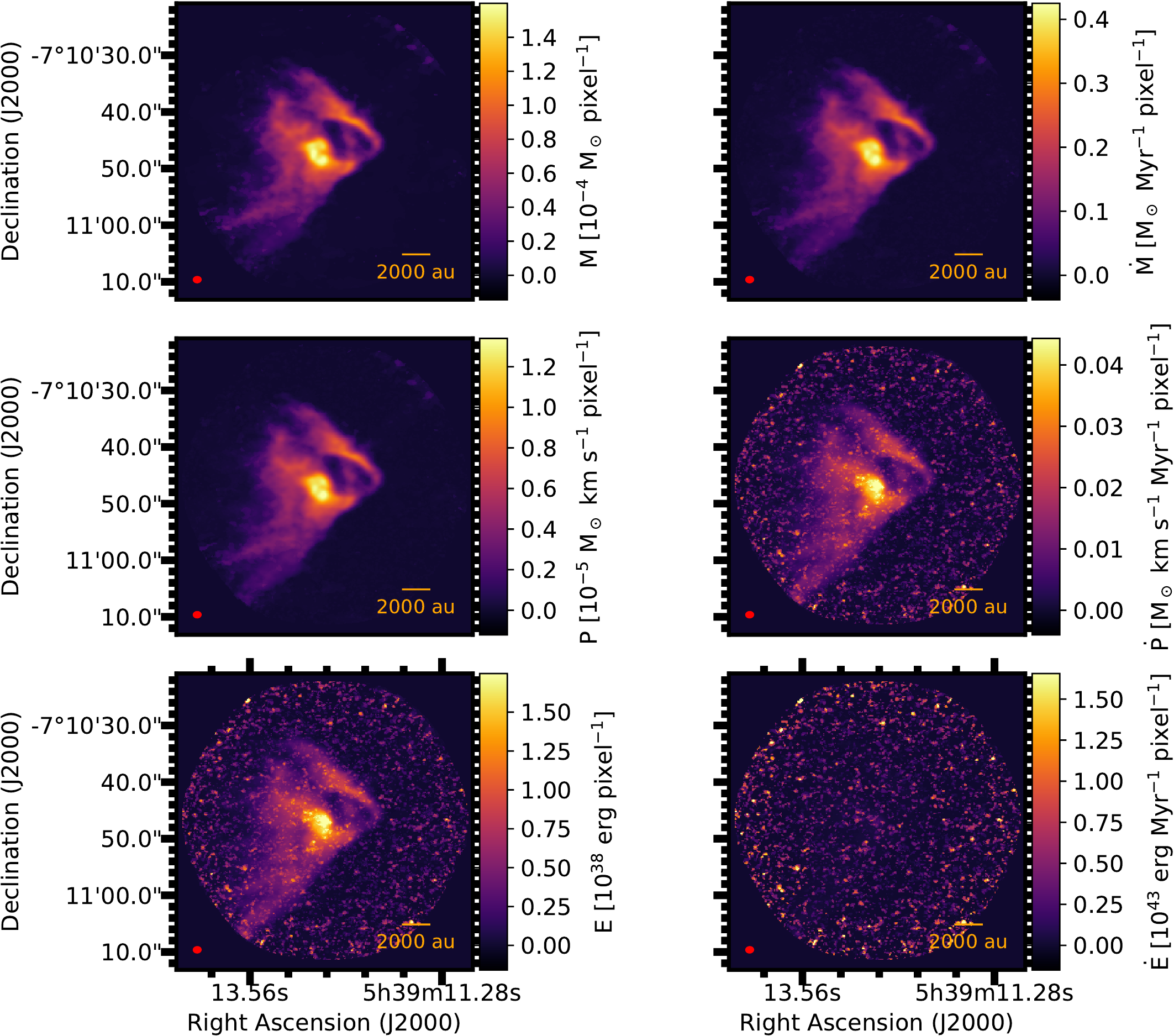}}
\caption {HOPS-129 outflow mass, momentum, energy  maps, and the corresponding instantaneous rate maps. 
{ Note that the pixel size is 0\as17 $\times$  0\as17. The red ellipses in the bottom left of plots represent the synthesized beam. There is no detection in the energy ejection rate map due to the high noise level.}}
\label{fig28}
\end{figure*}

\begin{figure*}[tbh]
\centering
\makebox[\textwidth]{\includegraphics[width=\textwidth]{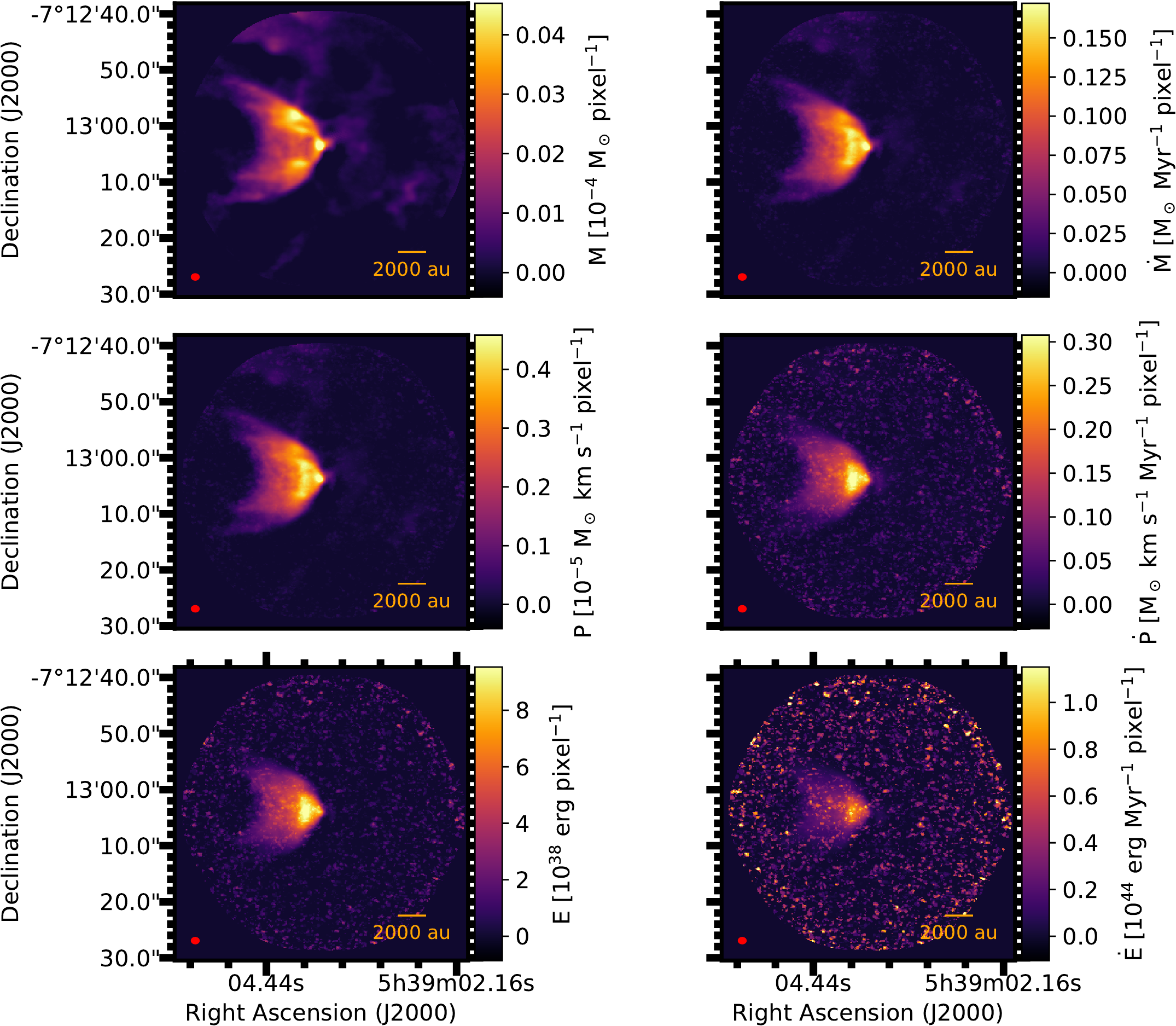}}
\caption {HOPS-130 outflow mass, momentum, energy  maps, and the corresponding instantaneous rate maps. 
{ Note that the pixel size is 0\as17 $\times$  0\as17. The red ellipses in the bottom left of plots represent the synthesized beam.}}
\label{fig29}
\end{figure*}

\begin{figure*}[tbh]
\centering
\makebox[\textwidth]{\includegraphics[width=\textwidth]{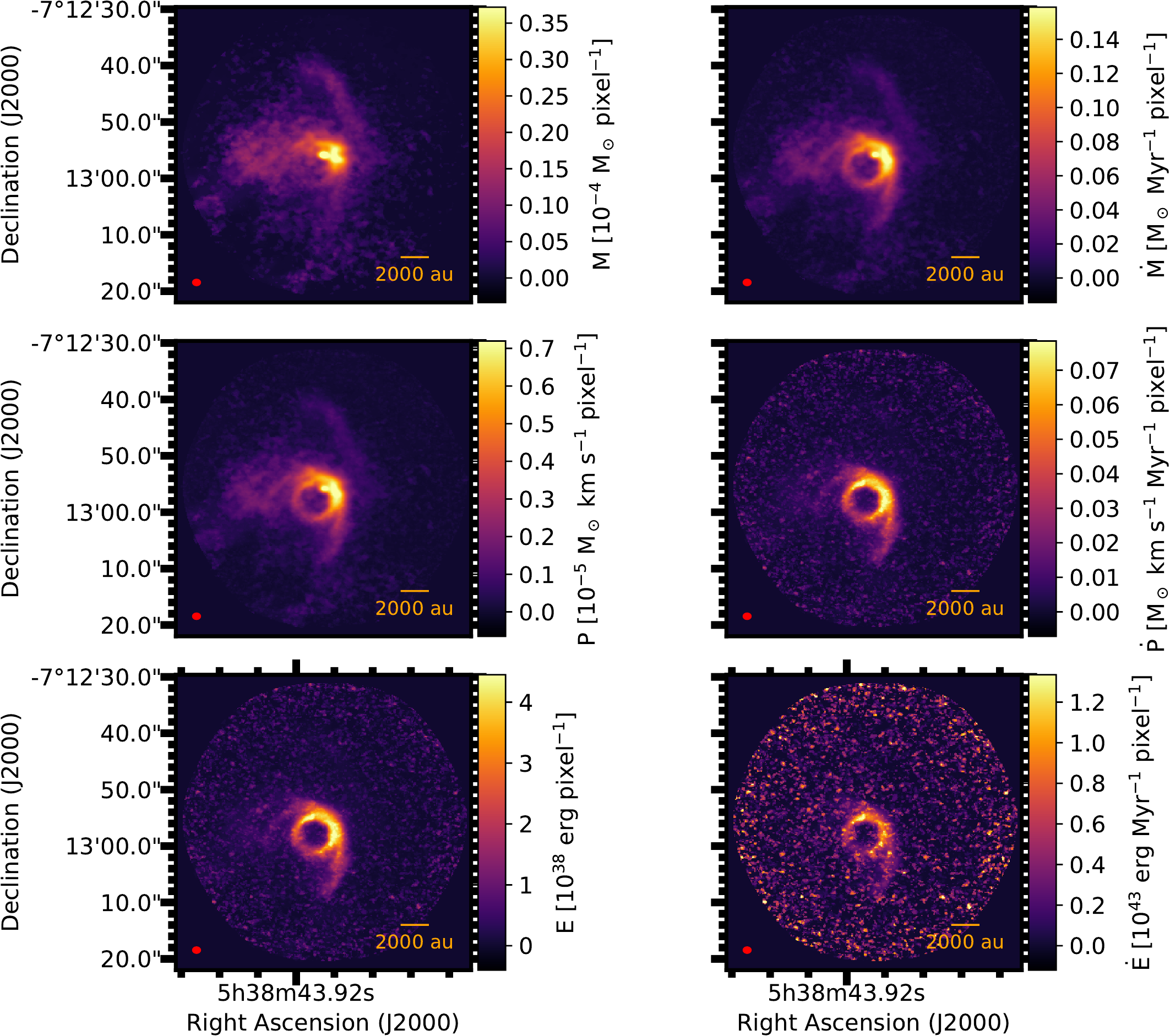}}
\caption {HOPS-134 outflow mass, momentum, energy  maps, and the corresponding instantaneous rate maps. 
{ Note that the pixel size is 0\as17 $\times$  0\as17. The red ellipses in the bottom left of plots represent the synthesized beam.}}
\label{fig29a}
\end{figure*}

\begin{figure*}[tbh]
\centering
\makebox[\textwidth]{\includegraphics[width=\textwidth]{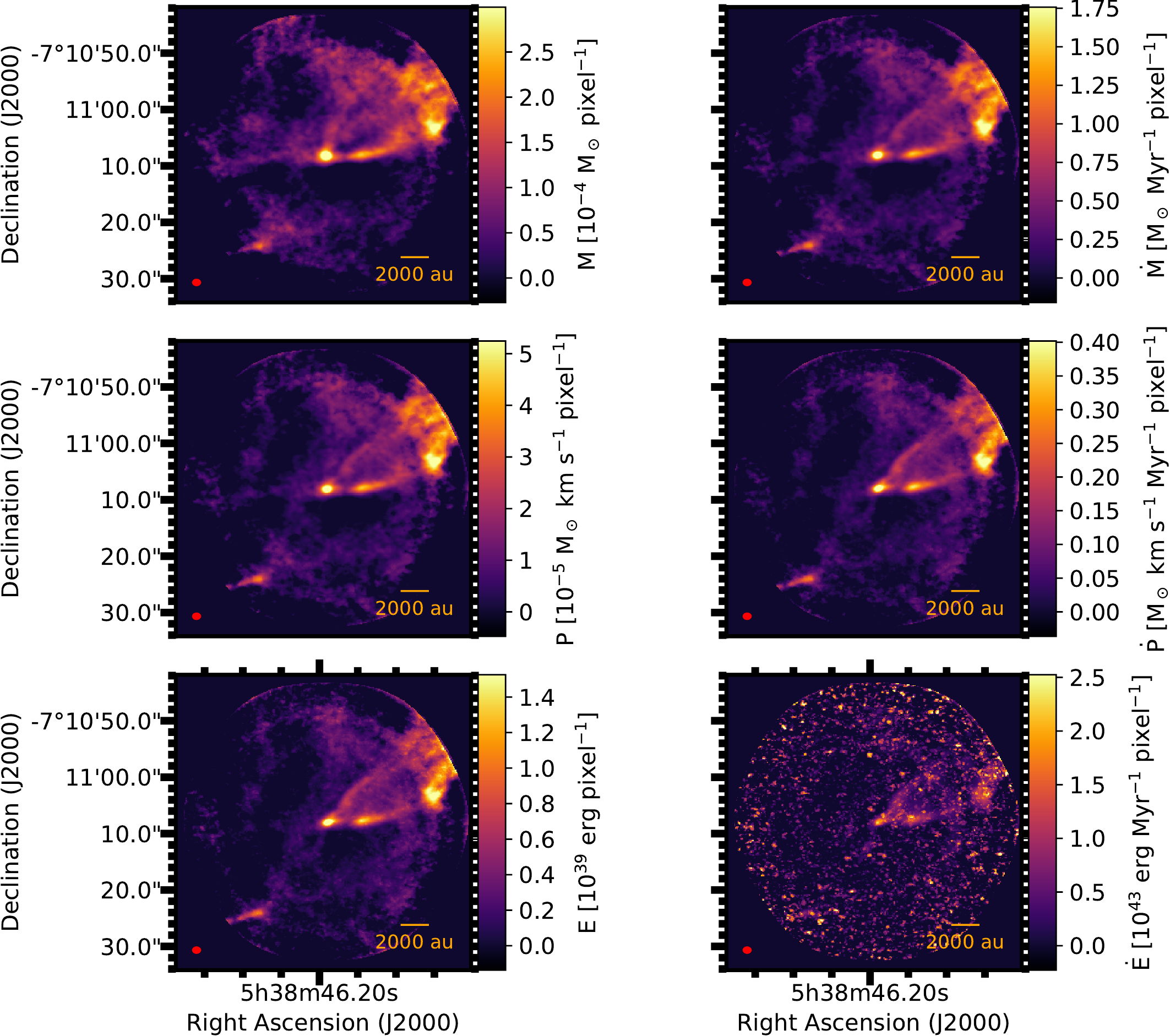}}
\caption {HOPS-135 outflow mass, momentum, energy  maps, and the corresponding instantaneous rate maps. 
{ Note that the pixel size is 0\as17 $\times$  0\as17. The red ellipses in the bottom left of plots represent the synthesized beam.}}
\label{fig30}
\end{figure*}

\begin{figure*}[tbh]
\centering
\makebox[\textwidth]{\includegraphics[width=\textwidth]{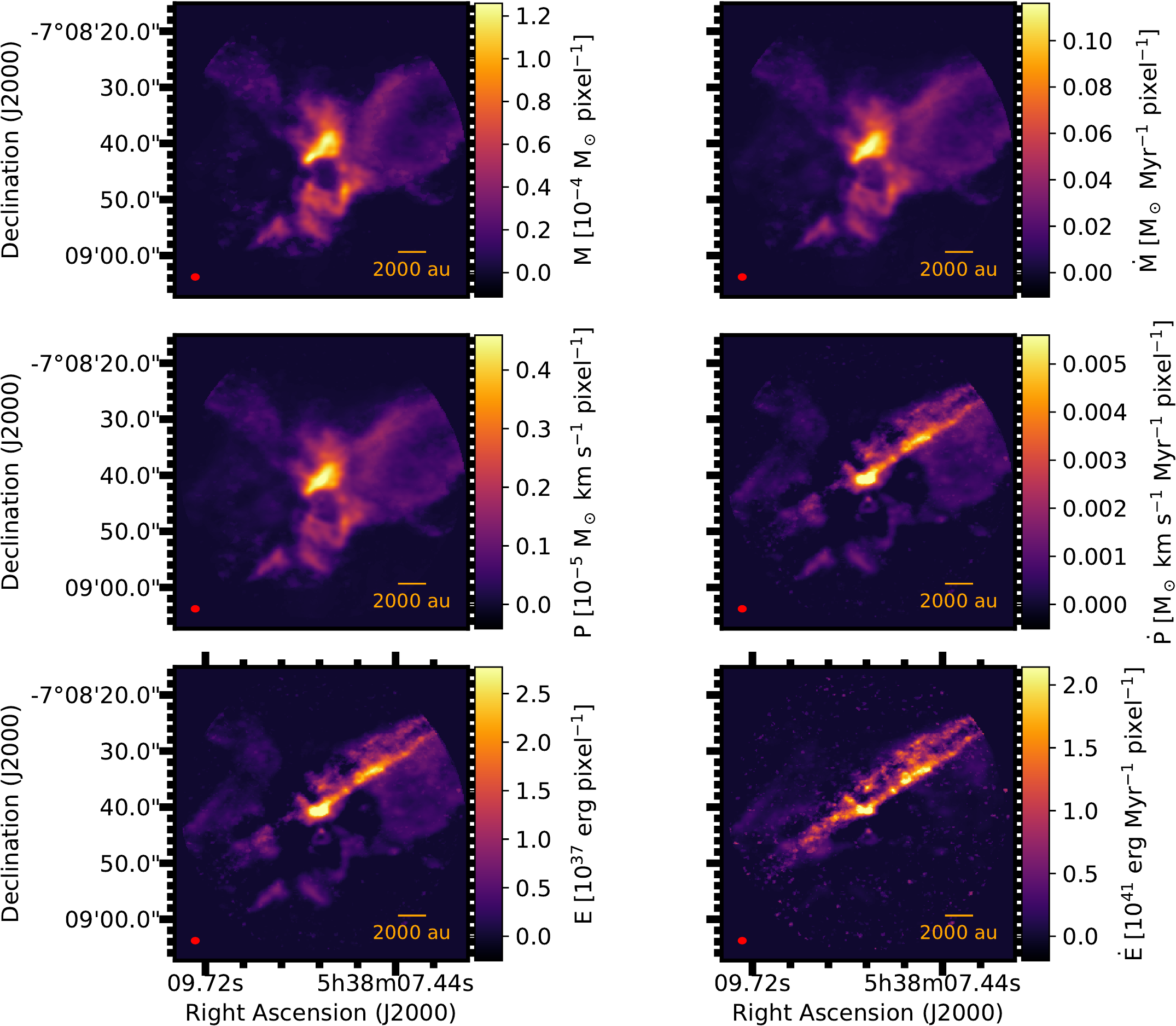}}
\caption {HOPS-150 outflow mass, momentum, energy  maps, and the corresponding instantaneous rate maps. 
{ Note that the pixel size is 0\as17 $\times$  0\as17. The red ellipses in the bottom left of plots represent the synthesized beam.} 2 jets with molecular bullets appear clearly in the energy ejection rate maps. The southern jet coincides spatially with HOPS-150 A, but the northern jet is originated from a nearby source outside the field of view.}
\label{fig31}
\end{figure*}

\begin{figure*}[tbh]
\centering
\makebox[\textwidth]{\includegraphics[width=\textwidth]{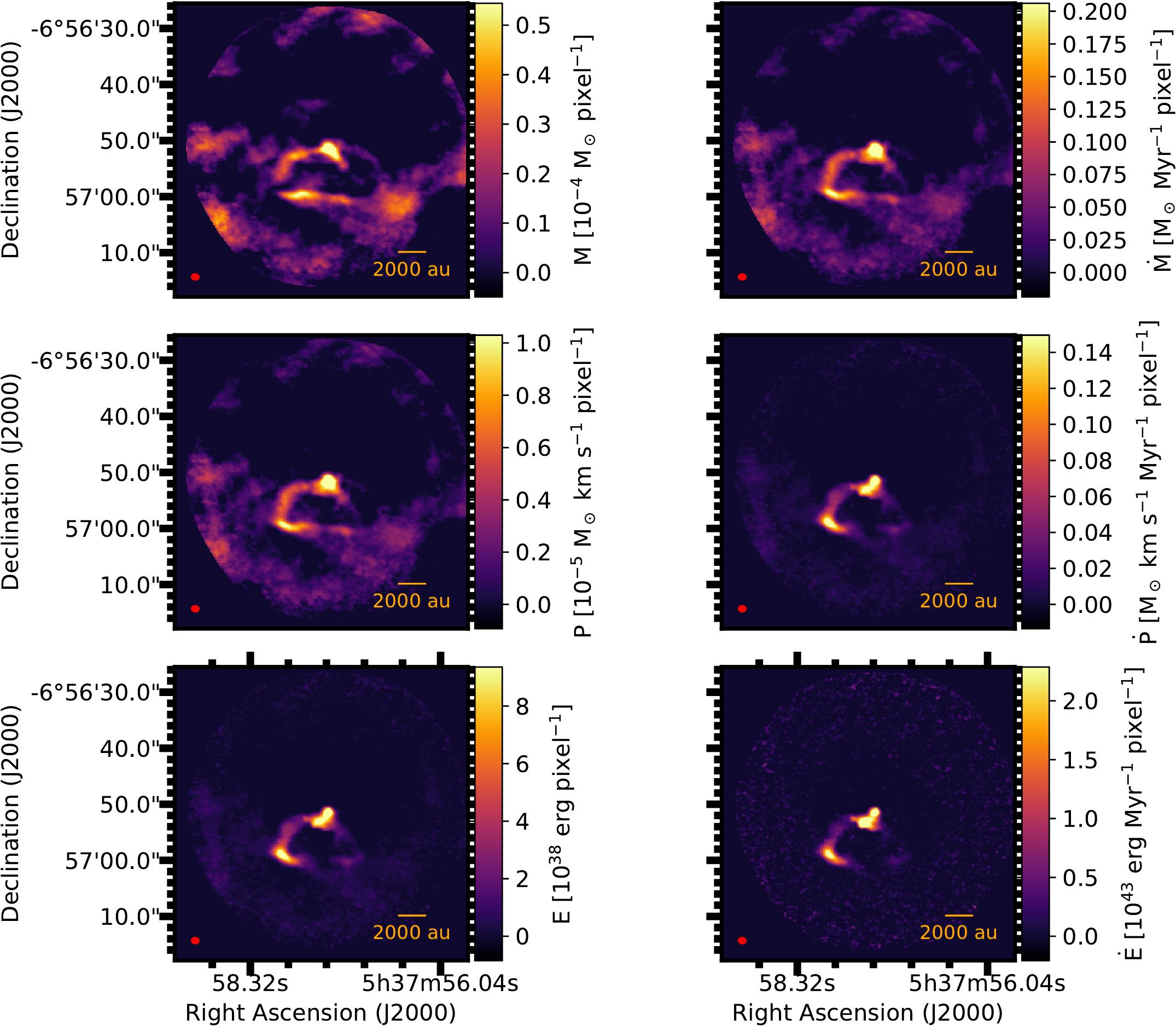}}
\caption {HOPS-157 outflow mass, momentum, energy  maps, and the corresponding instantaneous rate maps. 
{ Note that the pixel size is 0\as17 $\times$  0\as17. The red ellipses in the bottom left of plots represent the synthesized beam.}}
\label{fig32}
\end{figure*}

\begin{figure*}[tbh]
\centering
\makebox[\textwidth]{\includegraphics[width=\textwidth]{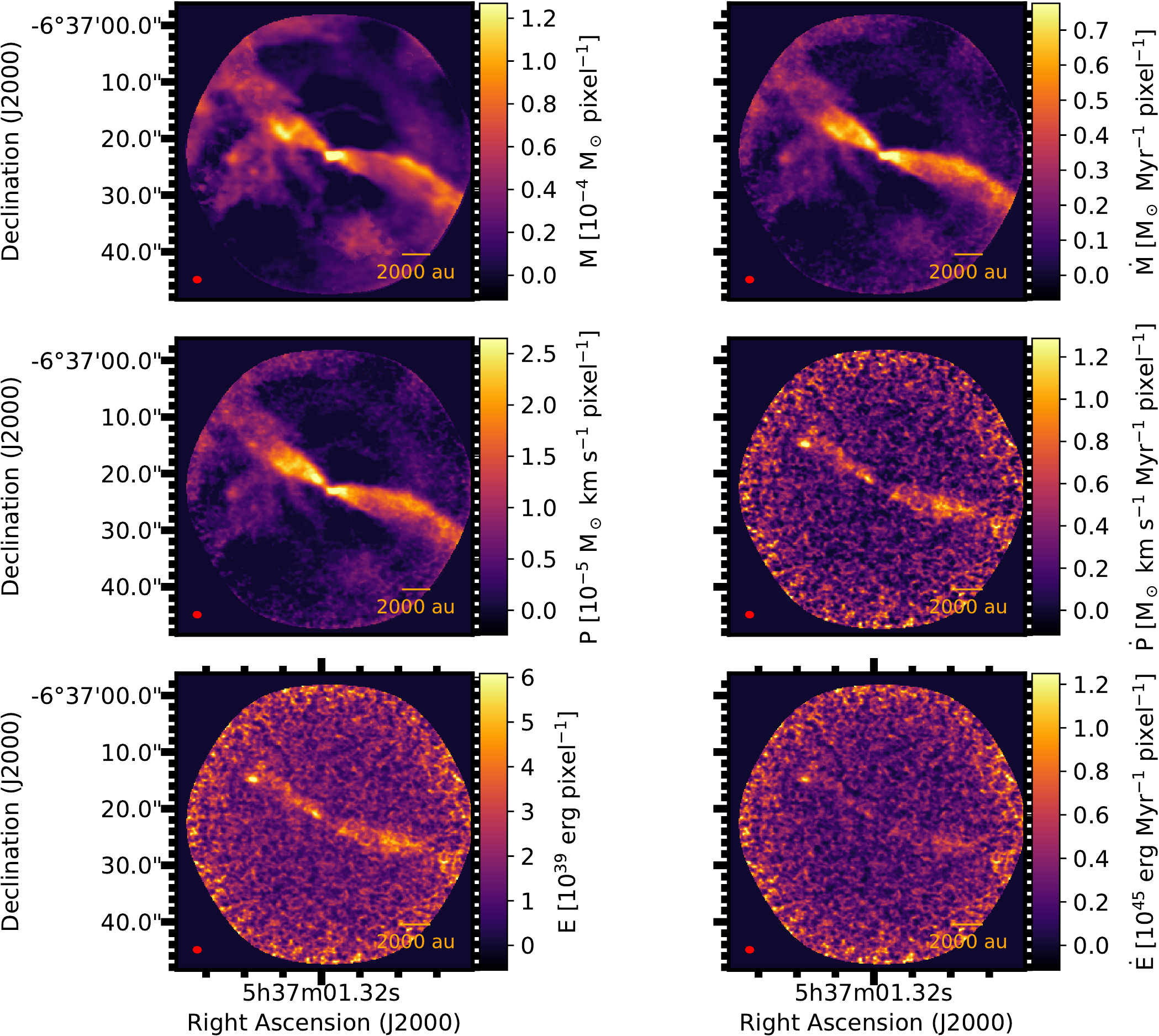}}
\caption {HOPS-164 outflow mass, momentum, energy  maps, and the corresponding instantaneous rate maps. 
{ Note that the pixel size is 0\as17 $\times$  0\as17. The red ellipses in the bottom left of plots represent the synthesized beam.}}
\label{fig33}
\end{figure*}

\begin{figure*}[tbh]
\centering
\makebox[\textwidth]{\includegraphics[width=\textwidth]{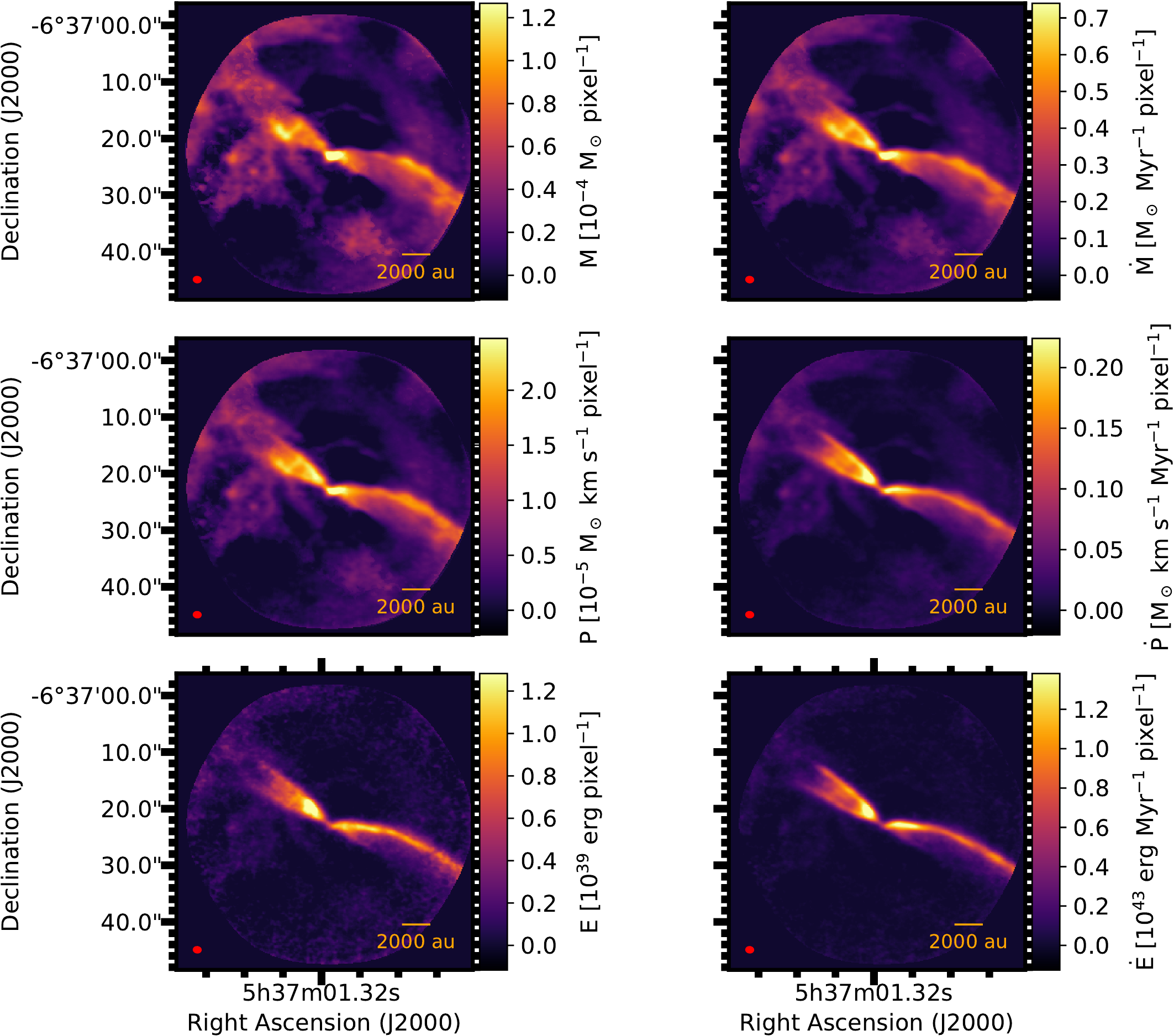}}
\caption {HOPS-164 outflow mass, momentum, energy  maps, and the corresponding instantaneous rate maps. Note for this figure set, we exclude the high-velocity jet components. { The pixel size is 0\as17 $\times$  0\as17. The red ellipses in the bottom left of plots represent the synthesized beam.}}
\label{fig34}
\end{figure*}

\begin{figure*}[tbh]
\centering
\makebox[\textwidth]{\includegraphics[width=\textwidth]{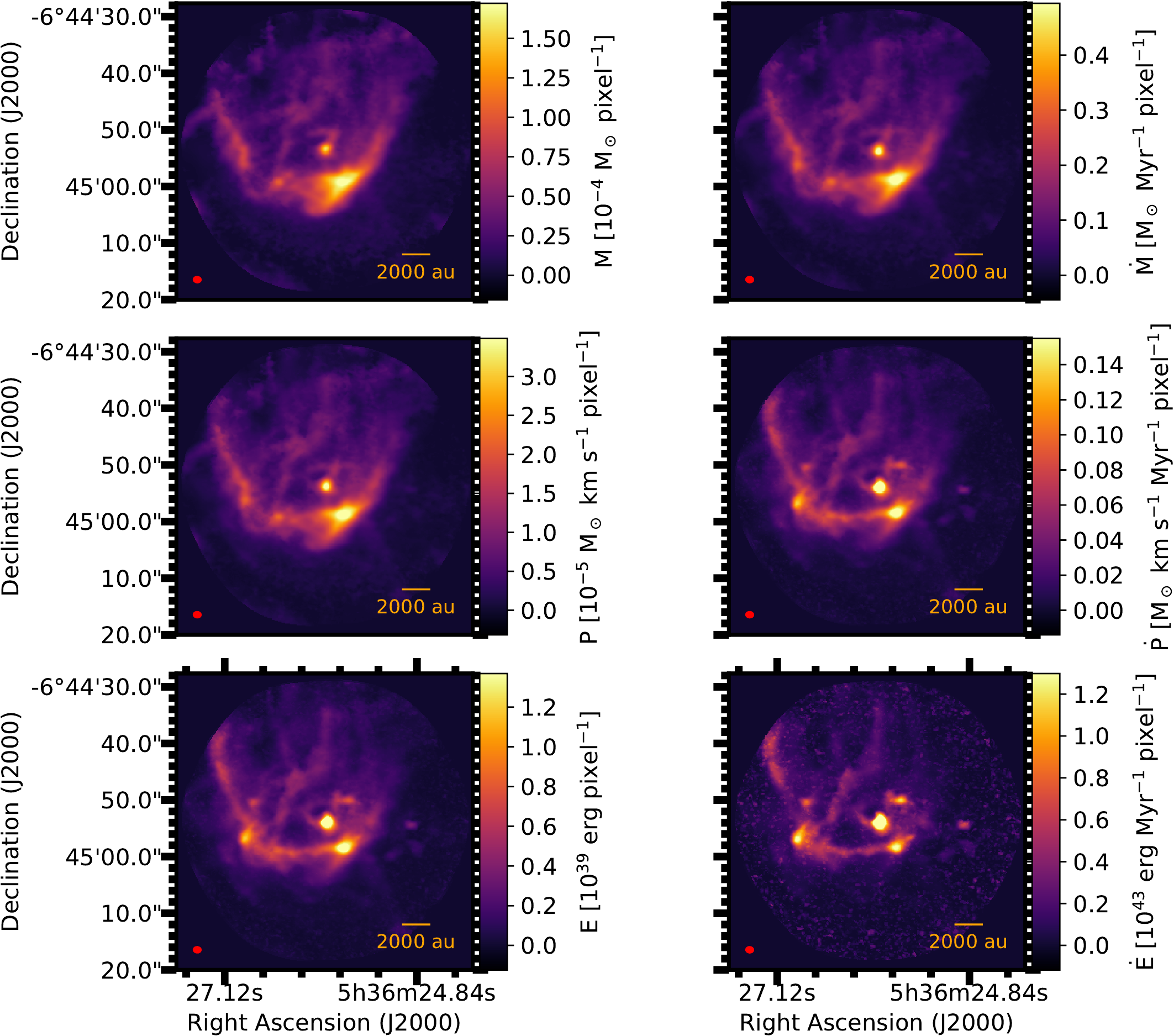}}
\caption {HOPS-166 outflow mass, momentum, energy  maps, and the corresponding instantaneous rate maps. 
{ Note that the pixel size is 0\as17 $\times$  0\as17. The red ellipses in the bottom left of plots represent the synthesized beam.}}
\label{fig35}
\end{figure*}

\begin{figure*}[tbh]
\centering
\makebox[\textwidth]{\includegraphics[width=\textwidth]{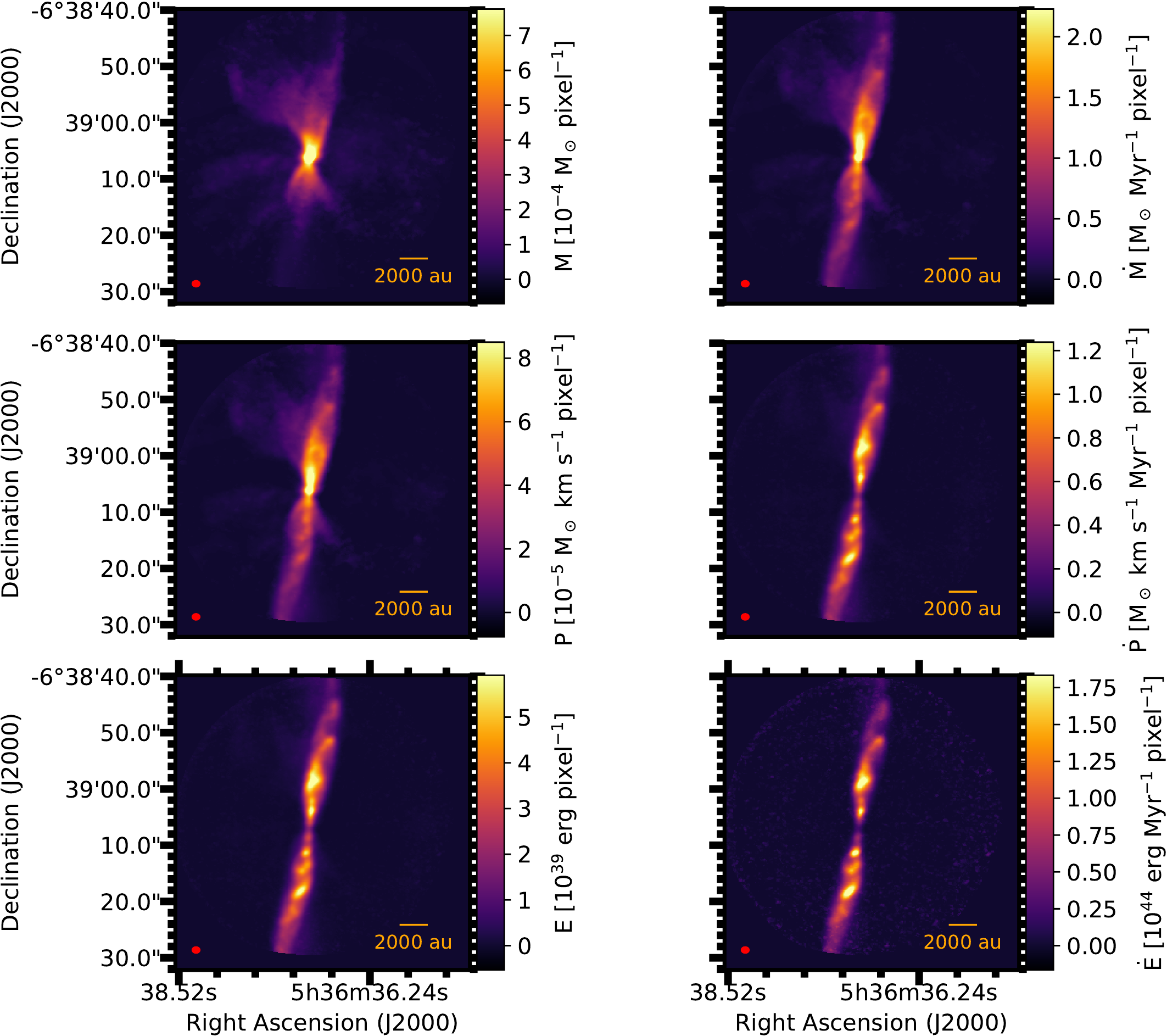}}
\caption {HOPS-169 outflow mass, momentum, energy  maps, and the corresponding instantaneous rate maps. 
{ Note that the pixel size is 0\as17 $\times$  0\as17. The red ellipses in the bottom left of plots represent the synthesized beam.}}
\label{fig36}
\end{figure*}

\begin{figure*}[tbh]
\centering
\makebox[\textwidth]{\includegraphics[width=\textwidth]{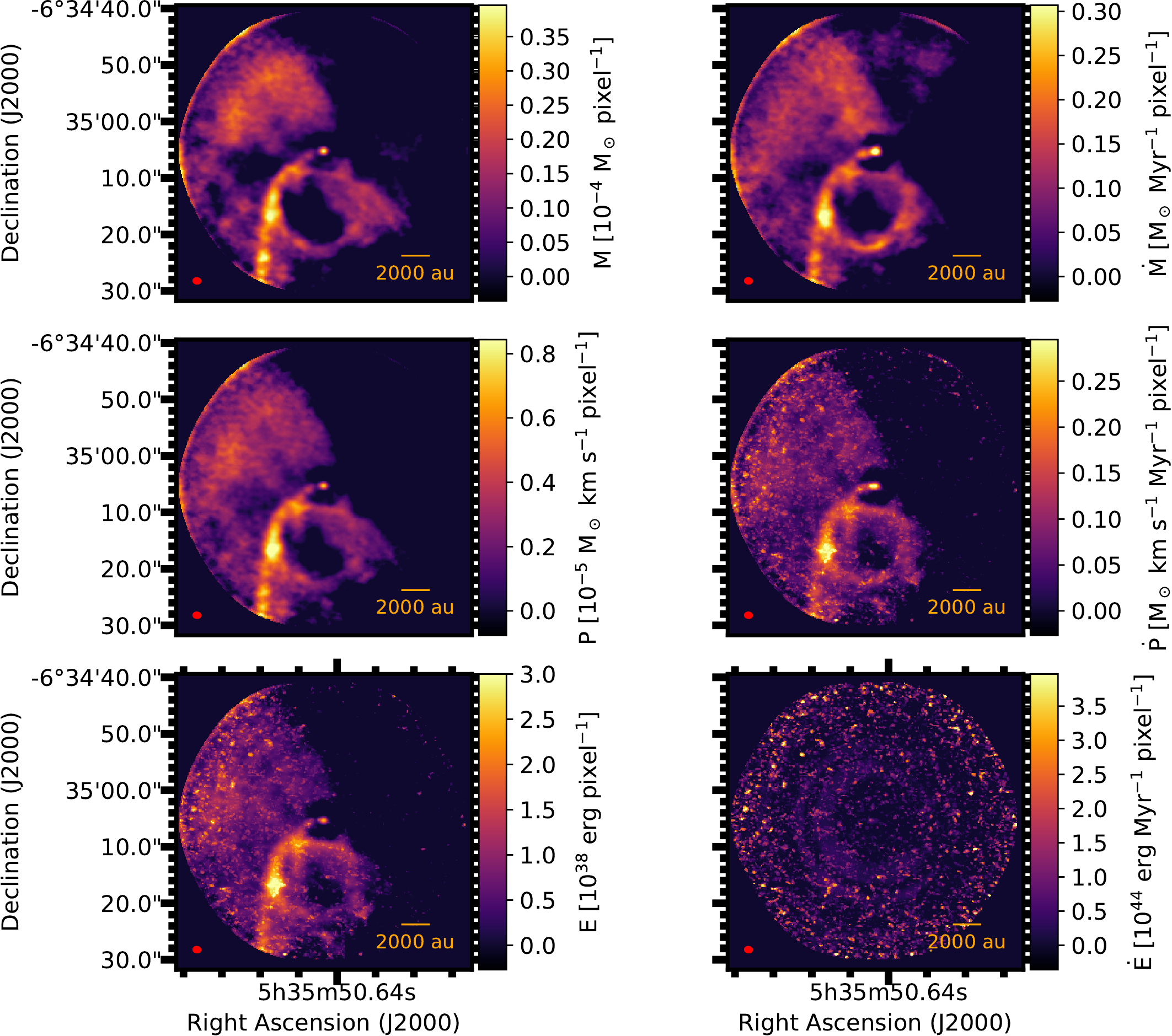}}
\caption {HOPS-177 outflow mass, momentum, energy  maps, and the corresponding instantaneous rate maps. 
{ Note that the pixel size is 0\as17 $\times$  0\as17. The red ellipses in the bottom left of plots represent the synthesized beam. There is no detection in the energy ejection rate map due to the high noise level.}}
\label{fig37}
\end{figure*}


\begin{figure*}[tbh]
\centering
\makebox[\textwidth]{\includegraphics[width=\textwidth]{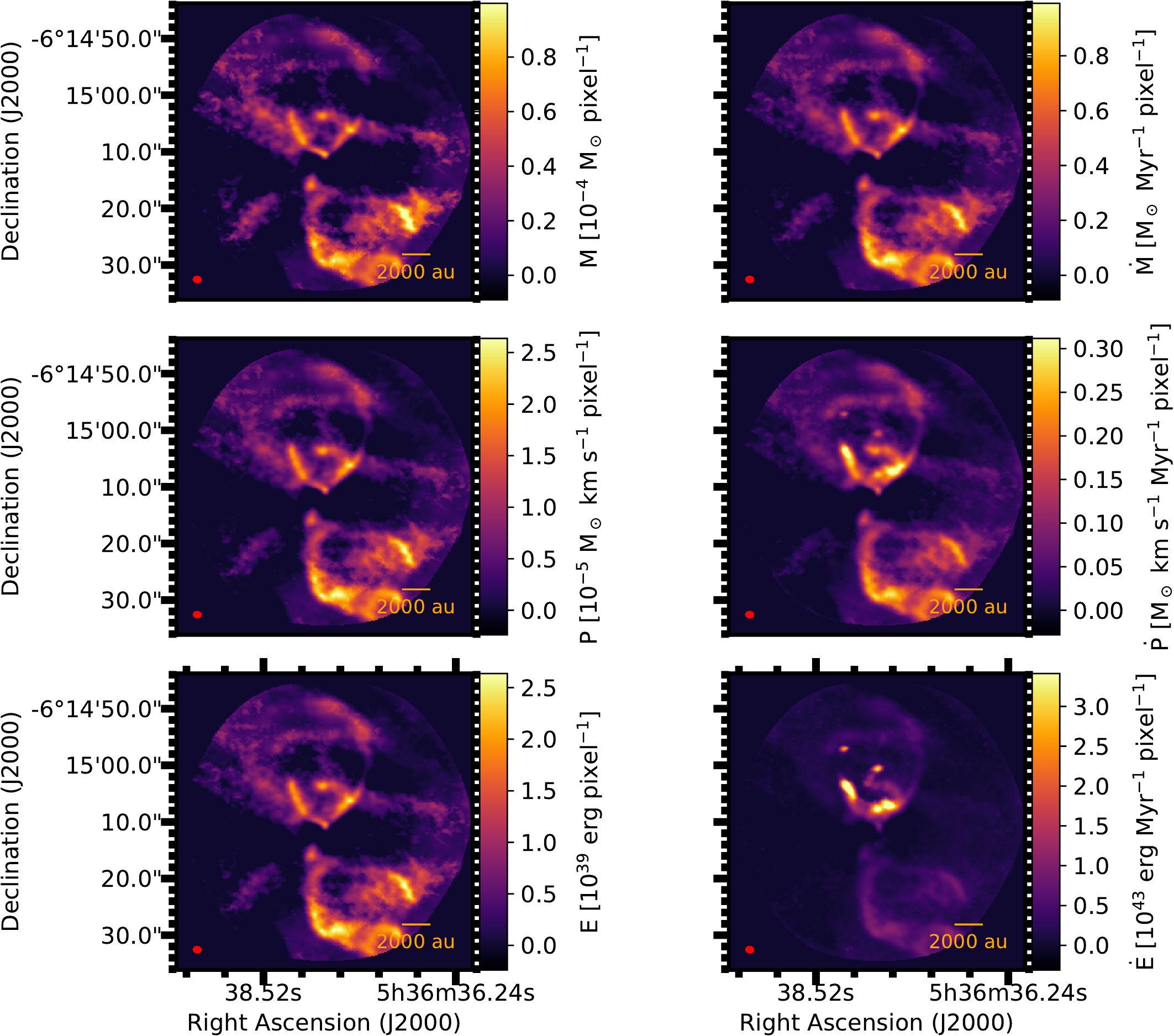}}
\caption {HOPS-185 outflow mass, momentum, energy  maps, and the corresponding instantaneous rate maps. 
{ Note that the pixel size is 0\as17 $\times$  0\as17. The red ellipses in the bottom left of plots represent the synthesized beam.}}
\label{fig39}
\end{figure*}

\begin{figure*}[tbh]
\centering
\makebox[\textwidth]{\includegraphics[width=\textwidth]{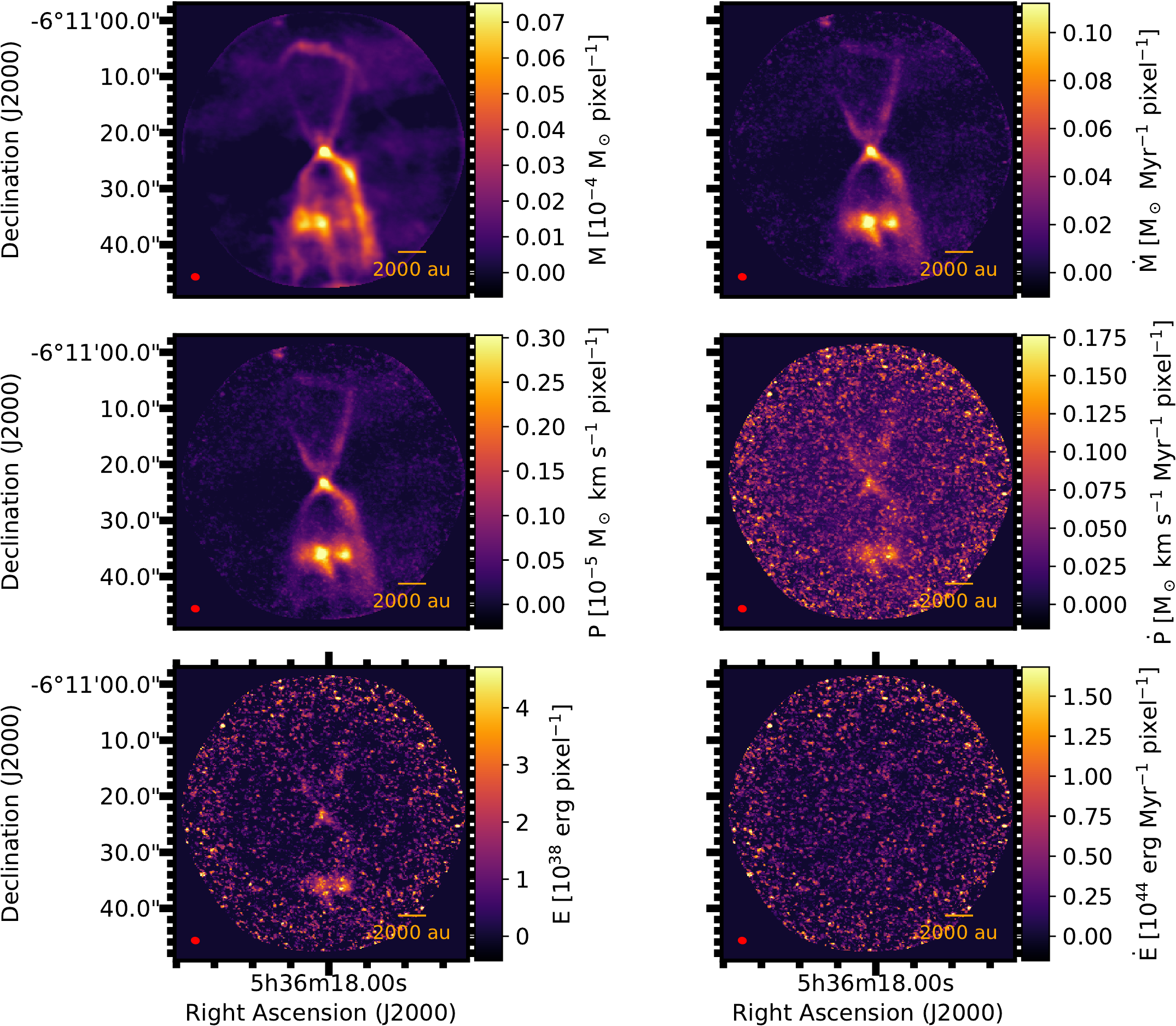}}
\caption {HOPS-191 outflow mass, momentum, energy  maps, and the corresponding instantaneous rate maps. 
{ Note that the pixel size is 0\as17 $\times$  0\as17. The red ellipses in the bottom left of plots represent the synthesized beam. There is no detection in the energy ejection rate map due to the high noise level.}}
\label{fig40}
\end{figure*}

\begin{figure*}[tbh]
\centering
\makebox[\textwidth]{\includegraphics[width=\textwidth]{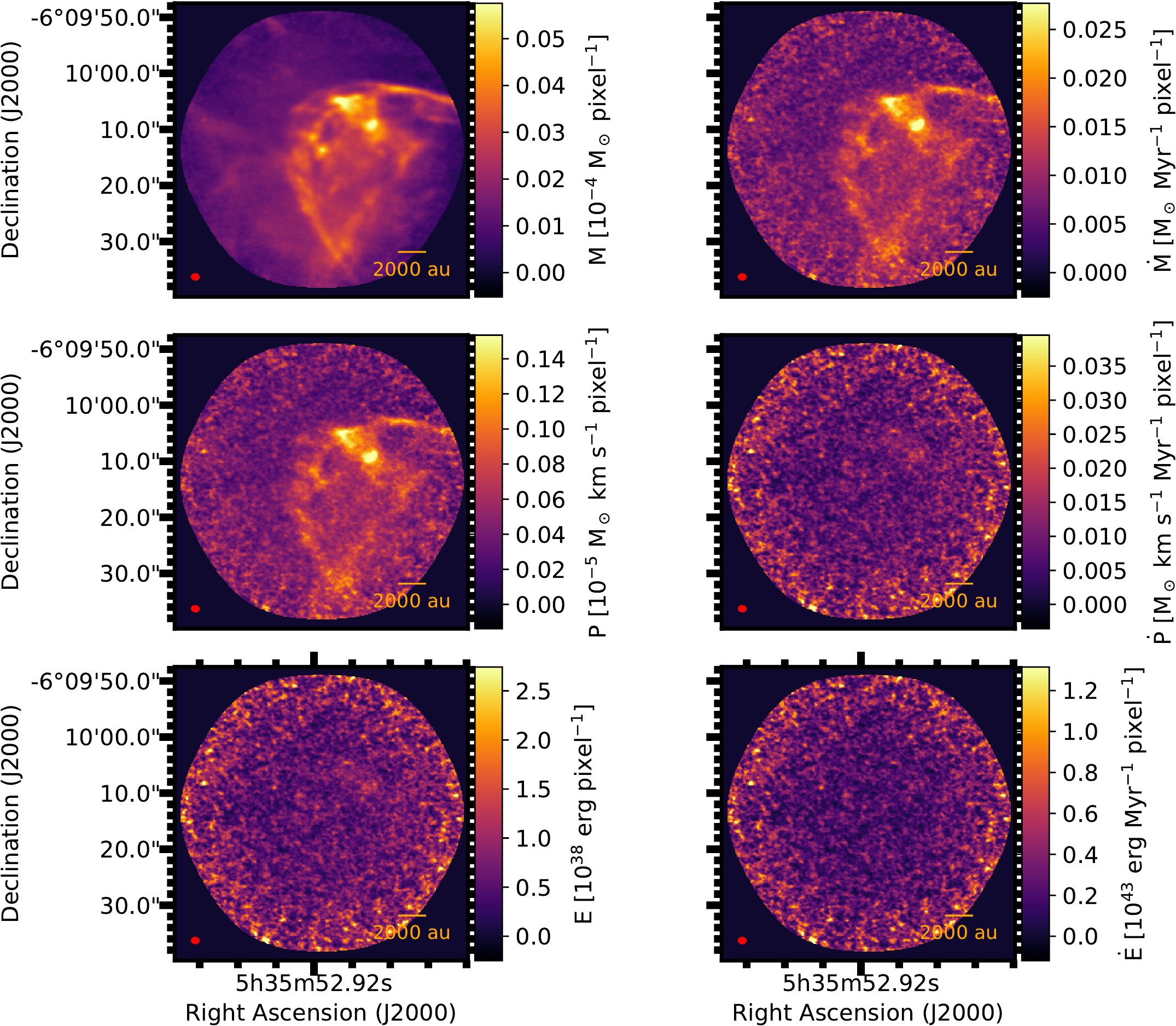}}
\caption {HOPS-194 outflow mass, momentum, energy  maps, and the corresponding instantaneous rate maps. 
{ Note that the pixel size is 0\as17 $\times$  0\as17. The red ellipses in the bottom left of plots represent the synthesized beam. No clear detection is seen in the energy map, and the energy and momentum ejection rate maps.}}
\label{fig41}
\end{figure*}

\begin{figure*}[tbh]
\centering
\makebox[\textwidth]{\includegraphics[width=\textwidth]{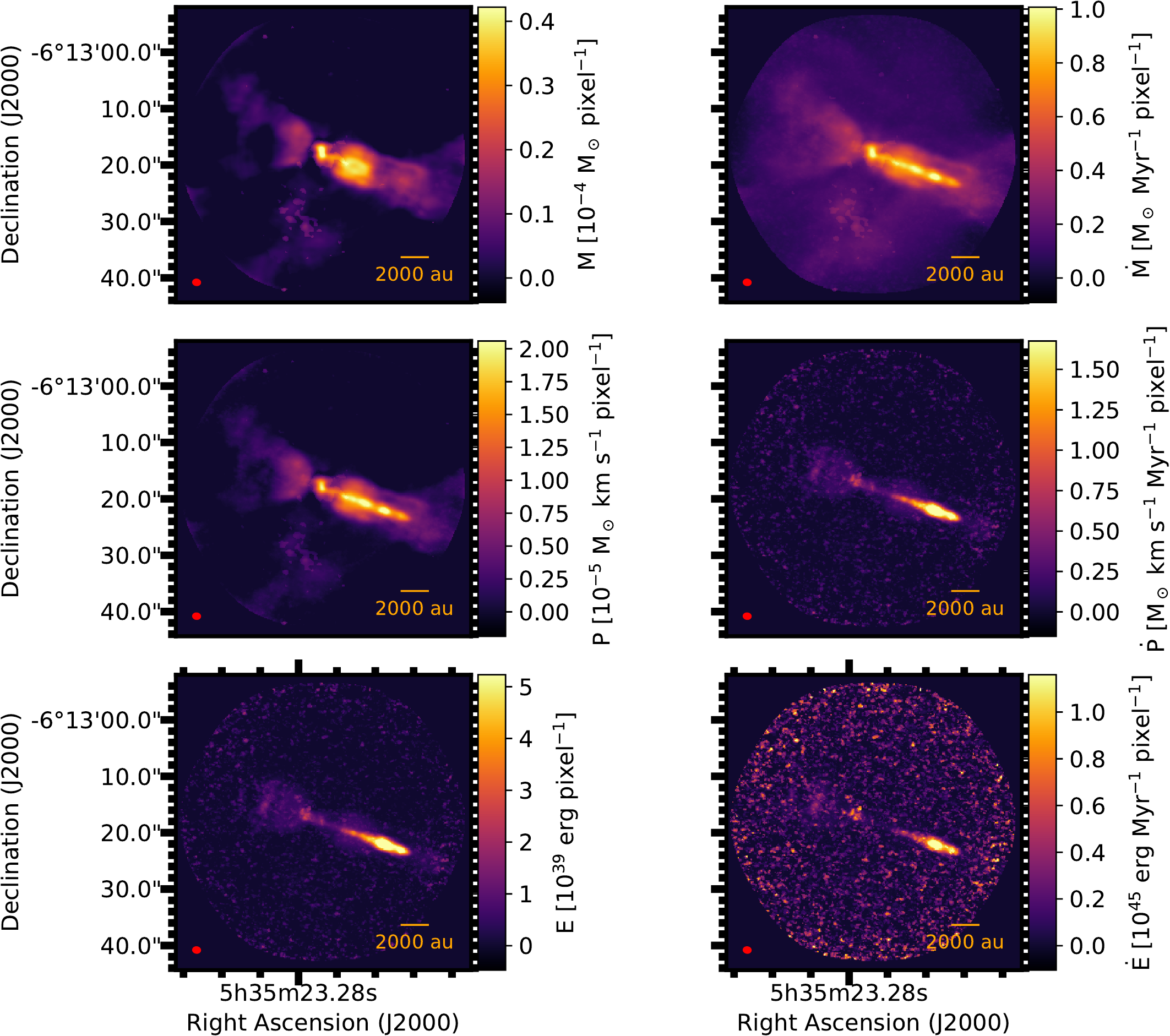}}
\caption {HOPS-198 outflow mass, momentum, energy  maps, and the corresponding instantaneous rate maps. 
{ Note that the pixel size is 0\as17 $\times$  0\as17. The red ellipses in the bottom left of plots represent the synthesized beam.}}
\label{fig42}
\end{figure*}

\begin{figure*}[tbh]
\centering
\makebox[\textwidth]{\includegraphics[width=\textwidth]{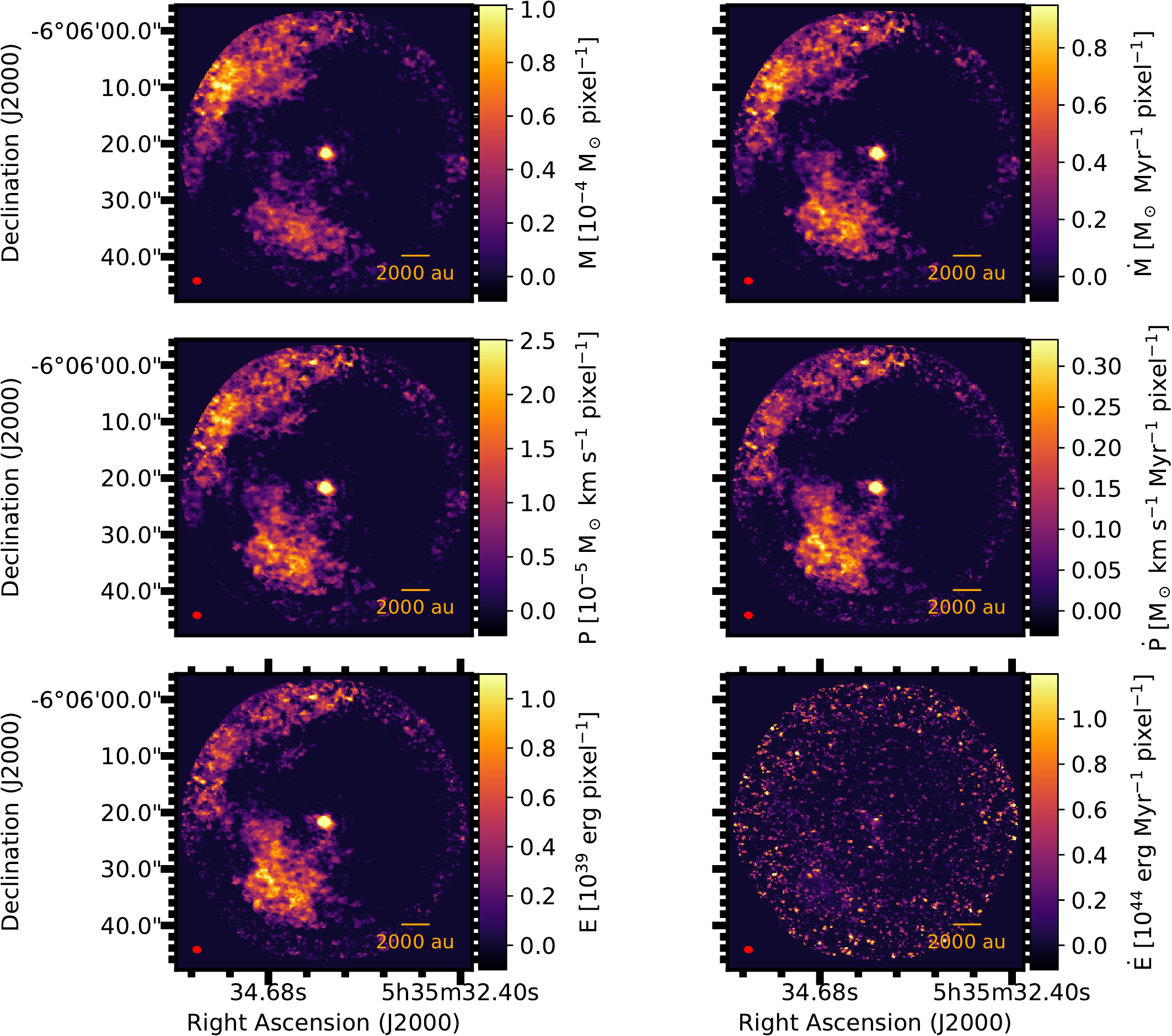}}
\caption {HOPS-200 outflow mass, momentum, energy  maps, and the corresponding instantaneous rate maps. 
{ Note that the pixel size is 0\as17 $\times$  0\as17. The red ellipses in the bottom left of plots represent the synthesized beam.}}
\label{fig43}
\end{figure*}

\begin{figure*}[tbh]
\centering
\makebox[\textwidth]{\includegraphics[width=\textwidth]{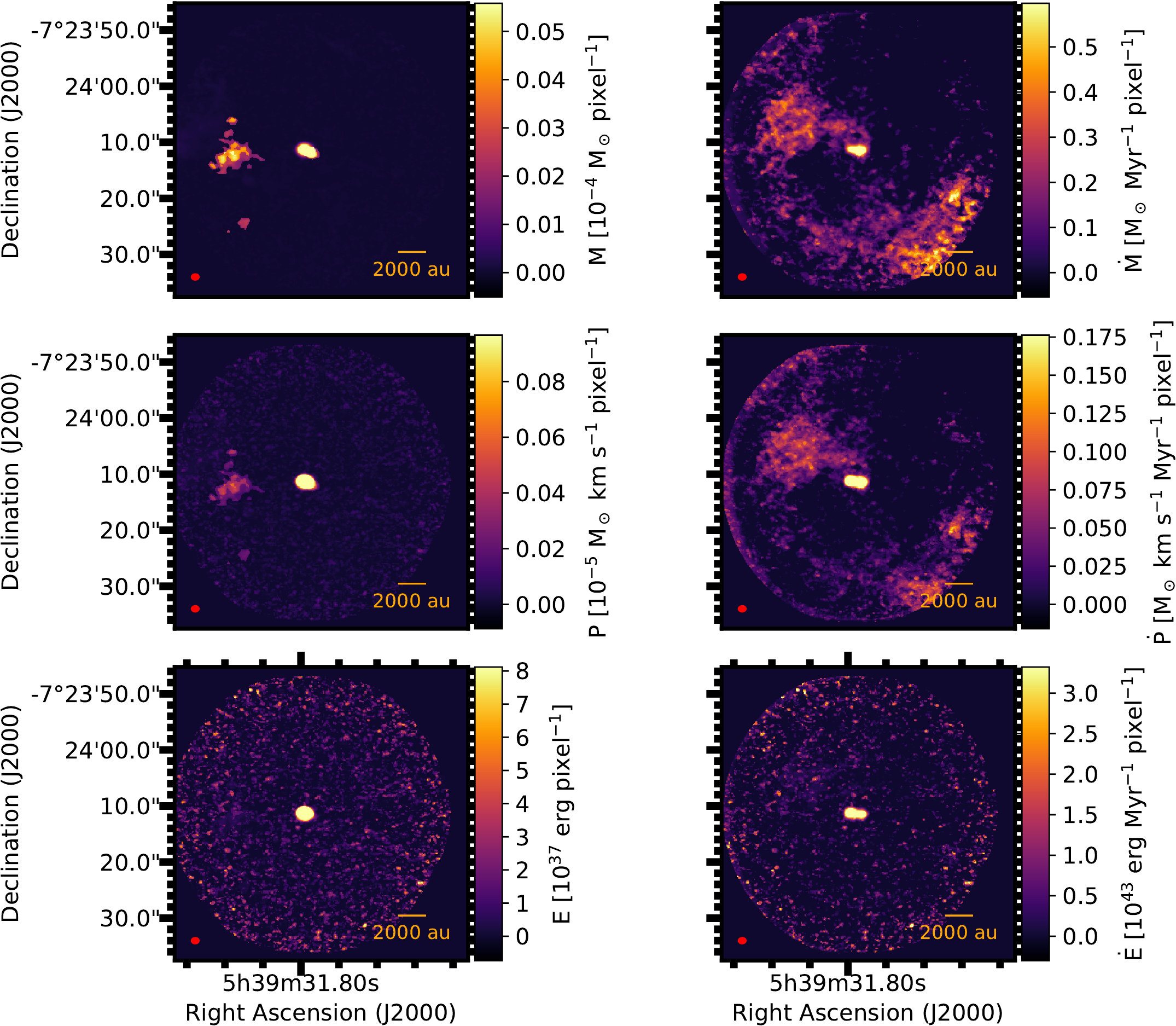}}
\caption {HOPS-408 outflow mass, momentum, energy  maps, and the corresponding instantaneous rate maps. 
{ Note that the pixel size is 0\as17 $\times$  0\as17. The red ellipses in the bottom left of plots represent the synthesized beam.}}
\label{fig:HOPS408}
\end{figure*}

\section{Outflow Opening Angle, Momentum opening angle, and Energy opening angle}
In this section, we presents all the measurements for the traditional outflow cavity opening angle, momentum opening angle and the energy opening angles.  


\begin{table}[H]
\setlength{\tabcolsep}{6pt} 
\caption{Summary of outflow opening angles without inclination correction}             
\label{table:opening_angles_noinc}      
\centering                          
\begin{tabular}{c c c c }        
\hline\hline                 
Source  & Traditional  & Momentum & Energy  \\    
& opening angle\tablenotemark{a} & opening angle\tablenotemark{a} & opening angle\tablenotemark{a} \\  
&  ($^\circ$)&  ($^\circ$) & ($^\circ$)  \\ 
\hline                        
HOPS-10 & 38.0 $\pm$ 3.4 / 62.6 $\pm$ 3.6 & 77.1 $\pm$ 12.7 / 48.6 $\pm$ 2.8 & 23.3 $\pm$ 2.6 / 64.4 $\pm$ 4.7\\
\hline                                   
HOPS-11 & 117.2 $\pm$ 23.7 / 118.6 $\pm$ 5.0 & 42.0 $\pm$ 3.8 / 40.6 $\pm$ 4.7 & 30.1 $\pm$ 2.8 / 84.6 $\pm$ 5.7\\
\hline
HOPS-13 & 50.0 $\pm$ 2.4 / 57.6 $\pm$ 5.6 & NA / NA & NA / NA \\
\hline
HOPS-127 & 77.9 $\pm$ 5.6 / 62.9 $\pm$ 4.3 & NA / 112.7 $\pm$ 9.1 & NA / 130.7 $\pm$ 32.5\\
\hline
HOPS-129 &  75.6 $\pm$ 6.4 / NA & 45.2 $\pm$ 8.0 / 82.9 $\pm$ 6.5 & 38.6 $\pm$ 5.6 / 56.7 $\pm$ 7.5\\
\hline
HOPS-130 & 68.3 $\pm$ 8.1 / NA & 91.9 $\pm$ 13.1 / NA & 143.6 $\pm$ 14.6 / NA\\
\hline
HOPS-134 & 131.5 $\pm$ 17.1 / NA & NA / NA & NA / NA \\
\hline
HOPS-135 & NA / 71.3 $\pm$ 5.3 & NA / 64.6 $\pm$ 4.2 & 40.1 $\pm$ 2.5 / 62.6 $\pm$ 4.3\\
\hline
HOPS-150\tablenotemark{b} & NA / NA & 106.1 $\pm$ 11.7 / NA & NA / NA\\
\hline
HOPS-157 & 110.2 $\pm$ 11.0 / NA &  NA / NA & NA / NA\\ 
\hline
HOPS-164 & 45.0 $\pm$ 3.0 / 29.3 $\pm$ 0.5 & 56.4 $\pm$ 3.3 / 32.4 $\pm$ 1.8 & 45.8 $\pm$ 1.8 / 26.0 $\pm$ 0.7 \\
\hline
HOPS-166 & NA / NA & { NA} / { NA} & NA / 155.5 $\pm$ 82.4 \\
\hline
HOPS-169 & 67.6 $\pm$ 5.9 / 45.6 $\pm$ 1.6 & 34.0 $\pm$ 2.2 / 35.5 $\pm$ 3.3 & 36.3 $\pm$ 2.7 / 23.2 $\pm$ 1.2\\
\hline
HOPS-177 & 134.9 $\pm$ 19.0 / 91.6 $\pm$ 12.0 & NA / NA & 112.0 $\pm$ 5.2 / NA\\
\hline
HOPS-185 & 112.2 $\pm$ 13.9 / 92.2 $\pm$ 11.3 & NA / 97.5 $\pm$ 8.0 & NA / 98.0 $\pm$ 7.2\\
\hline
HOPS-191 & 113.2 $\pm$ 8.7 / 77.6 $\pm$ 8.5 & 83.8 $\pm$ 4.7 / 71.6 $\pm$ 5.7 & NA / NA \\
\hline
HOPS-194 & NA / 110.0 $\pm$ 11.1 & NA / NA & NA / NA \\
\hline
HOPS-198 & 69.9 $\pm$ 4.5 / NA & 60.0 $\pm$ 3.7 / 66.3 $\pm$ 3.7 & 34.4 $\pm$ 4.1 / NA \\
\hline
HOPS-200 & 137.2 $\pm$ 20.7 / NA & NA / NA & NA / NA \\
\hline
HOPS-355 & 24.3 $\pm$ 1.3 / 34.3 $\pm$ 1.5 & 31.2 $\pm$ 1.6 / 27.1 $\pm$ 1.8 & 40.8 $\pm$ 5.4 / 26.3 $\pm$ 3.3 \\
\hline
HOPS-408 & NA / NA & 37.2 $\pm$ 2.4 / 24.5 $\pm$ 1.3 & 26.4 $\pm$ 4.5 / 24.0 $\pm$ 8.9 \\
\hline

\end{tabular}
\begin{flushleft}
\tablenotetext{a}{For each source we give two estimates. The first value is the estimate for the blue-shifted  lobe and the second value is for  the red-shifted lobe. NA indicates the  Gaussian fit used to derive the opening angle was not good, it did not converge or there is simply no outflow lobe for that source. The opening angle is defined as the full-width quarter maximum (FWQM) of the fitted Gaussian.} \tablenotetext{b} {HOPS-150 is a binary system and it is contaminated by two collimated molecular outflows from two nearby sources. The Gaussian fits performed poorly for HOPS-150. \tablenotetext{c}{HOPS-194 opening angle is measured directly from the C$^{18}$O outflow cavity as shown in \autoref{fig:outflow_gallery}. The uncertainty is adopted to be 10\% of the measurement.}}
\end{flushleft}
\end{table}

\begin{table}[H]
\setlength{\tabcolsep}{6pt} 
\caption{Summary of the outflow opening angles with inclination correction}             
\label{table:opening_angles}      
\centering                          
\begin{tabular}{c c c c }        
\hline\hline                 
Source  & Traditional  & Momentum & Energy  \\    
& opening angle\tablenotemark{a} & opening angle\tablenotemark{a} & opening angle\tablenotemark{a} \\  
&  ($^\circ$)&  ($^\circ$) & ($^\circ$)  \\ 
\hline                        
HOPS-10 & 33.2 $\pm$ 2.9 / 55.5 $\pm$ 3.1 & 69.2 $\pm$ 11.0 / 42.7 $\pm$ 2.4 & 20.2 $\pm$ 2.3 / 57.2 $\pm$ 4.1 \\
\hline                                   
HOPS-11 & 64.4 $\pm$ 9.2 / 65.9 $\pm$ 1.9 & 16.8 $\pm$ 1.5 / 16.2 $\pm$ 1.8 & 11.8 $\pm$ 1.1 / 38.6 $\pm$ 2.2 \\
\hline
HOPS-13 & 44.7 $\pm$ 2.1 / 51.7 $\pm$ 4.9 & NA / NA & NA / NA \\
\hline
HOPS-127 & 65.8 $\pm$ 4.5 / 52.1 $\pm$ 3.4 & NA / 100.5 $\pm$ 7.3 & NA / 120.3 $\pm$ 26.3 \\
\hline
HOPS-129 & 63.6 $\pm$ 5.1 / NA & 36.8 $\pm$ 6.4 / 70.5 $\pm$ 5.2 & 31.3 $\pm$ 4.5 / 46.7 $\pm$ 6.0 \\
\hline
HOPS-130 & 65.3 $\pm$ 7.7 / NA & 88.6 $\pm$ 12.4 / NA & 141.6 $\pm$ 13.8 / NA \\
\hline
HOPS-134 & 101.6 $\pm$ 9.5 / NA & NA / NA & NA / NA \\
\hline
HOPS-135 & NA / 62.0 $\pm$ 4.4 & NA / 55.8 $\pm$ 3.5 & 34.0 $\pm$ 2.1 / 54.0 $\pm$ 3.6 \\
\hline
HOPS-150\tablenotemark{b} & NA / NA & 79.8 $\pm$ 7.4 / NA & NA / NA \\
\hline
HOPS-157 &  70.9 $\pm$ 5.5 / NA & NA / NA & NA / NA \\
\hline
HOPS-164 & 35.4 $\pm$ 2.3 / 22.8 $\pm$ 0.4 & 44.9 $\pm$ 2.5 / 25.3 $\pm$ 1.4 & 36.1 $\pm$ 1.4 / 20.2 $\pm$ 0.5 \\
\hline
HOPS-166 & NA / NA & { NA} / { NA} & NA / 116.1 $\pm$ 33.9 \\
\hline
HOPS-169 & 46.7 $\pm$ 3.8 / 30.3 $\pm$ 1.0 & 22.3 $\pm$ 1.4 / 23.3 $\pm$ 2.1 & 23.9 $\pm$ 1.7 / 15.1 $\pm$ 0.8 \\
\hline
HOPS-177 & 127.5 $\pm$ 16.0 / 81.7 $\pm$ 10.1 & NA / NA & 102.6 $\pm$ 4.4 / NA \\
\hline
HOPS-185 & 108.2 $\pm$ 12.9 / 88.0 $\pm$ 10.5 & NA / 93.3 $\pm$ 7.4 & NA / 93.8 $\pm$ 6.7 \\
\hline
HOPS-191 & 108.5 $\pm$ 8.0 / 72.8 $\pm$ 7.8 & 78.9 $\pm$ 4.3 / 66.9 $\pm$ 5.2 & NA / NA \\
\hline
HOPS-194 & NA / 65.1 $\pm$ 5.0 & NA / NA & NA / NA \\
\hline
HOPS-198 & 69.1 $\pm$ 4.4 / NA & 59.3 $\pm$ 3.6 / 65.5 $\pm$ 3.6 & 33.9 $\pm$ 4.0 / NA \\
\hline
HOPS-200 & 134.6 $\pm$ 19.4 / NA & NA / NA & NA / NA \\
\hline
HOPS-355 & 11.9 $\pm$ 0.6 / 17.0 $\pm$ 0.7 & 15.4 $\pm$ 0.8 / 13.3 $\pm$ 0.9 & 20.4 $\pm$ 2.6 / 12.9 $\pm$ 1.6 \\
\hline
HOPS-408 & NA / NA & 32.2 $\pm$ 2.1 / 21.1 $\pm$ 1.1 & 22.7 $\pm$ 3.9 / 20.7 $\pm$ 7.6 \\
\hline

\end{tabular}
\begin{flushleft}
\tablenotetext{a}{For each source we give two estimates. The first value is the estimate for the blue-shifted  lobe and the second value is for the red-shifted lobe. NA indicates the  Gaussian fit used to derive the opening angle was not good, it did not converge or there is simply no outflow lobe for that source. The opening angle is defined as the full-width quarter maximum (FWQM) of the fitted Gaussian. The errors are from the uncertainty of the Gaussian fit.} \tablenotetext{b} {HOPS-150 is a binary system and it is contaminated by two collimated molecular outflows from two nearby sources. The Gaussian fits performed poorly for HOPS-150. \tablenotetext{c}{HOPS-194 opening angle is measured directly from the C$^{18}$O outflow cavity as shown in \autoref{fig:outflow_gallery}. The uncertainty is adopted to be 10\% of the measurement.}}
\end{flushleft}
\end{table}

\end{appendix}

\end{document}